%% file: main.tex
\documentclass[11pt,letterpaper]{article}
\usepackage{graphicx} 
\usepackage{auxlib}
\usepackage{microtype}
\allowdisplaybreaks[1]
\pdfoutput=1

\title{Santa Claus and Weighted Nash Social Welfare with Subadditive Valuations}
\author{Yuhang Chen\thanks{School of Computer Science, Nanjing University, yuhangchen@smail.nju.edu.cn} \and Yuda Feng\thanks{School of Computer Science, Nanjing University, yudafeng@smail.nju.edu.cn} \and Weijiang Hu\thanks{School of Computer Science, Nanjing University, huweijiang@smail.nju.edu.cn} \and Shi Li\thanks{School of Computer Science, Nanjing University, shili@nju.edu.cn} \and Ruilong Zhang\thanks{Department of Computer Science, City University of Hong Kong (Dongguan), ruilong.zhang@cityu-dg.edu.cn}}
\date{}

\begin{document}

\maketitle

\begin{abstract}
\input{abstract}
\end{abstract}


\clearpage
{\small
\tableofcontents
}
\clearpage

\input{intro}

\input{prelim}

\input{max-min}

\input{nsw}
\input{oracle-lower-bounds}

\clearpage
\section*{AI Disclosure}
\addcontentsline{toc}{section}{AI Disclosure}
The main algorithmic ideas in this paper were contributed by GPT-5.6, GPT-6, GPT-5.6 Sol, and GPT-6-Astra (OpenAI), referred to below as AI. The human authors' only direct contribution to the mathematical constructions was a hard instance for the configuration LP with XOS valuations; AI subsequently extended the analysis to subadditive valuations.
The authors initially used AI to obtain an $O(\sqrt n\,\mathrm{polylog}\,n)$-approximation for weighted Nash social welfare with XOS valuations. From this algorithm, the authors observed a connection between the techniques for XOS Nash social welfare and the Santa Claus problem. They then used AI to extend the approach to Santa Claus and to subadditive valuations. Through long-term interactions with AI, the authors finally obtained the current $O(\sqrt n)$-approximation guarantees for both problems.
Afterward, the authors began to understand, simplify, and build connections to the literature's techniques; this led to the current paper.
In addition to these technical contributions, GPT-6-Astra (OpenAI) and Claude-Opus-5.5 (Anthropic) assisted with drafting and revising the manuscript.
The authors take full responsibility for the correctness, originality, references, and attribution of all content.

\section*{Acknowledgment}
\addcontentsline{toc}{section}{Acknowledgment}
The work of Ruilong Zhang is supported by the National Natural Science Foundation of China (No. 62602540)
and City University of Hong Kong (Dongguan) start-up funding (No. R002026G003).
The work of Yuhang Chen, Yuda Feng, Weijiang Hu and Shi Li is supported by State Key Laboratory for Novel Software Technology, New Cornerstone Science Foundation, and Fundamental and Interdisciplinary Disciplines Breakthrough Plan of the Ministry of Education of China (No.JYB2025XDXM118).

\clearpage
\phantomsection
\addcontentsline{toc}{section}{References}
\bibliographystyle{alpha}
\bibliography{main}

\clearpage

\appendix
\input{appendix}

\end{document}

%% file: abstract.tex
We study two fairness objectives for allocating $m$ indivisible items among $n$ agents with subadditive valuations: the max-min objective, also known as the Santa Claus problem, which maximizes the smallest value of an agent, and the weighted Nash social welfare, which maximizes the weighted geometric mean of the agents' values.

For the max-min allocation problem, we give a randomized polynomial-time $O(\sqrt n)$-approximation algorithm for subadditive valuations with demand queries, improving the $2n-1$ bound of Khot and Ponnuswami~\cite{KP07}.
For additive valuations and for submodular valuations with value queries, we give simpler $O(\sqrt n)$-approximation algorithms. For additive valuations, this improves the $O(\sqrt{n\log n})$ bound of Saha and Srinivasan~\cite{SS18}.
For these two classes, $O(m^\varepsilon)$-approximation algorithms are also known for every constant $\varepsilon>0$~\cite{CCK09,BMR24}.
Our ratios depend only on $n$ and match, up to constants, the $\Omega(\sqrt n)$ integrality gap of the configuration LP~\cite{BS06}.

For the weighted Nash social welfare problem with subadditive valuations and demand queries, we give an $O(\Gamma(w))$-approximation algorithm, where $\Gamma(w)=\exp\bigl((\ln n-H(w))/2\bigr)$ and $H(w)$ is the entropy of the weight vector $w$.
The factor $\Gamma(w)$ equals $1$ for equal weights and is at most $\sqrt n$ for all weights.
We show that this ratio is tight for algorithms analyzed against the configuration LP: for every weight vector, the configuration LP has an integrality gap of $\Theta(\Gamma(w))$, even for XOS valuations, and this gap is $\Omega(\sqrt n)$ when one agent has almost all the weight.
For submodular valuations, the configuration LP has a constant integrality gap for all weights~\cite{BFHLZ26}.

We also prove information-theoretic oracle query lower bounds, already for XOS valuations and allowing value, demand, and XOS supporting-clause queries together. For every fixed query exponent $D>0$, at most $n^D$ adaptive queries cannot guarantee approximation factors better than $\Omega_D(\sqrt{n/\ln n})$ for max-min allocation and $\Omega_D(\Gamma(w)/\sqrt{\ln n})$ for each prescribed weight vector, either with success probability at least $2/3$ or in expected welfare. The hard instances have $m=\Theta(n)$ items. The weighted lower bound is vacuous below one, and in particular does not rule out constant approximations for equal weights. Thus the query bounds leave a factor $O(\sqrt{\ln n})$, while the configuration-LP gaps match our upper bounds up to constants.

Both algorithms build on the random matching idea for large items, which has been used previously for the two problems: max-min allocation in \cite{AS10,SS18} and weighted Nash social welfare in \cite{FHLZ25}. An item matched as a large item can also belong to configurations that use it as a small item. We bound the resulting loss across these configurations by charging it to the at most $n$ matched items.

%% file: intro.tex
\section{Introduction}
\label{sec:intro}

We study the problem of allocating a set $M$ of $m$ indivisible items to a set $N$ of $n$ agents.
Each agent $i\in N$ has a normalized monotone valuation $v_i:2^M\to\R_{\ge0},$\footnote{We say $v_i$ is normalized if $v_i(\emptyset) = 0$.} and an allocation $A=(A_i)_{i\in N}$ consists of pairwise disjoint bundles $A_i\subseteq M$; without loss of generality, it is a partition of $M$.
We consider two well-known welfare objectives: 
\begin{itemize}
    \item \emph{Santa Claus}, where we aim to maximize the value received by the worst-off agent $\min_{i\in N}v_i(A_i)$.
    \item \emph{Weighted Nash social welfare (NSW)}, where, given positive weights $(w_i)_{i\in N}$ with $\sum_{i\in N}w_i=1$, we aim to maximize the weighted geometric mean of the agents' values \(\mathrm{NSW}_w(A):=\prod_{i\in N}v_i(A_i)^{w_i}.\)
\end{itemize}


With equal weights, this objective reduces to the usual Nash social welfare, the geometric mean of the agents' values.
This unweighted objective has axiomatic foundations in the bargaining work of Nash~\cite{Nash50} and the social welfare framework of Kaneko and Nakamura~\cite{KN79}.
Unequal weights allow agents to have different entitlements, for example when they represent groups of different sizes.
For any fixed weights, multiplying an agent's valuation by a positive constant leaves the ranking of allocations unchanged, so the choice of an optimal allocation does not depend on the units used to measure the agents' values.

We focus on subadditive valuations, which satisfy $v_i(S\cup T)\le v_i(S)+v_i(T)$ for all $S,T\subseteq M$.
Since such valuations may not have succinct descriptions, algorithms access them through value and demand oracles.
With only value-oracle access, both objectives have strong approximation lower bounds. Barman, Bhaskar, Krishna, and Sundaram~\cite[Section~5]{BBKS20} proved that, for every fixed $0<\varepsilon<1$, an $O(n^{1-\varepsilon})$-approximation requires exponentially many value queries, even for XOS valuations and, in the NSW case with equal weights.
We therefore use demand queries for general subadditive valuations. 
Note that a demand oracle can also simulate value queries~\cite{BN09}.
For the special cases of additive and submodular valuations, simpler algorithms are possible, and they may use different oracles; see \cref{sec:prelim}.

\paragraph{Max-min allocation.}
For additive valuations, Bansal and Sviridenko~\cite{BS06} showed that the configuration LP has an integrality gap of $\Omega(\sqrt n)$, where $n$ is the number of agents.
Asadpour and Saberi~\cite{AS10} rounded the configuration LP to obtain an $O(\sqrt n\log^3 n)$-approximation. Saha and Srinivasan~\cite{SS18} improved the ratio to $O(\sqrt{n\log n})$ using bipartite dependent rounding.
Chakrabarty, Chuzhoy, and Khanna~\cite{CCK09} gave an $O(m^\varepsilon)$-approximation in time $m^{O(1/\varepsilon)}$ for each fixed $\varepsilon>0$, where $m$ is the number of items. Their tradeoff also gives a polylogarithmic approximation in quasi-polynomial time.
Bamas, Morell, and Rohwedder~\cite{BMR24} obtained the corresponding guarantees for submodular valuations with value queries.%
\footnote{Chakrabarty, Chuzhoy, and Khanna~\cite{CCK09} denote the number of items by $n$. For fixed $\varepsilon$, the layered graph of Bamas, Morell, and Rohwedder~\cite{BMR24} has $O(1/\varepsilon)$ layers and $O(m+n)$ vertices per layer in our notation. Positive optimum requires $m\ge n$, so rescaling $\varepsilon$ gives the stated $O(m^\varepsilon)$ bound.}
Goemans, Harvey, Iwata, and Mirrokni~\cite{GHIM09} obtained an earlier bound for submodular valuations by approximating each valuation by the square root of an additive function.
Khot and Ponnuswami~\cite{KP07} gave a $(2n-1)$-approximation for subadditive valuations.
Barman, Bhaskar, Krishna, and Sundaram~\cite[Theorem~9]{BBKS20} obtained an $O(n)$-approximation using only value queries, as part of their result for $p$-mean welfare.

On the negative side, Bez{\'a}kov{\'a} and Dani~\cite{BD05} proved that approximating max-min allocation within any constant factor better than $2$ is NP-hard, even for additive valuations.
For subadditive valuations, the construction of Ezra, Feldman, Neyman, Talgam-Cohen, and Weinberg~\cite{EFNTW19} also implies that any such approximation with success probability at least $2/3$ requires $2^{\Omega(\sqrt m)}$ demand queries, even for two agents.%

When all agents share the same valuation function, the Santa Claus optimum equals their common \emph{maximin share} (MMS)~\cite{Budish11}, the largest value an agent can guarantee by partitioning the items into $n$ bundles and receiving a least-valued bundle. Thus, approximating Santa Claus is equivalent to finding an approximate MMS allocation in this setting.
For identical additive valuations, Woeginger~\cite{Woe97} gave a polynomial-time approximation scheme.
For identical submodular valuations, Boehmer~\cite{Boe26} gave a polynomial-time $5/2$-approximation using value queries.
For identical subadditive valuations, Barman and Sundaram~\cite{BS20} gave a polynomial-time $40$-approximation using demand queries, which also applies to identical XOS valuations.

\paragraph{Weighted Nash social welfare.}
With equal weights, constant-factor approximations are known for additive~\cite{CG18,BKV18}, submodular~\cite{LV21,GHLVV26,BFHLZ26}, and subadditive valuations~\cite{DLRV24}, the last result uses demand queries.
For XOS valuations and equal weights, Kulkarni, Murhekar, Rivkin, and Khuller~\cite{KMRK27} gave an $O(1)$-approximation using both demand and XOS queries.
For general weights and additive valuations, Brown, Laddha, Pittu, and Singh~\cite{BLPS24} gave a $5\exp(2D_{\mathrm{KL}}(w\,\|\,u))$-approximation, where $u$ is the uniform distribution on $N$, this ratio can be polynomial in $n$.
Feng and Li~\cite{FL24} then gave an $(e^{1/e}+\varepsilon)$-approximation for additive valuations using the configuration LP.
For submodular valuations with arbitrary weights, Feng, Hu, Li, and Zhang~\cite{FHLZ25} gave the first polynomial-time constant-factor approximation. Bei, Feng, Hu, Li, and Zhang~\cite{BFHLZ26} improved the ratio from $233+\varepsilon$ to $3.56+\varepsilon$.
For subadditive valuations and arbitrary weights, Barman, Bhaskar, Krishna, and Sundaram~\cite{BBKS20} gave an $O(n)$-approximation using value queries.

On the hardness side, Lee~\cite{Lee17} proved that maximizing NSW is APX-hard even for additive valuations and equal weights.
Garg, Kulkarni, and Kulkarni~\cite[Theorem~4.1]{GKK23} proved that approximating NSW within any constant factor better than $e/(e-1)$ is NP-hard, even for equal weights and identical submodular valuations.
For XOS valuations and equal weights, Barman, Krishna, Kulkarni, and Narang~\cite[Theorem~2]{BKKN24} proved that any constant factor better than $e/(e-1)$ requires exponentially many queries, even when both demand and XOS queries are available.
For subadditive valuations, the factor-$2$ query lower bound derived above from~\cite{EFNTW19} also applies to NSW, even for two agents with equal weights.

\subsection{Our Results}
\label{sec:results}

All our algorithms are based on \emph{Configuration LP} techniques.
We use the configuration LP for max-min allocation~\cite{BS06} and the logarithmic configuration LP used for weighted Nash social welfare with additive~\cite{FL24} and submodular~\cite{FHLZ25} valuations. Both assign each agent a probability distribution over bundles, called \emph{configurations}, with total mass at most one on the configurations containing any given item.
\cref{tab:results} summarizes the algorithmic guarantees and configuration-LP gaps; \cref{tab:oracle-lower-bounds} records the separate oracle query lower bounds.
For max-min allocation, we obtain the following.

\begin{theorem}
\label{thm:intro-maxmin}
There are randomized polynomial-time $O(\sqrt n)$-approximation algorithms for max-min allocation with
\begin{enumerate}[label=(\roman*),ref=\roman*,leftmargin=*]
  \item\label[resultpart]{part:mm-additive} additive valuations,
  \item\label[resultpart]{part:mm-submodular} submodular valuations using value queries, and
  \item\label[resultpart]{part:mm-subadditive} subadditive valuations using demand queries.
\end{enumerate}
\end{theorem}

The approximation ratios do not depend on the number $m$ of items. 
Bansal and Sviridenko~\cite{BS06} showed that the configuration LP has an integrality gap of $\Omega(\sqrt{n})$, implying that, relative to the configuration LP of Bansal and Sviridenko~\cite{BS06}, our approximation is optimal up to a constant factor.

For subadditive valuations, \cref{thm:intro-maxmin} improves the previous ratio of $2n-1$~\cite{KP07}.
For additive valuations, our algorithm removes the remaining $\sqrt{\log n}$ factor from the LP rounding bound of Saha and Srinivasan~\cite{SS18}.
For additive and submodular valuations, the $O(m^\varepsilon)$-approximations of~\cite{CCK09,BMR24} give better guarantees when $m$ is polynomial in $n$ and $\varepsilon$ is a sufficiently small constant.
When $m$ is superpolynomial in $n$, $m^\varepsilon$ exceeds $\sqrt n$ by more than any constant factor for every fixed $\varepsilon>0$, so our ratio is better among polynomial-time algorithms.%

\medskip

For the weighted Nash social welfare problem, our ratio depends on the entropy $H(w)=-\sum_{i\in N}w_i\ln w_i$ of the weights%
\footnote{Throughout the paper, $\ln$ denotes the natural logarithm. Inside asymptotic notation such as $O(\cdot)$ and $\Omega(\cdot)$, the base of a logarithm only affects constant factors, and we write $\log$.}
through
\[
  \Gamma(w):=\prod_{i\in N}(nw_i)^{w_i/2}=\exp\left(\frac{\ln n-H(w)}{2}\right).
\]

\begin{theorem}
\label{thm:intro-nsw}
For subadditive valuations and positive rational weights $w$ with $\sum_{i\in N}w_i=1$, there is a randomized polynomial-time algorithm using demand queries that returns an allocation $A$ with
\[
  \E\bigl[\mathrm{NSW}_w(A)\bigr]\ge\frac{\mathrm{OPT}_{\mathrm{NSW}}}{O(\Gamma(w))},
\]
where $\mathrm{OPT}_{\mathrm{NSW}}$ is the maximum weighted Nash social welfare.
\end{theorem}

Since $0\le H(w)\le\ln n$, we have $1\le\Gamma(w)\le\sqrt n$, and also $\Gamma(w)\le\sqrt{n\max_iw_i}$.
When all weights are $\frac1n$, we have $\Gamma(w)=1$, and \cref{thm:intro-nsw} gives a constant-factor approximation for subadditive valuations with demand queries, as in~\cite{DLRV24}.
More generally, the ratio is constant whenever $H(w)\ge\ln n-O(1)$, for example when all weights are within constant factors of $1/n$.
Equivalently, $\Gamma(w)=\exp\bigl(D_{\mathrm{KL}}(w\,\|\,u)/2\bigr)$, so for subadditive valuations our ratio is, up to a constant factor, the fourth root of the ratio $5\exp(2D_{\mathrm{KL}}(w\,\|\,u))$ that~\cite{BLPS24} obtained for additive valuations.

Our algorithm also compares with the configuration LP, which maximizes the weighted logarithmic value $L(y)=\sum_{i,S}w_iy_{i,S}\ln v_i(S)$; see \cref{subsec:nsw-application}.
The dependence on the weights is necessary for this LP:

\begin{theorem}
\label{thm:intro-gap}
Let $L_{\mathrm{LP}}$ be the optimum of the configuration LP for weighted Nash social welfare.
For every positive weight vector $w$, the worst-case ratio $e^{L_{\mathrm{LP}}}/\mathrm{OPT}_{\mathrm{NSW}}$ over XOS instances with positive optimum is $\Theta(\Gamma(w))$.
The same holds for subadditive instances.
\end{theorem}

Thus \cref{thm:intro-nsw} is best possible up to a constant factor among algorithms that compare with the configuration LP, for every weight vector.
In particular, the integrality gap can be as large as $\Omega(\sqrt n)$: if one agent has weight $1-1/n$ and every other agent has weight $1/(n(n-1))$, then $\Gamma(w)=(n-1)^{1/2-1/n}\ge\sqrt{n-1}/2$.
Over all weight vectors, the worst-case integrality gap is therefore $\Theta(\sqrt n)$, and our worst-case ratio $O(\sqrt n)$ is tight for this LP, as it is for max-min allocation.
For submodular valuations, Bei, Feng, Hu, Li, and Zhang~\cite{BFHLZ26} bounded the integrality gap of this LP by $3.45$ for every weight vector. Thus the worst-case gap over weights is constant for submodular valuations and $\Theta(\sqrt n)$ for XOS valuations.

\begin{table}[htbp]
\centering
\small
\renewcommand{\arraystretch}{1.2}
\begin{tabular}{@{}llllll@{}}
\hline
\textbf{Problem} & \textbf{Valuations} & \textbf{Oracles} & \textbf{Prior bounds} & \textbf{This paper} & \textbf{LP gap}\\
\hline
Max-min & Additive & Explicit & \begin{tabular}[t]{@{}l@{}}$O(\sqrt{n\log n})$~\cite{SS18}\\ $O(m^\varepsilon)$~\cite{CCK09}\end{tabular} & $O(\sqrt n)$ (\cref{thm:intro-maxmin}) & $\Theta(\sqrt n)$\\
Max-min & Submodular & Value & \begin{tabular}[t]{@{}l@{}}$2n-1$~\cite{KP07}\\ $O(m^\varepsilon)$~\cite{BMR24}\end{tabular} & $O(\sqrt n)$ (\cref{thm:intro-maxmin}) & $\Theta(\sqrt n)$\\
Max-min & Subadditive & Demand & $2n-1$~\cite{KP07} & $O(\sqrt n)$ (\cref{thm:intro-maxmin}) & $\Theta(\sqrt n)$\\
Weighted NSW & Subadditive & Demand & $O(n)$~\cite{BBKS20} & $O(\Gamma(w))$ (\cref{thm:intro-nsw}) & $\Theta(\Gamma(w))$\\
\hline
\end{tabular}
\caption{Our results and prior approximation guarantees. The oracle column describes our algorithms. Both subadditive objectives admit prior $O(n)$-approximations using only value queries~\cite{BBKS20}.
For max-min allocation with additive and submodular valuations, the $O(m^\varepsilon)$ ratios of~\cite{CCK09,BMR24} hold in time $m^{O(1/\varepsilon)}$ for every constant $\varepsilon>0$; they are incomparable with ours.
The last column gives the integrality gap of the configuration LP.
Its lower bounds are due to~\cite{BS06} for max-min allocation, already with additive valuations, and to \cref{thm:nsw-gap} for weighted Nash social welfare, already with XOS valuations; the upper bounds follow from our algorithms.
Since $\Gamma(w)\le\sqrt n$, with equality up to a constant factor when one agent has almost all the weight, the worst-case gap over all weights is $\Theta(\sqrt n)$ for weighted Nash social welfare as well.
For weighted Nash social welfare with additive and submodular valuations, constant-factor approximations are known~\cite{FL24,FHLZ25,BFHLZ26}.}
\label{tab:results}
\end{table}

\paragraph{Oracle query lower bounds.}
The LP gaps above concern a particular relaxation. We also prove limitations on every algorithm with a bounded number of queries, independently of its computation or use of that LP. The following overview is made precise in \cref{thm:oracle-maxmin,thm:oracle-weighted-nsw}.

\begin{theorem}\label{thm:intro-oracle-lower-bounds}
  Fix $D>0$. There are $c_D>0$ and $n_D$, depending only on $D$, such that for every $n\ge n_D$, no randomized algorithm using at most $n^D$ adaptive value, demand, and XOS supporting-clause queries in total can guarantee, on all normalized monotone XOS instances,
  \[
    c_D\sqrt{n/\ln n}\quad\text{for max-min allocation},\qquad
    c_D\Gamma(w)/\sqrt{\ln n}\quad\text{for weighted NSW}.
  \]
  Both assertions hold for success probability at least $2/3$ and for guarantees in expected welfare. The weighted assertion holds for every prescribed positive rational weight vector, which the algorithm may know. For each algorithm, the construction gives a fixed instance with exactly $n$ agents, $n\le m\le20n/19$, positive optimum, and legal baseline-preferring oracles. Demand prices may even be arbitrary real numbers.
\end{theorem}

The displayed lower bound is vacuous below one. In particular, equal weights have $\Gamma(w)=1$, so this result does not exclude their constant-factor approximation. For bounded-entropy weights, it gives $\Omega_D(\sqrt{n/\ln n})$, matching the worst-case dependence on $n$ up to a factor $O(\sqrt{\ln n})$. Since XOS valuations are subadditive, these lower bounds also apply to subadditive instances with ordinary demand queries. This implication does not provide or require a supporting-clause oracle for general subadditive valuations. The query budget is polynomial in $n$ and $m=\Theta(n)$; its relation to the encoding length of the prescribed weights is discussed in \cref{subsec:oracle-polynomial}.

\begin{table}[htbp]
\centering
\small
\renewcommand{\arraystretch}{1.2}
\begin{tabular}{@{}lll@{}}
\hline
\textbf{Objective} & \textbf{Query lower bound} & \textbf{Hard valuation class}\\
\hline
Max-min & $\Omega_D(\sqrt{n/\ln n})$ & XOS\\
Weighted NSW, prescribed $w$ & $\Omega_D(\Gamma(w)/\sqrt{\ln n})$ & XOS\\
\hline
\end{tabular}
\caption{Information-theoretic approximation lower bounds for a total budget of $n^D$ queries, for fixed $D>0$ and sufficiently large $n$. Value, demand (even at arbitrary real prices), and XOS supporting-clause queries are all allowed. The bounds hold for success probability $2/3$ and in expected welfare, using legal baseline-preferring oracles. The weighted expression is vacuous below one. These are query bounds, separate from the LP gaps in \cref{tab:results}; they imply the corresponding lower bounds for subadditive valuations with ordinary demand queries.}
\label{tab:oracle-lower-bounds}
\end{table}

\subsection{Technical Overview}
\label{sec:overview}

Our basic rounding procedures first assign individual items to agents through a random matching and then allocate bundles that exclude the matched items. Both steps are guided by a configuration-LP solution, where $y_{i,S}$ is the probability assigned to configuration $S$ of agent $i$ (\cref{sec:prelim}). For max-min allocation, we follow Asadpour and Saberi~\cite{AS10} and Saha and Srinivasan~\cite{SS18} in separating configurations that can satisfy an agent with one large item from those that require a bundle of small items. For weighted Nash social welfare, we follow Feng, Hu, Li, and Zhang~\cite{FHLZ25} in selecting a most valuable singleton from each configuration as its head and treating the remaining items as its tail (see \eqref{eq:nsw-heads}). The matching assigns the designated singletons with their LP marginals (\cref{subsec:mm-algorithm,subsec:nsw-matching}).

Allocating bundles requires resolving overlaps between configurations while preserving their value. For subadditive valuations, we use a conditional form of Feige's guiding-tree rounding~\cite{Feige09} (\cref{lem:mm-service}). After scaling the configuration masses by a constant factor and adding empty configurations, each run selects a configuration for every agent and produces disjoint subbundles. Conditional on any selected configuration, an agent receives a constant fraction of its value with constant probability. Following Feige and Huang~\cite{FH25}, we combine independent runs by assigning random colors to items and retaining each item only in the run corresponding to its color. This gives agents repeated opportunities to obtain valuable bundles while preserving feasibility. The color probabilities reflect the objective: the subadditive max-min algorithm uses uniform probabilities to reach a common value threshold (\cref{lem:mm-colored-service}), whereas the NSW algorithm uses geometrically decreasing probabilities to control expected logarithmic welfare (\cref{lem:nsw-geometric}).

The analysis exploits the capacity constraint linking singleton assignments to the remaining configurations. Let $b_j$ be the total singleton mass of item $j$ and let $f_j=1-b_j$. The total LP mass using $j$ in small configurations or tails is at most $f_j$ (see \eqref{eq:mm-residual} and \eqref{eq:nsw-tail-capacity}, respectively). Thus, an item that is likely to be matched can occur in only a small amount of residual configuration mass. Charging $b_j/f_j$ to each such occurrence, weighted by its LP mass, costs at most $b_j$ per item and at most $\sum_j b_j\le n$ overall (\eqref{eq:mm-penalty-budget} for max-min and \eqref{eq:nsw-odds-budget} for NSW); positive-mass occurrences necessarily have $f_j>0$. For NSW, this budget controls the aggregate loss in the configuration benchmark caused by removing matched items (\cref{lem:nsw-aggregate}). For max-min allocation, it controls the sum of conditional failure probabilities weighted by the LP masses (see \eqref{eq:mm-budget}). We first develop the rounding lemma that makes this latter guarantee precise.

\paragraph{Max-min allocation: the rounding lemma.}
Suppose every configuration $S$ of agent $i$ comes with an additive function $a^{i,S}:2^S \to \R_{\geq 0}$ dominated by $v_i$, and is either a single item with $a^{i,S}$-value at least $\tau$ (the item is a large item), or consists of items with $a^{i,S}$-values less than $\tau$ (the items are small items).
Our rounding lemma (\cref{lem:mm-rounding}) selects configurations with their LP marginals and gives every agent coefficient value at least $\tau$ with probability at least $3/4$, provided that
\[
  \sum_{i\in N}\sum_{S\text{ small}}y_{i,S}\min\left\{1,\left(\frac{32\tau}{\mu_{i,S}}\right)^3\right\}\le\frac14,
  \qquad
  \text{where } \mu_{i,S}=\sum_{j\in S}a^{i,S}_jf_j.
\]
The baseline $\mu_{i,S}$ is the $a^{i,S}$-value of $S$ that is expected to survive the singleton large-item matching.
We use bipartite dependent rounding~\cite{GKPS06} and its same-side negative correlation~\cite{SS18} to couple the assignment of large items with configuration selection.
After the matching, the agents that select other configurations share the surviving items equally, and Shmoys--Tardos rounding~\cite{ST93} makes the allocation integral.
Saha and Srinivasan~\cite[Section~4.1]{SS18} also use a large-item matching followed by an additive rounding step, but discard items with low survival probability or high realized congestion and scale the remaining fractional allocation by a common factor.
Writing the large-item threshold as $T/\lambda$ and the survival-probability cutoff as $\eta$, a sufficient condition for controlling failures caused by discarding items is $\eta=O(\lambda/n)$.
Obtaining polynomially small failure bounds from concentration of the surviving value, followed by a union bound over agents, uses $\eta\lambda=\Omega(\log n)$.
Balancing these conditions gives $\lambda=\Theta(\sqrt{n\log n})$, so their threshold is chosen to be $\Theta(T/\sqrt{n\log n})$.
Our allocation is feasible directly: each surviving item is divided among the agents whose selected configurations contain it.
This avoids both the discarding thresholds and the common scaling factor.
A Chernoff bound for the surviving value and a third-moment bound for the overlap with other selected configurations show that, conditional on selecting $S$, it fails with probability at most $\min\{1,(32\tau/\mu_{i,S})^3\}$ (see \eqref{eq:mm-local-tail}).
The key analytical difference is that we bound the total failure probability after weighting these conditional bounds by the LP marginals.
Thus, configurations with small surviving baseline may have large conditional failure probabilities, provided that their total LP mass is small.
For submodular valuations, the marginal-value construction of Dobzinski, Nisan, and Schapira~\cite{DNS10} gives an additive lower bound that is tight on each configuration (\cref{subsec:mm-exact}).
With supporting clauses scaled to total $T$, the Cauchy--Schwarz inequality gives $\mu_{i,S}\ge T/(1+P(S)/(128\lceil\sqrt n\rceil))$ for the penalty $P(S)=\sum_{j\in S}b_j/f_j$ (\cref{subsec:mm-exact}).
Using the penalty budget above, the threshold $\tau=T/(128\lceil\sqrt n\rceil)$ balances configurations with small and large penalties.
In particular, a small surviving baseline requires a large penalty, so the penalty bound controls the total LP mass of configurations with large failure probabilities.
This accounting avoids the uniform per-agent concentration requirement above and permits a large-item threshold of order $T/\sqrt n$, removing the additional $\sqrt{\log n}$ loss in the threshold used by Saha and Srinivasan~\cite{SS18}.
This gives the ratio $O(\sqrt n)$ for additive and submodular valuations (\cref{thm:mm-exact}).

\paragraph{Max-min allocation: extending to subadditive valuations.}
We extend the preceding approach using the same rounding lemma; the new task is to construct additive lower bounds whose baselines satisfy its budget condition.
A subadditive valuation need not have an additive lower bound that is tight on a given configuration.
Dobzinski~\cite{Dobzinski07} obtained additive lower bounds for subadditive valuations from demand queries, losing a logarithmic factor. We compute such a bound on each configuration using the geometric price grid of Balcan, Blum, and Mansour~\cite{BBM08} (\cref{lem:mm-log-support}). Applying the preceding argument to these lower bounds gives only an $O(\sqrt n\log m)$ ratio.
Instead, we scale the price of each item $j$ by $1/f_j$ (\cref{lem:mm-weighted-support}).
The resulting baseline satisfies \eqref{eq:mm-envelope}, which has a logarithmic term and a square-root term.
Either the logarithmic term gives a sufficiently large baseline, or the square-root term forces a large penalty; the cubic failure bound and penalty budget handle these two cases.
These bounds verify the budget condition of \cref{lem:mm-rounding} and give an $O(\sqrt n)$-approximation when $m\le128n^2$ (\cref{cor:mm-local}).

For larger $m$, we select a core of $128n^2$ items in $128n$ rounds, in each of which every agent takes its most valuable remaining item, and we set aside one random round as a backup (\cref{subsec:mm-general}).
Agents whose backup item is valuable take it, and every other agent values each item outside the core at most $O(T/\sqrt n)$.
We apply this bounded-item rounding procedure inside the remaining core, preserving the original configuration marginals to determine which agents need an allocation outside it.
Outside the core, we use the guiding-tree rounding and uniform coloring described above. We set aside a random half of the outside items to repair agents whose first allocation falls short of the target (\cref{thm:mm-subadditive}).

\paragraph{Weighted Nash social welfare.}
We round a solution of the configuration LP, which maximizes $L(y)=\sum_{i,S}w_iy_{i,S}\ln v_i(S)$ and can be solved up to an additive constant with demand queries (\cref{lem:nsw-producer}).
We match every agent to a head, then we realize the LP head marginals as a limit of Gibbs matching distributions, using the maximum-entropy formulation~\cite{SV14}, and prove the reference-set coupling in \cref{lem:nsw-matching}.
Independently, we allocate the tails using the repeated rounding described above, with geometrically decreasing color probabilities. We then remove the matched heads from the tail bundles.
Let $B$ be the set of matched heads, and let $D_{i,S}=v_i(\text{head of }S)+v_i(T_{i,S}\setminus B)$ be the buffered value of $S$ after the deletion of $B$.
The analysis has two estimates.
First, the deletion loss satisfies $\sum_{i,S}y_{i,S}\E\bigl[(v_i(S)/D_{i,S})^2\bigr]\le196n$ (\cref{lem:nsw-aggregate}), by the capacity budget above.
Second, the actual allocation achieves $\E[\ln\mathrm{NSW}_w(A)]\ge\sum_{i,S}w_iy_{i,S}\E[\ln D_{i,S}]-O(1)$ (\cref{lem:nsw-realize}).
The weight-dependent loss arises when we combine the two estimates.
We sample $(i,S)$ with probability $w_iy_{i,S}$ and apply Jensen's inequality (see \eqref{eq:nsw-entropy-average}):
\[
  L(y)-\sum_{i,S}w_iy_{i,S}\E[\ln D_{i,S}]
  =\frac12\left(\E\ln\frac{v_i(S)^2}{w_iD_{i,S}^2}-H(w)\right)
  \le\frac12\bigl(\ln(196n)-H(w)\bigr)
  =\ln\Gamma(w)+O(1).
\]
The factor $1/w_i$ cancels the sampling weight $w_i$, so only the unweighted bound $196n$ is needed.

\paragraph{The integrality gap.}
Our lower bound in \cref{thm:nsw-gap} adapts the group of unit-demand agents in the submodular gap construction of Bei, Feng, Hu, Li, and Zhang~\cite{BFHLZ26}, described in their full version~\cite[Section~6, Figure~4]{BFHLZfull}. We give the competing agent an XOS valuation and choose the group sizes as a function of the prescribed weights.
A special agent has $\ell$ disjoint rows of $\ell$ items each, and values a set by the largest number of its items in a single row, divided by $\ell$.
Each row also has $\ell-1$ blockers that value any single item of the row.
The configuration LP gives every agent value one, but in every allocation with positive welfare, the special agent receives at most one item from each row and hence value at most $1/\ell$.
Choosing $\ell\approx\sqrt{nw_i}$ for every agent $i$ of large weight, and using the remaining agents as blockers, gives a gap of at least $\Gamma(w)/\sqrt8$.
When one agent has weight $1-1/n$, a single gadget with $\ell\approx\sqrt{n/2}$ for this agent already gives the gap $\Omega(\sqrt n)$.

\paragraph{Hiding a useful additive clause.}
For the query lower bounds, we replace each square row construction by a rectangle with $R$ rows of $L$ items. The special agent values a set by its largest intersection with a row, divided by $L$. We then independently choose one item in each row and add a hidden additive clause supported on this transversal. The blockers force any positive-welfare allocation to leave at most one item per row for the special agent. Giving it the hidden transversal attains constant value, whereas a fixed output is unlikely to collect enough hidden items.

We choose oracle answers to agree with the baseline whenever possible. A fixed value or supporting-clause query reveals the hidden clause only if it strictly exceeds the baseline value. For demand queries, we compare the optimized profits of the hidden and row clauses; subtracting the surplus from negative prices gives the same exponential tail bound. Fixing an algorithm's random choices determines its entire baseline transcript. Before the first revelation, its execution on the hidden instance follows that transcript, allowing a union bound over the fixed queries. A separate tail bound on its fixed output handles success without revelation. This argument avoids assuming that the hidden items remain independent after conditioning on the transcript.

Taking $R=\Theta_D(L\ln n)$ makes revelations unlikely for $n^D$ queries. One gadget yields the max-min bound. For weighted NSW, we assign a gadget to each sufficiently heavy agent, with $L_i$ proportional to $\sqrt{nw_i/\ln n}$, and use other agents as distinct blockers without changing any weight. Multiplying the resulting losses gives the entropy expression $\Gamma(w)/\sqrt{\ln n}$ and a more detailed weight-profile bound in \cref{eq:oracle-weight-profile}.

\subsection{Other Related Works}
\label{sec:related}

\paragraph{Oracle lower bounds.}
The query bounds in \cref{sec:oracle-lower-bounds} are different from both the configuration-LP gaps and the value-only limitations discussed above. They continue to hold when demand and XOS supporting-clause queries are also available, using XOS instances and prescribed weights. The weighted bound depends on the full weight profile and is vacuous for equal weights; it does not replace the constant-factor demand-and-XOS lower bound for equal weights cited above from~\cite{BKKN24}.

\paragraph{Restricted max-min allocation.}
In the restricted assignment case with additive valuations, every item $j$ has a value $p_j$, and every agent values it at either $p_j$ or $0$.
Bansal and Sviridenko~\cite{BS06} gave an $O(\log\log n/\log\log\log n)$-approximation for this case. Feige~\cite{Feige08} proved a constant integrality gap for the configuration LP. Haeupler, Saha, and Srinivasan~\cite{HSS11} made the local lemma argument constructive to obtain a randomized constant-factor approximation. Annamalai, Kalaitzis, and Svensson~\cite{AKS17} later gave a combinatorial constant-factor approximation.
Bamas, Garg, and Rohwedder~\cite{BGR21} extended this setting to a common monotone submodular valuation $v$ with agent-specific eligibility sets $\Gamma_i\subseteq M$, so that $v_i(S)=v(S\cap\Gamma_i)$.
They gave a randomized polynomial-time $O(\log\log m)$-approximation by reducing the configuration-LP rounding problem to weighted hypergraph matching and applying the Lov{\'a}sz Local Lemma.
The algorithm uses value queries, its approximate LP separation is described in the full version~\cite[Appendix~B.1]{BGR20full}.

\paragraph{Utilitarian welfare.}
The utilitarian objective maximizes the total value $\sum_{i\in N}v_i(A_i)$.
For subadditive valuations with demand queries, Feige~\cite{Feige09} rounded the configuration LP to obtain a randomized $2$-approximation. Our algorithms adapt his guiding-tree rounding procedure.
Feige also showed that the integrality gap can approach $2$ and that it is NP-hard to approximate the problem within any constant factor better than $2$, even when all agents have the same valuation.
Feige and Vondr{\'a}k~\cite{FV10} showed that for every fixed $0<\varepsilon<1$, a $(2-\varepsilon)$-approximation is NP-hard even when each valuation depends on only $O_\varepsilon(1)$ items, so demand queries can be answered in polynomial time.
Ezra, Feldman, Neyman, Talgam-Cohen, and Weinberg~\cite{EFNTW19} showed that even for two agents, any fixed improvement over the factor of $2$ requires $2^{\Omega(\sqrt m)}$ bits of randomized communication.
With polynomially many value queries alone, Qiu and Weinberg~\cite{QW24} gave an $O(\sqrt{m/\log m})$-approximation. Their comparison gives a matching lower bound for deterministic algorithms already for XOS valuations, from the construction of Dobzinski, Nisan, and Schapira~\cite{DNS10}.

\paragraph{Maximin shares.}
For agents with different valuations, an $\alpha$-MMS allocation gives each agent at least an $\alpha$ fraction of that agent's own maximin share.
For additive valuations, Kurokawa, Procaccia, and Wang~\cite{KPW18} showed that exact MMS allocations need not exist, but $2/3$-MMS allocations always exist.
Garg and Taki~\cite{GT21} gave a strongly polynomial-time algorithm for $3/4$-MMS allocations, and Huang and Zhou~\cite{HZ25} obtained $(7/9-\varepsilon)$-MMS allocations for any $\varepsilon>0$ in time polynomial in $n$, $m$, and $1/\varepsilon$.
For submodular valuations, Ben Uziahu and Feige~\cite{BUF23} gave a polynomial-time algorithm for $10/27$-MMS allocations using value queries.
For XOS valuations, Feige and Grinberg~\cite{FG25} proved the existence of $4/17$-MMS allocations, their later work~\cite{FG26} proves the existence of $\alpha$-MMS allocations for some constant $\alpha>11/40$ when $n$ is sufficiently large.

For subadditive valuations, Seddighin and Seddighin~\cite{SS24MMS} proved the existence of $\Omega(1/(\log n\log\log n))$-MMS allocations.
Feige and Huang~\cite{FH25} improved this guarantee to $1/(14\log_2 n)$ by sharpening concentration bounds under independent item sampling, their sampling lemma gives the bound used in \cref{subsec:mm-tools}.
Seddighin and Seddighin~\cite{SS25MMS} subsequently obtained an $\Omega(1/(\log\log n)^2)$-MMS guarantee, which Feige~\cite{Feige25MMS} improved to $\Omega(1/\log\log n)$ by converting allocations with bounded item multiplicity into disjoint allocations with a smaller loss.
These general subadditive guarantees are existence results and do not establish polynomial-time demand-query algorithms.

\subsection{Roadmap}
\label{sec:roadmap}

\cref{sec:prelim} defines the model, the oracles, and the configuration LPs for both problems.
\cref{sec:max-min} presents our max-min algorithms: the rounding lemma and its proof (\cref{subsec:mm-setting,subsec:mm-algorithm,subsec:mm-analysis}), the algorithms for valuations with supporting clauses (\cref{subsec:mm-exact}), and the algorithm for subadditive valuations (\cref{subsec:mm-quadratic,subsec:mm-tools,subsec:mm-general}).
\cref{sec:nsw} presents our algorithm for weighted Nash social welfare and the integrality gap of its configuration LP.
\cref{sec:oracle-lower-bounds} proves the strong-oracle query lower bounds, including the prescribed-weight construction and its consequences.
The appendix contains the computation of the configuration solutions and the finite implementation of the Nash social welfare rounding.

%% file: prelim.tex
\section{Preliminaries}\label{sec:prelim}

We are given a set $N$ of $n$ agents and a set $M$ of $m$ items. Each agent $i$ has a valuation $v_i:2^M\to\mathbb R_{\ge0}$, which we always assume to be normalized, that is, $v_i(\emptyset)=0$, and monotone, that is, $v_i(S)\le v_i(S')$ whenever $S\subseteq S'$. An allocation is a collection $(A_i)_{i\in N}$ of pairwise disjoint bundles $A_i\subseteq M$, some items may remain unassigned. Since the valuations are monotone, giving the unassigned items to arbitrary agents does not decrease any agent's value. Hence, we may assume without loss of generality that every allocation is a partition of $M$.

The goal of max-min allocation is to maximize $\min_{i\in N}v_i(A_i)$. We use $\mathrm{OPT}_{\mathrm{MM}}$ to denote the optimal value.

In the weighted Nash social welfare problem, each agent $i\in N$ has a positive rational weight $w_i$, with $\sum_{i\in N}w_i=1$. For an allocation $A=(A_i)_{i\in N}$, its weighted Nash social welfare is
\begin{equation}\label{eq:nsw-welfare}
  \mathrm{NSW}_w(A)=\prod_{i\in N}v_i(A_i)^{w_i}.
\end{equation}
The goal is to maximize this objective, and we denote its optimal value by $\mathrm{OPT}_{\mathrm{NSW}}$. When $w_i=1/n$ for every agent, this is the unweighted Nash social welfare, the geometric mean of the agents' values.

\paragraph{Valuations and oracles.}
A valuation $v$ is \emph{additive} if $v(S)=\sum_{j\in S}v(\{j\})$ for every $S\subseteq M$. It is \emph{submodular} if $v(S\cup\{j\})-v(S)\ge v(S'\cup\{j\})-v(S')$ whenever $S\subseteq S'$ and $j\notin S'$. It is \emph{XOS} if it can be written as the maximum of finitely many nonnegative additive functions, called \emph{clauses}. It is \emph{subadditive} if $v(S\cup S')\le v(S)+v(S')$ for all $S,S'\subseteq M$. Every additive valuation is submodular, every submodular valuation is XOS, and every XOS valuation is subadditive~\cite{LLN06}.

A \emph{value query} for a set $S$ returns $v_i(S)$. Given prices $p\in\mathbb R_{\ge0}^M$, a \emph{demand query} returns a set $D$ maximizing $v_i(D)-\sum_{j\in D}p_j$, with ties broken arbitrarily. A \emph{restricted demand query} also specifies a set $S$ and maximizes the same objective over subsets of $S$.

For an XOS valuation, an \emph{XOS supporting-clause query} on $S$ returns a nonnegative additive clause $a$ with $a(S)=v_i(S)$ and $a(Q)\le v_i(Q)$ for every $Q\subseteq M$. In the lower-bound constructions, the returned clause is one of the clauses defining the valuation; the full representation is not supplied. All algorithms in the upper-bound sections use the nonnegative-price demand model above. Only \cref{sec:oracle-lower-bounds} allows negative demand prices, thereby proving a lower bound in a stronger model.

\paragraph{Approximation and query guarantees.}
For either welfare objective $W$ and a positive optimum $\mathrm{OPT}$, an $\alpha$-approximation with success probability $p$ means $\Pr[W(A)\ge\mathrm{OPT}/\alpha]\ge p$ on every instance. An approximation in expectation means $\E W(A)\ge\mathrm{OPT}/\alpha$; it is distinct from a bound on $\E\ln W(A)$. Approximation factors are at least one; a displayed lower bound below one is vacuous. In the query lower bounds, the budget is the total number of queries across all agents and all allowed query types, with adaptivity permitted and computation between queries unrestricted. A guarantee must hold for every legal oracle implementation, so it is enough to exhibit a fixed hard instance with one legal implementation on which it fails.

For running-time bounds, we assume that valuation values are rational with a known polynomial bound on their encoding length, and that oracle-returned additive coefficients have polynomial encoding length. Blumrosen and Nisan~\cite{BN09} showed how to simulate value queries by demand queries, \cref{app:demand-values} gives the implementation of value and restricted demand queries under this convention.

\paragraph{Configuration LPs.}
A configuration for agent $i$ is a bundle $S\subseteq M$. Given sets $\mathcal C_i\subseteq2^M$ of configurations, a \emph{feasible configuration solution} assigns nonnegative masses $y_{i,S}$ satisfying
\begin{align}
  \sum_{S\in\mathcal C_i}y_{i,S}&=1 &&\forall i\in N,\label{eq:mm-row}\\
  \sum_{i\in N}\ \sum_{S\in\mathcal C_i:\,j\in S}y_{i,S}&\le1 &&\forall j\in M.\label{eq:mm-capacity}
\end{align}
Thus, each agent's masses form a probability distribution over configurations, and the total mass on each item is at most one. A solution is \emph{explicit} if its positive masses are rational and are given together with their configurations as a list of triples $(i,S,y_{i,S})$.

For max-min allocation, we use the target-based configuration LP of Bansal and Sviridenko~\cite{BS06}, replacing the additive value of a configuration by $v_i(S)$. For a target $t\ge0$, let $\mathcal C_i(t)=\{S\subseteq M:v_i(S)\ge t\}$. The LP asks for a feasible configuration solution with $\mathcal C_i=\mathcal C_i(t)$ for every agent. We call such a solution a \emph{configuration solution with objective value $t$}. Let $T_{\mathrm{LP}}$ be the largest target for which this LP is feasible.

For Nash social welfare, Feng and Li~\cite{FL24} introduced the logarithmic configuration LP for additive valuations, and Feng, Hu, Li, and Zhang~\cite{FHLZ25} used it for submodular valuations. For $\mathrm{OPT}_{\mathrm{NSW}}>0$, set $\mathcal C_i=\{S\subseteq M:v_i(S)>0\}$. Subject to \eqref{eq:mm-row} and \eqref{eq:mm-capacity}, the LP maximizes
\begin{equation}\label{eq:nsw-benchmark}
  L(y):=\sum_{i,S}w_i y_{i,S}\ln v_i(S).
\end{equation}
Here the sum ranges over $i\in N$ and $S\in\mathcal C_i$. For an explicit feasible configuration solution, we also use $L(y)$ to denote its logarithmic benchmark. Let $L_{\mathrm{LP}}$ be the optimum over all positive-valued configurations. For monotone valuations, requiring equality in the item constraints, as in Bei, Feng, Hu, Li, and Zhang~\cite{BFHLZ26}, gives the same optimum: any slack can be filled by adding the item to suitable configurations, splitting their masses if necessary, without changing other item marginals or decreasing the objective.

%% file: max-min.tex
\section{Max-min allocation}\label{sec:max-min}
In this section, we present $O(\sqrt n)$-approximation algorithms for the max-min allocation problem with additive, submodular, and subadditive valuations. We first prove a general rounding lemma and apply it to additive and submodular valuations. We then develop additional tools to extend this approach to subadditive valuations.

\subsection{The configuration setting}\label{subsec:mm-setting}
Asadpour and Saberi~\cite{AS10} and Saha and Srinivasan~\cite{SS18} separate large singleton configurations from bundles of small items in the additive case. We use this separation with additive coefficients that may depend on the configuration. Let $N$ be a set of $n\ge1$ agents and $M$ a set of $m$ items. We are given an explicit feasible configuration solution $(\mathcal C_i,y_{i,S})_{i\in N}$ as defined in \cref{sec:prelim}, with zero-mass configurations omitted.

Moreover, for every agent $i$, every configuration $S \in \mathcal C_i$ has coefficients
$a^{i,S}_j \geq 0$ for every $j\in S$, which defines an additive function:
\[
  a^{i,S}(Q):=\sum_{j\in Q}a^{i,S}_j\qquad \forall Q\subseteq S.
\]
It is possible that $a^{i,S}_j \neq a^{i,S'}_{j}$ for two different configurations $S, S' \in \mathcal C_i$ containing $j$.

Let $\tau>0$. We assume that for every
agent $i$, there is a partition
$\mathcal C_i = \mathcal L_i\uplus \mathcal S_i$ such that
\begin{itemize}
	\item every $S\in\mathcal L_i$ is a singleton with $a^{i,S}(S)\ge\tau$, and
	\item every $S\in\mathcal S_i$ has $a^{i,S}_j<\tau$ for all $j\in S$.
\end{itemize}

We say configurations in $\mathcal L_i$ and $\mathcal S_i$ are large and small respectively.  For each item $j \in M$, define
\begin{equation}\label{eq:mm-residual}
b_j=\sum_{i\in N:\{j\}\in\mathcal L_i}y_{i,\{j\}},
\qquad f_j=1-b_j \geq \sum_{i \in N, S \in \mathcal{S}_i: j \in S} y_{i, S}.
\end{equation}
The inequality follows from the capacity constraint \eqref{eq:mm-capacity}.

For a small configuration $S \in \mathcal S_i$,  define the \emph{baseline value}
\begin{equation}\label{eq:mm-baseline}
  \mu_{i,S}:=\sum_{j \in S}a^{i, S}_j f_j.
\end{equation}

\begin{lemma}\label{lem:mm-rounding}
  In the setting above, suppose that
  \begin{equation}\label{eq:mm-budget}
    \sum_{i\in N}\sum_{S\in\mathcal S_i}y_{i,S}
    \min\left\{1,\left(\frac{32\tau}{\mu_{i,S}}\right)^3\right\}
    \le\frac14.
  \end{equation}
 Here and below, we interpret \(\min\{1,(32\tau/\mu_{i,S})^3\}\) as \(1\) when \(\mu_{i,S}=0\).Then there is a randomized polynomial-time algorithm that selects
  a configuration $S_i\in\mathcal C_i$ for each agent $i\in N$ and
  returns an allocation $A_i\subseteq S_i$ such that,
  with probability at least $3/4$,
  \[
    a^{i,S_i}(A_i)\ge\tau \qquad \forall i\in N.
  \]
  Moreover, for every $i\in N$ and $S\in\mathcal C_i$, we have
  $\Pr[S_i=S]=y_{i,S}$.
\end{lemma}

\subsection{The rounding algorithm}\label{subsec:mm-algorithm}

In this subsection, we describe the algorithm for \cref{lem:mm-rounding} and establish the properties needed for its analysis.
\paragraph{Singleton matching and configuration selection.}
We construct a bipartite graph with the agents in $N$ on the left. On the right, we include a vertex for each item $j\in M$ with $b_j>0$ and a private vertex $u_{i,S}$ for each $i\in N$ and $S\in\mathcal S_i$. For each $\{j\}\in\mathcal L_i$, we add an edge $(i,j)$ with mass $y_{i,\{j\}}$; for each private vertex $u_{i,S}$, we add an edge $(i,u_{i,S})$ with mass $y_{i,S}$. The fractional degree of a vertex is the total mass of its incident edges. By \eqref{eq:mm-row} and \eqref{eq:mm-capacity}, every left vertex has fractional degree $1$, and every right vertex has fractional degree at most $1$.

We apply the bipartite dependent rounding of Gandhi, Khuller, Parthasarathy, and Srinivasan~\cite{GKPS06}. Its degree-preservation and edge-marginal properties give a matching that covers every agent and includes each edge with probability equal to its initial mass. The private vertices place configuration choices and reserved items on the same side of the graph, so the negative correlation of matched vertices proved by Saha and Srinivasan~\cite{SS18} applies to both.

\begin{lemma}\label{lem:mm-matching}
  For every right vertex $v$, let $\chi_v\in\{0,1\}$ indicate that $v$ is matched, and let $d_v$ be its fractional degree in the initial graph. Then $\Pr[\chi_v=1]=d_v$ for every right vertex $v$. Moreover, for every set $Q$ of right vertices and every right vertex $u\notin Q$ with $d_u>0$,
  \begin{equation}\label{eq:mm-cylinder}
    \E\left[\prod_{v\in Q}\chi_v\right]\le\prod_{v\in Q}d_v,
    \qquad
    \E\left[\prod_{v\in Q}\chi_v\ \middle|\ \chi_u=1\right]
         \le\prod_{v\in Q}d_v.
  \end{equation}
\end{lemma}

\begin{proof}
  Summing the marginals of the mutually exclusive edges incident to $v$ gives $\Pr[\chi_v=1]=d_v$. The first inequality is the same-side negative correlation of Saha and Srinivasan~\cite[Theorem~4.4]{SS18} applied to the right vertices in $Q$. Vertices of fractional degree one are matched with certainty and can be omitted from the product.

  For the second inequality, we have $\Pr[\chi_u=1]=d_u$. Applying the first inequality to $Q\cup\{u\}$ gives
  \[
    \E\left[\prod_{v\in Q}\chi_v\ \middle|\ \chi_u=1\right]
    =\frac{\E[\prod_{v\in Q\cup\{u\}}\chi_v]}{d_u}
    \le\prod_{v\in Q}d_v.
  \]
\end{proof}

An agent $i$ matched to an item $j$ selects $S_i:=\{j\}$ and is permanently assigned $A_i:=\{j\}$. Since $a^{i,\{j\}}_j\ge\tau$, we call $i$ happy and remove both $i$ and $j$ from further consideration. An agent $i$ matched to a private vertex $u_{i,S}$ selects $S_i:=S$ and is called unhappy. Since the matching preserves the edge marginals, we have $\Pr[S_i=S]=y_{i,S}$ for every $i\in N$ and $S\in\mathcal C_i$, this proves the final statement of \cref{lem:mm-rounding}.

\paragraph{Fractional assignment of remaining items.}
Let $F_j\in\{0,1\}$ indicate that item $j$ was \emph{not} permanently assigned as a singleton, then we have $\E[F_j]=1-b_j=f_j$, and let $C_j$ be the number of unhappy agents $i$ with $j\in S_i$. For each item $j$ with $F_j=1$ and $C_j>0$, we allocate a $1/C_j$ fraction of $j$ to each unhappy agent whose selected configuration contains $j$. This gives a feasible fractional allocation. For each unhappy agent $i$, define its total fractional value by
\begin{align*}
  X_i := \sum_{j\in S_i} a^{i,S_i}_j \frac{F_j}{C_j}.
\end{align*}

\paragraph{Integral assignment via Shmoys--Tardos rounding.}
Bez\'{a}kov\'{a} and Dani~\cite{BD05} bounded the additive loss per agent in max-min allocation by the value of one item. We use the slot construction of Shmoys and Tardos~\cite{ST93} on the support of the fractional assignment. For each unhappy agent $i$, we sort the items in its fractional support in nonincreasing order of $a^{i,S_i}_j$ and pack their assigned fractions into slots of unit capacity in this order. An item's fraction may be split between two adjacent slots, and the last slot may be partially filled. This gives a fractional matching between slots and unassigned items. We find a matching in its support that covers all full slots and assign each matched item to the agent owning the corresponding slot. The resulting bundles $A_i\subseteq S_i$ are pairwise disjoint and contain only items $j$ with $F_j=1$. For each agent, the item matched to a full slot is worth at least the fractional value of the next slot, since that slot has mass at most one and contains only items of weakly smaller value. Summing over full slots bounds the loss by the fractional value of the first slot, which is at most $\tau$. Thus,
\begin{equation}\label{eq:mm-slot}
  a^{i,S_i}(A_i)\ge X_i-\tau \qquad \text{for every unhappy agent }i.
\end{equation}
This completes the description of the rounding algorithm.

In the applications below, configurations may be replaced by subsets before rounding, and equal resulting sets are merged by adding their masses. After rounding, the \emph{proportional rule} recovers an original configuration for each agent, choosing among those mapped to its selected set with probability proportional to their masses; this preserves the original marginals and does not change the allocation.

\subsection{Analysis of the failure probability}\label{subsec:mm-analysis}
In this subsection, we analyze the failure probability of the algorithm. Fix an agent $i\in N$ and a small configuration $S\in\mathcal S_i$, we use $\Pr_S[\cdot]$ and $\E_S[\cdot]$ to denote probability and expectation conditional on $S_i=S$, respectively. For brevity, write $a_j=a^{i,S}_j$, $a(Q)=a^{i,S}(Q)$, $\mu=\mu_{i,S}$, and $X=X_i$. Define
\[
  Z=\sum_{j\in S}a_jF_j,
\]
since $\E[F_j]=f_j$, the unconditional expectation of $Z$ is $\E[Z]=\sum_{j\in S}a_jf_j=\mu$.

\begin{lemma}\label{lem:mm-bound-Ai}
	For every agent $i\in N$ and every small configuration $S\in\mathcal S_i$,
	\begin{equation}\label{eq:mm-local-tail}
		\Pr_S[a^{i,S}(A_i)<\tau]
		\le \Pr_S[X_i<2\tau]
		\le\min\left\{1,\left(\frac{32\tau}{\mu_{i,S}}\right)^3\right\}.
	\end{equation}
\end{lemma}
\begin{proof}
  The first inequality follows from \eqref{eq:mm-slot}. The second inequality is trivial when $0\le\mu\le32\tau$, so we may assume that $\mu>32\tau$. We bound $\Pr_S[X<2\tau]$ by controlling the value $Z$ that remains after singleton assignment and the loss from sharing the remaining items with other agents.

  We first show that $Z$ is unlikely to fall below $\mu/2$ under the conditional law $\Pr_S$. The event $S_i=S$ is equivalent to $\chi_{u_{i,S}}=1$, and the initial fractional degree of $u_{i,S}$ is $y_{i,S}>0$. For an item $j$ with $b_j>0$, the indicator of its reservation as a singleton is $1-F_j=\chi_j$, and its initial fractional degree is $b_j$. Items with $b_j=0$ have $1-F_j=0$ deterministically. Thus, \cref{lem:mm-matching} gives
  \[
    \E_S\left[\prod_{j\in J}(1-F_j)\right]\le\prod_{j\in J}b_j
    \qquad\text{for every }J\subseteq S.
  \]
  Since $F_j\in\{0,1\}$, we have $2^{-a_jF_j/\tau}=2^{-a_j/\tau}+(1-2^{-a_j/\tau})(1-F_j)$. Expanding the product of these expressions gives nonnegative coefficients, so the preceding bound applies to each term. Consequently,
  \begin{align*}
    \E_S[2^{-Z/\tau}]
    &=\E_S\left[\prod_{j\in S}
      \left(2^{-a_j/\tau}+(1-2^{-a_j/\tau})(1-F_j)\right)\right]
     \le\prod_{j\in S}\left(2^{-a_j/\tau}+b_j(1-2^{-a_j/\tau})\right)\\
    &=\prod_{j\in S}\left(1-f_j(1-2^{-a_j/\tau})\right)
     \le\exp\left(-\sum_{j\in S}f_j(1-2^{-a_j/\tau})\right)
     \le e^{-\mu/(2\tau)}.
  \end{align*}
  The last two inequalities use $1-x\le e^{-x}$ and $1-2^{-q}\ge q/2$ for $q\in[0,1]$, together with $0\le a_j\le\tau$. Applying Markov's inequality to the nonnegative random variable $2^{-Z/\tau}$, we obtain
  \begin{align}
    \Pr_S[Z<\mu/2]
    &=\Pr_S\left[2^{-Z/\tau}>2^{-\mu/(2\tau)}\right]
     \le 2^{\mu/(2\tau)}\E_S[2^{-Z/\tau}]\nonumber\\
    &\le \exp\left(-\frac{(1-\ln2)\mu}{2\tau}\right)
      \le e^{-\mu/(8\tau)},\label{eq:mm-survival}
  \end{align}
  where the last inequality uses $\ln2<3/4$.

  We next consider the case $Z\ge\mu/2$. For each $i'\ne i$ and $S'\in\mathcal S_{i'}$, let $I_{i',S'}$ indicate that agent $i'$ selects $S'$, and all sums over $(i',S')$ below range over these pairs. Under $S_i=S$, agent $i$ contributes one to $C_j$ for every $j\in S$. Hence $C_j\ge1$ and
  \[
    C_j-1=\sum_{i',S':\,j\in S'}I_{i',S'}
    \qquad \forall j\in S.
  \]
  Define
  \[
    D=\sum_{j\in S}a_jF_j(C_j-1),
  \]
  thus $D$ counts the value of each remaining item once for every other selected small configuration containing it. Since $Z+D=\sum_{j\in S}a_jF_jC_j$, the Cauchy--Schwarz inequality gives
  \begin{align*}
    Z^2
    &=\left(\sum_{j\in S}
      \sqrt{\frac{a_jF_j}{C_j}}\sqrt{a_jF_jC_j}\right)^2\\
    &\le\left(\sum_{j\in S}\frac{a_jF_j}{C_j}\right)
      \left(\sum_{j\in S}a_jF_jC_j\right)
      =X(Z+D).
  \end{align*}

  To control $D/Z$ when $Z\ge\mu/2$, define
  \[
    c_{i',S'}=\min\left\{1,\frac{2a(S\cap S')}{\mu}\right\},
    \qquad
    Y=\sum_{i',S'}c_{i',S'}I_{i',S'}.
  \]
  The coefficients $c_{i',S'}$ are deterministic and lie in $[0,1]$. For each pair $(i',S')$, the remaining items in $S\cap S'$ have total value $\sum_{j\in S\cap S'}a_jF_j$, which is at most $Z$ because $S\cap S'\subseteq S$, and at most $a(S\cap S')$ because $F_j\le1$. Therefore, on the event $Z\ge\mu/2>0$,
  \[
    \frac{1}{Z}\sum_{j\in S\cap S'}a_jF_j
    \le\min\left\{1,\frac{a(S\cap S')}{Z}\right\}
    \le c_{i',S'}.
  \]
  Multiplying by $I_{i',S'}$ and summing, we get
  \[
    \frac{D}{Z}
    =\sum_{i',S'}I_{i',S'}\frac{1}{Z}
      \sum_{j\in S\cap S'}a_jF_j
    \le Y.
  \]
  Combining this with the bound from Cauchy--Schwarz yields
  \[
    X\ge\frac{Z^2}{Z+D}
      \ge\frac{Z}{1+Y}
      \ge\frac{\mu}{2(1+Y)}.
  \]
  In particular, if $X<2\tau$ and $Z\ge\mu/2$, then
  \[
    Y>\frac{\mu}{4\tau}-1>\frac{\mu}{8\tau},
  \]
  where the last inequality follows from $\mu>32\tau$. It remains to bound the probability that $Y>\mu/(8\tau)$.

  We do so by bounding the third moment of $Y$ under $\Pr_S$. First, let
  \[
    \rho:=\E[Y]=\sum_{i',S'}c_{i',S'}y_{i',S'},
  \]
  Using $c_{i',S'}\le2a(S\cap S')/\mu$ and the capacity bound \eqref{eq:mm-residual}, we have
  \[
    \rho
    \le\frac{2}{\mu}\sum_{i',S'}a(S\cap S')y_{i',S'}
    =\frac{2}{\mu}\sum_{j\in S}a_j
      \sum_{i',S':\,j\in S'}y_{i',S'}
    \le\frac{2}{\mu}\sum_{j\in S}a_jf_j
      =2.
  \]

  For brevity, let $\alpha,\beta,\gamma$ denote pairs $(i',S')$ with $i'\ne i$ and $S'\in\mathcal S_{i'}$. Expanding $Y^3$ and using $I_\alpha^2=I_\alpha$, we obtain
  \begin{align}
    \E_S[Y^3]
    &=\sum_\alpha c_\alpha^3\E_S[I_\alpha]
      +3\sum_{\alpha\ne\beta}c_\alpha^2c_\beta\E_S[I_\alpha I_\beta]
      +\sum_{\substack{\alpha,\beta,\gamma\text{ distinct}}}
      c_\alpha c_\beta c_\gamma
      \E_S[I_\alpha I_\beta I_\gamma]\nonumber\\
    &\le\sum_\alpha c_\alpha^3y_\alpha
      +3\sum_{\alpha\ne\beta}c_\alpha^2c_\beta y_\alpha y_\beta\nonumber
      +\sum_{\substack{\alpha,\beta,\gamma\text{ distinct}}}
      c_\alpha c_\beta c_\gamma
      y_\alpha y_\beta y_\gamma\nonumber\\
    &\le\sum_\alpha c_\alpha y_\alpha
      +3\sum_{\alpha,\beta}(c_\alpha y_\alpha)(c_\beta y_\beta)
      +\sum_{\alpha,\beta,\gamma}
      (c_\alpha y_\alpha)(c_\beta y_\beta)(c_\gamma y_\gamma)\nonumber\\
    &=\rho+3\rho^2+\rho^3
      \le2+3\cdot2^2+2^3=22.\label{eq:mm-third}
  \end{align}
  The first inequality follows from the conditional product bound in \cref{lem:mm-matching}, since $I_{i',S'}=\chi_{u_{i',S'}}$ and distinct pairs correspond to distinct private vertices, all different from $u_{i,S}$. The second inequality uses $0\le c_\alpha\le1$ and extends the sums to all ordered pairs and triples. The final inequality follows from $\rho\le2$.

  We have shown that $X<2\tau$ implies either $Z<\mu/2$ or $Y>\mu/(8\tau)$. Combining \eqref{eq:mm-survival} with the union bound and Markov's inequality applied to $Y^3$, we obtain
  \begin{align*}
    \Pr_S[X<2\tau]
    &\le\Pr_S[Z<\mu/2]+\Pr_S[Y>\mu/(8\tau)]
     \le e^{-\mu/(8\tau)}
      +\frac{\E_S[Y^3]}{(\mu/(8\tau))^3}\\
    &\le e^{-\mu/(8\tau)}+22(8\tau/\mu)^3
     \le28(8\tau/\mu)^3
      \le(32\tau/\mu)^3.
  \end{align*}
  Here the third inequality uses \eqref{eq:mm-third}, and the fourth uses $e^{-q}\le6/q^3$ for $q=\mu/(8\tau)>4$. This proves \eqref{eq:mm-local-tail}, and hence \cref{lem:mm-bound-Ai}.
\end{proof}

We now combine the bounds for individual agents to prove \cref{lem:mm-rounding}.

\begin{proof}[Proof of \cref{lem:mm-rounding}]
  Fix an agent $i\in N$. If $S_i\in\mathcal L_i$, then $a^{i,S_i}(A_i)\ge\tau$ by construction. Thus, using $\Pr[S_i=S]=y_{i,S}$ and \cref{lem:mm-bound-Ai}, we obtain
  \begin{align*}
    \Pr[a^{i,S_i}(A_i)<\tau]
    &=\sum_{S\in\mathcal S_i}y_{i,S}\Pr_S[a^{i,S}(A_i)<\tau]\\
    &\le\sum_{S\in\mathcal S_i}y_{i,S}\min\left\{1,\left(\frac{32\tau}{\mu_{i,S}}\right)^3\right\}.
  \end{align*}
  The union bound over all agents, together with \eqref{eq:mm-budget}, shows that the algorithm fails with probability at most $1/4$.
\end{proof}

\subsection{Applications to additive and submodular valuations}
\label{subsec:mm-exact}

In this subsection, we apply \cref{lem:mm-rounding} to additive and submodular valuations.
The configuration LP has exponentially many variables, so we do not solve it exactly. Since we only aim for an $O(\sqrt n)$-approximation, we do not optimize constant factors. Losing a constant factor in the target, we may assume that we are given an explicit configuration solution with polynomial support. \cref{app:configuration-producer} proves the following lemma by combining ellipsoid feasibility and primal recovery~\cite{GLS88} with the submodular knapsack algorithm of Sviridenko~\cite{Sviridenko04} or the demand-query algorithm of Badanidiyuru, Dobzinski, and Oren~\cite{BDO19}, according to the valuation class.

\begin{lemma}\label{lem:mm-lp-solution}
  There is a polynomial-time algorithm that either detects that $T_{\mathrm{LP}}=0$, in which case every allocation is optimal, or computes an explicit configuration solution with objective value $T$,  supported on polynomially many configurations with $T>T_{\mathrm{LP}}/6$. For submodular valuations, the algorithm uses only value queries. For subadditive valuations, it uses only ordinary demand queries.
\end{lemma}

We apply \cref{lem:mm-lp-solution} to obtain a configuration solution with objective value $T > T_{\mathrm{LP}}/6$. A \emph{supporting clause}\footnote{The name comes from the XOS representation~\cite{DNS10} of valuations as maxima of additive functions called clauses (\cref{sec:prelim}): restricted to $S$, a clause attaining the maximum at $S$ is a supporting clause for $(i,S)$. Conversely, a monotone valuation is XOS if and only if every set has a supporting clause.} for $(i,S)$ is a nonnegative vector $w^{i,S} \in \R_{\geq 0}^S$ satisfying
\begin{equation}\label{eq:mm-exact-support}
  w^{i,S}(Q)\le v_i(Q)\quad(Q\subseteq S),\qquad
  w^{i,S}(S)=v_i(S).
\end{equation}

To apply \cref{lem:mm-rounding}, we first assume we are given supporting clauses for all pairs $(i,S)$ in the support of the configuration solution.  After stating the resulting guarantee in \cref{thm:mm-exact}, we show how to obtain these clauses for each valuation class.

\paragraph{Preparing the configuration solution.}
Normalize the clause coefficients and set the rounding threshold by
\[
  a^{i,S}_j=\frac{T w^{i,S}_j}{v_i(S)},
  \qquad
  \tau=\frac{T}{128\lceil\sqrt n\rceil}.
\]
Since $v_i(S)\ge T$, the normalized coefficients satisfy $a^{i,S}(Q)\le v_i(Q)$ for every $Q\subseteq S$, and $a^{i,S}(S)=T$.

If a configuration $S$ contains an item $j$ with $a^{i,S}_j\ge\tau$, we choose one such item and replace $S$ by $\{j\}$, retaining the coefficient $a^{i,S}_j$. We merge repeated singletons by adding their masses and keeping the largest coefficient. Replacing configurations by subsets and merging equal sets preserve \eqref{eq:mm-row} and cannot increase the total mass on any item. We continue to write $y_{i,S}$ for the resulting masses.

The resulting singletons form the large configurations $\mathcal L_i$. All configurations left unchanged have additive total $T$ and every coefficient below $\tau$, these form the small configurations $\mathcal S_i$. Since $\tau<T$, no small configuration is a singleton. It remains to verify the budget condition \eqref{eq:mm-budget}.

\paragraph{Bounding the baseline values.}
Define $b_j$ and $f_j=1-b_j$ by \eqref{eq:mm-residual} for the resulting configuration solution. For every small configuration $S$, define
\[
  P(S)=\sum_{j\in S}\frac{b_j}{f_j}.
\]
This penalty is well defined: every item in $S$ lies in the small configuration $S$ of positive mass, so \eqref{eq:mm-residual} implies $f_j>0$. Exchanging the order of summation and applying \eqref{eq:mm-residual}, we obtain
\begin{equation}\label{eq:mm-penalty-budget}
  \sum_i\sum_{S\in\mathcal S_i}y_{i,S}P(S)
  =\sum_{j:f_j>0}\frac{b_j}{f_j}
    \sum_i\sum_{\substack{S\in\mathcal S_i\\j\in S}}y_{i,S}
  \le\sum_j b_j\le n.
\end{equation}
The last inequality follows from \eqref{eq:mm-row}, since $\sum_j b_j$ is the total mass of large singleton configurations.

Now fix a small configuration $S\in\mathcal S_i$. Its additive total is $T$, and the Cauchy--Schwarz inequality gives
\[
  T^2
  \le\left(\sum_{j\in S}f_ja^{i,S}_j\right)
      \left(\sum_{j\in S}\frac{a^{i,S}_j}{f_j}\right)
  =\mu_{i,S}\sum_{j\in S}\frac{a^{i,S}_j}{f_j}.
\]
Using $f_j=1-b_j$ and $a^{i,S}_j\le\tau$, we also have
\[
  \sum_{j\in S}\frac{a^{i,S}_j}{f_j}
  =T+\sum_{j\in S}a^{i,S}_j\frac{b_j}{f_j}
  \le T+\tau P(S)
  =T\left(1+\frac{P(S)}{128\lceil\sqrt n\rceil}\right).
\]
Combining these inequalities yields
\[
  \mu_{i,S}\ge\frac{T}{1+\frac{P(S)}{128\lceil\sqrt n\rceil}}.
\]

\paragraph{Verifying the budget condition.}
The baseline bound and the choice of $\tau$ imply
\[
  \frac{32\tau}{\mu_{i,S}}
  \le\frac{32}{128\lceil\sqrt n\rceil}
    \left(1+\frac{P(S)}{128\lceil\sqrt n\rceil}\right).
\]
We consider two cases according to the value of $P(S)$.

If $P(S)\le128\lceil\sqrt n\rceil$, then
\[
  \min\left\{1,\left(\frac{32\tau}{\mu_{i,S}}\right)^3\right\}
  \le\left(\frac{64}{128\lceil\sqrt n\rceil}\right)^3
  \le\frac{1}{8n^{3/2}}.
\]
If $P(S)>128\lceil\sqrt n\rceil$, then $1+P(S)/(128\lceil\sqrt n\rceil)\le2P(S)/(128\lceil\sqrt n\rceil)$, so
\begin{align*}
  \min\left\{1,\left(\frac{32\tau}{\mu_{i,S}}\right)^3\right\}
  &\le\min\left\{1,\left(\frac{64P(S)}{(128\lceil\sqrt n\rceil)^2}\right)^3\right\}
  \le\frac{64P(S)}{(128\lceil\sqrt n\rceil)^2}
   \le\frac{P(S)}{256n},
\end{align*}
where the second inequality uses $\min\{1,u^3\}\le u$ for $u\ge0$.

Applying these bounds in their respective cases and extending both sums to all small configurations, we obtain
\begin{align*}
  \sum_i\sum_{S\in\mathcal S_i}y_{i,S}
    \min\left\{1,\left(\frac{32\tau}{\mu_{i,S}}\right)^3\right\}
  &\le\frac{1}{8n^{3/2}}
      \sum_i\sum_{S\in\mathcal S_i}y_{i,S}
    +\frac{1}{256n}
      \sum_i\sum_{S\in\mathcal S_i}y_{i,S}P(S)\\
  &\le n\frac{1}{8n^{3/2}}+n\frac{1}{256n}
    \le\frac18+\frac1{256}<\frac14.
\end{align*}
Here we used \eqref{eq:mm-row} to bound the total mass of small configurations by $n$, and \eqref{eq:mm-penalty-budget} to bound their weighted penalty. This verifies \eqref{eq:mm-budget} and gives the following theorem.

\begin{theorem}\label{thm:mm-exact}
  Given a feasible configuration solution with objective value $T$ and supporting clauses, there is a randomized polynomial-time algorithm that finds an allocation in which every agent receives value at least $T/(128\lceil\sqrt n\rceil)$, with probability at least $3/4$.
\end{theorem}
\begin{proof}
  If $T=0$, assigning the empty set to every agent suffices. For $T>0$, the construction above satisfies the setting of \cref{lem:mm-rounding}, including the budget condition \eqref{eq:mm-budget}. Thus, with probability at least $3/4$, every agent $i$ receives a bundle $A_i$ with $a^{i,S_i}(A_i)\ge\tau$. Since the coefficients are dominated by the valuations, we also have $v_i(A_i)\ge\tau$.
\end{proof}

Now we discuss how to construct the supporting clauses for different valuation functions.

\paragraph{Additive valuations.}
For additive valuations $v_i(Q)=\sum_{j\in Q}v_{ij}$, the coefficients $w^{i,S}_j=v_{ij}$ satisfy \eqref{eq:mm-exact-support}. Given a feasible configuration solution with objective value $T$, \cref{thm:mm-exact} yields an allocation giving every agent value at least $T/(128\lceil\sqrt n\rceil)$ with probability at least $3/4$. This requires no valuation queries beyond reading the values $v_{ij}$. Since additive valuations are submodular, \cref{lem:mm-lp-solution} produces a configuration solution with objective value $T>T_{\mathrm{LP}}/6$. Combining this solution with the rounding algorithm gives an $O(\sqrt n)$ approximation for additive valuations.

\paragraph{Submodular valuations.}
Dobzinski, Nisan, and Schapira~\cite{DNS10} construct a supporting clause by ordering the items in $S$ and assigning each item its marginal value relative to the preceding items. For a normalized monotone submodular valuation, the resulting nonnegative coefficients satisfy \eqref{eq:mm-exact-support} and require $O(|S|)$ value queries.

\cref{lem:mm-lp-solution} produces a configuration solution with objective value $T>T_{\mathrm{LP}}/6$ using value queries. Applying \cref{thm:mm-exact} yields an $O(\sqrt n)$ approximation for submodular valuations in the value oracle model.



\subsection{Additive lower bounds for subadditive valuations}

The supporting clauses required by \cref{thm:mm-exact} may not exist for subadditive valuations. We instead construct nonnegative additive coefficients satisfying $a^{i,S}(Q)\le v_i(Q)$ for every $Q\subseteq S$, which may not satisfy $a^{i, S}(S) = v_i(S)$. To apply \cref{lem:mm-rounding}, we also need a sufficiently large baseline $\mu_{i,S}=\sum_{j\in S}a^{i,S}_j f_j$.

Dobzinski~\cite{Dobzinski07} showed that demand queries suffice to find prices supporting a subadditive valuation on a given set within a logarithmic factor. In \cref{lem:mm-log-support} we use a geometric grid of uniform prices, as in the single-buyer revenue bound of Balcan, Blum, and Mansour~\cite{BBM08}, and derive the quantitative bound that we extend below. \cref{lem:mm-weighted-support} extends this construction by scaling item prices by $1/f_j$ to control the baseline $\mu_{i,S}$. We use the resulting coefficients in the subsequent rounding analysis.

\begin{lemma}\label{lem:mm-log-support}
  Let $v$ be a normalized monotone subadditive valuation on a nonempty set $S$. Using $O(1+\log|S|)$ restricted demand queries and $O(|S|)$ value queries, we can construct an additive function $a(Q)=\sum_{j\in Q}a_j$ with nonnegative coefficients satisfying
  \[
    a(Q)\le v(Q)\quad(Q\subseteq S),\qquad
    a(S)\ge\frac{v(S)}{2\lceil\log_2(2|S|)\rceil}.
  \]
\end{lemma}
\begin{proof}
  Let $W=v(S)$ and $s=|S|$. If $W=0$, we return zero coefficients. Henceforth, assume $W>0$, and let $\nu=\max_{j\in S}v(\{j\})$. Monotonicity and subadditivity give $0<\nu\le W\le s\nu$.

  For a uniform item price $t>0$, let $D_t\subseteq S$ be the set returned by a restricted demand query on $S$, and define
  \[
    a_j^{(t)}=t\mathbf1_{\{j\in D_t\}}.
  \]
  We first show that these coefficients give an additive lower bound. For any $Q\subseteq S$, the optimality of $D_t$ implies
  \[
    v(D_t)-t|D_t|
    \ge v(D_t\setminus Q)-t|D_t\setminus Q|.
  \]
  Rearranging and then applying subadditivity and monotonicity, we obtain
  \[
    a^{(t)}(Q)=t|D_t\cap Q|
    \le v(D_t)-v(D_t\setminus Q)
    \le v(D_t\cap Q)\le v(Q).
  \]

  We now choose a price for which the additive total is sufficiently large. Set
  \[
    L=\left\lceil\log_2\frac{2s\nu}{W}\right\rceil,
    \qquad
    t_\ell=\nu2^{-\ell}\quad(0\le\ell\le L).
  \]
  The bounds on $W$ imply $1\le L\le\lceil\log_2(2s)\rceil$. We make a restricted demand query at each price $t_1,\ldots,t_L$ and return a candidate $a^{(t_\ell)}$ with the largest total value on $S$. Denote the returned additive function by $a$.

  To analyze this choice, define the maximum utility at price $t$ by
  \[
    U(t)=\max_{D\subseteq S}\{v(D)-t|D|\}.
  \]
  At the initial price $t_0=\nu$, subadditivity gives $v(D)\le\sum_{j\in D}v(\{j\})\le\nu|D|$ for every $D\subseteq S$. Thus, every set has nonpositive utility, while the empty set has utility zero, so $U(t_0)=0$. At the final price, the definition of $L$ gives $t_L\le W/(2s)$. The set $S$ itself therefore shows that
  \[
    U(t_L)\ge v(S)-t_L|S|\ge W/2.
  \]

  For each $\ell=1,\ldots,L$, the set $D_{t_\ell}$ is also a feasible choice at the preceding price $t_{\ell-1}=2t_\ell$. Hence,
  \[
    U(t_{\ell-1})
    \ge v(D_{t_\ell})-2t_\ell|D_{t_\ell}|
    =U(t_\ell)-t_\ell|D_{t_\ell}|.
  \]
  Equivalently, the increase in maximum utility when the price is halved satisfies
  \[
    U(t_\ell)-U(t_{\ell-1})
    \le t_\ell|D_{t_\ell}|=a^{(t_\ell)}(S).
  \]
  Summing over the price levels gives
  \[
    \frac W2
    \le U(t_L)-U(t_0)
    \le\sum_{\ell=1}^L a^{(t_\ell)}(S).
  \]
  Since we return the candidate with the largest total, its value is at least the average of these totals. Therefore,
  \[
    a(S)\ge\frac1L\sum_{\ell=1}^L a^{(t_\ell)}(S)
    \ge\frac{W}{2L}
    \ge\frac{v(S)}{2\lceil\log_2(2|S|)\rceil}.
  \]

  Computing $W$ and $\nu$ uses at most $s+1=O(|S|)$ value queries. The construction makes $L=O(1+\log|S|)$ restricted demand queries, and the candidate totals $t_\ell|D_{t_\ell}|$ are known from their returned sets, so selecting the largest requires no further value queries.
\end{proof}

Using these additive lower bounds in the proof of \cref{thm:mm-exact} yields only an $O(\sqrt n(1+\log m))$ approximation. To avoid this additional logarithmic factor, we choose the price of each item $j$ according to $f_j$ and verify \eqref{eq:mm-budget} directly.

\begin{lemma}\label{lem:mm-weighted-support}
  Let $v$ be a normalized, monotone, and subadditive valuation with
  $v(S)=W>0$.  Let $\nu=\max_{j\in S}v(\{j\})$, and let $0<f_j\le1$
  for every $j\in S$.  Define $P=\sum_{j\in S}(1-f_j)/f_j$ and
  $s=|S|$.  Then, using polynomially many restricted demand queries,
  we can find an additive lower bound vector $a \in [0, \nu]^S$, whose
  baseline $\mu=\sum_{j\in S}f_ja_j$ is positive and satisfies
  \begin{equation}\label{eq:mm-envelope}
    \frac{3W}{4}
    \le \mu\left[2+\log_2\frac{s\nu}{\mu}\right]
       +4\sqrt{\nu P\mu}.
  \end{equation}
\end{lemma}
\begin{proof}
  Monotonicity and subadditivity give $0<\nu\le W\le s\nu$. Set $t_\ell=\nu2^{-\ell}$, and let $L$ be the first nonnegative integer for which $t_L(s+P)\le W/4$. For each $\ell=0,\ldots,L$, make a restricted demand query on $S$ with item prices $t_\ell/f_j$, and let $D_\ell$ be the returned set. Define
  \[
    a_j^{(\ell)}=\frac{t_\ell}{f_j}\mathbf1_{\{j\in D_\ell\}},
    \qquad
    \mu_\ell=\sum_{j\in S}f_ja_j^{(\ell)}=t_\ell|D_\ell|.
  \]
  As in the proof of \cref{lem:mm-log-support}, comparing the utilities of $D_\ell$ and $D_\ell\setminus Q$ gives, for every $Q\subseteq S$,
  \begin{align*}
    a^{(\ell)}(Q)
    &=t_\ell\sum_{j\in D_\ell\cap Q}\frac1{f_j}
    \le v(D_\ell)-v(D_\ell\setminus Q)
     \le v(D_\ell\cap Q)\le v(Q).
  \end{align*}
  Thus, every candidate is an additive lower bound. Taking $Q=\{j\}$ also gives $a_j^{(\ell)}\le v(\{j\})\le\nu$. We return a candidate maximizing $\mu_\ell$, denote it by $a$, and write $\mu=\sum_{j\in S}f_ja_j$.

  To analyze $\mu$, define the maximum utility at price parameter $t$ by
  \[
    U(t)=\max_{D\subseteq S}\left\{v(D)-t\sum_{j\in D}\frac1{f_j}\right\}.
  \]
  Since $v(D)\le\nu|D|$ and $f_j\le1$, every set has nonpositive utility at $t_0=\nu$. The empty set has utility zero, so $U(t_0)=0$. Using $\sum_{j\in S}f_j^{-1}=s+P$ and $t_L(s+P)\le W/4$, we obtain
  \[
    U(t_L)\ge v(S)-t_L\sum_{j\in S}\frac1{f_j}
    =W-t_L(s+P)\ge\frac{3W}{4}.
  \]
  In particular, $D_L$ is nonempty, so $\mu\ge\mu_L>0$. The coefficient bounds and $f_j\le1$ also give $\mu\le s\nu$.

  As in \cref{lem:mm-log-support}, we compare a demanded set at two consecutive prices. For $1\le\ell\le L$, evaluating $D_\ell$ at $t_{\ell-1}=2t_\ell$ gives
  \[
    U(t_{\ell-1})
    \ge v(D_\ell)-2t_\ell\sum_{j\in D_\ell}\frac1{f_j}
    =U(t_\ell)-t_\ell\sum_{j\in D_\ell}\frac1{f_j}.
  \]
  Therefore,
  \begin{align*}
    U(t_\ell)-U(t_{\ell-1})
    &\le t_\ell\sum_{j\in D_\ell}\frac1{f_j} =t_\ell|D_\ell|+t_\ell\sum_{j\in D_\ell}\frac{1-f_j}{f_j}.
  \end{align*}
  Summing over $\ell=1,\ldots,L$, we obtain
  \[
    \frac{3W}{4}\le U(t_L)-U(t_0)\le A+B,
  \]
  where
  \[
    A=\sum_{\ell=1}^L t_\ell|D_\ell|,
    \qquad
    B=\sum_{\ell=1}^L t_\ell\sum_{j\in D_\ell}\frac{1-f_j}{f_j}.
  \]
  We bound these two sums separately.

  Each term of $A$ is at most $\mu$ by the choice of the output, and at most $s\nu2^{-\ell}$ because $|D_\ell|\le s$. Let $k=\lfloor\log_2(s\nu/\mu)\rfloor\ge0$, then
  \[
    A
    \le\sum_{\ell\ge1}\min\{\mu,s\nu2^{-\ell}\}
    \le k\mu+\sum_{\ell=k+1}^{\infty}s\nu2^{-\ell}
     =k\mu+s\nu2^{-k}
    \le\mu\left[2+\log_2\frac{s\nu}{\mu}\right],
  \]
  where the last inequality uses $k\le\log_2(s\nu/\mu)$ and $s\nu2^{-k}<2\mu$.

  It remains to bound $B$. If $P=0$, then $B=0$. Suppose $P>0$, and fix a cutoff $t^*>0$. For a level with $t_\ell\le t^*$, its contribution to $B$ is at most $t_\ell P$. For a level with $t_\ell>t^*$, the bound $a_j^{(\ell)}\le\nu$ implies $t_\ell/f_j\le\nu$ for every $j\in D_\ell$. Hence its contribution satisfies
  \[
    t_\ell\sum_{j\in D_\ell}\frac{1-f_j}{f_j}
    \le\nu|D_\ell|
    \le\frac{\nu\mu}{t_\ell}.
  \]
  Since the price is halved at each step, the corresponding geometric sums satisfy
  \[
    \sum_{\substack{1\le\ell\le L\\t_\ell\le t^*}}t_\ell\le2t^*,
    \qquad
    \sum_{\substack{1\le\ell\le L\\t_\ell>t^*}}\frac1{t_\ell}\le\frac2{t^*}.
  \]
  Combining the bounds for the two groups of levels gives
  \[
    B\le2t^*P+\frac{2\nu\mu}{t^*}.
  \]
  Choosing $t^*=\sqrt{\nu\mu/P}$ yields $B\le4\sqrt{\nu P\mu}$, which also holds when $P=0$. Together with the bound on $A$ and $3W/4\le A+B$, this proves \eqref{eq:mm-envelope}.

  Finally, the number of price levels satisfies
  \[
    L=\left\lceil\log_2\frac{4\nu(s+P)}{W}\right\rceil
    \le\left\lceil\log_2\bigl(4(s+P)\bigr)\right\rceil.
  \]
  Thus, the number of queries and the encodings of the prices are polynomial in the encodings of the values and weights. By the value-evaluation reduction in \cref{app:demand-values}, we can compute $W$ and $\nu$ exactly using polynomially many restricted demand queries. Selecting the candidate with the largest baseline requires no further value queries, since each $\mu_\ell=t_\ell|D_\ell|$ is known from the returned set.
\end{proof}

\cref{lem:mm-weighted-support} gives the baseline bound needed for rounding. In \cref{subsec:mm-quadratic}, we use it to verify \eqref{eq:mm-budget} and obtain an $O(\sqrt n)$ approximation for instances with at most $128n^2$ items.

\subsection{Subadditive rounding with at most $128n^2$ items}
\label{subsec:mm-quadratic}

In this subsection, we consider subadditive instances with $m\le128n^2$ and establish the rounding guarantee in \cref{cor:mm-local}. Combined with the configuration solution of \cref{lem:mm-lp-solution}, this gives an $O(\sqrt n)$ approximation using demand queries. The rounding procedure will also be used as a component of the algorithm for an arbitrary number of items in \cref{subsec:mm-general}.

\paragraph{Preparing the configuration solution.}
Given a feasible configuration solution with objective value $T$, set
\[
  \kappa=2^{20}\lceil\sqrt n\rceil,\qquad \tau=T/\kappa.
\]
When $T=0$, we can select configurations with their prescribed marginals and assign no items. We therefore assume $T>0$ and will apply \cref{lem:mm-rounding} with threshold $\tau$.

For each original configuration $(i,S)$, define $u^{i,S}(Q)=T v_i(Q)/v_i(S)$ for $Q\subseteq S$. Since $v_i(S)\ge T$, we have $u^{i,S}(S)=T$ and $u^{i,S}(Q)\le v_i(Q)$ for every $Q\subseteq S$. The valuation $u^{i,S}$ remains normalized, monotone, and subadditive. A demand query for $u^{i,S}$ at prices $p_j$ can be implemented by a restricted demand query for $v_i$ on $S$ at prices $v_i(S)p_j/T$.

We apply the singleton replacement and merging procedure of \cref{subsec:mm-exact}, using $u^{i,S}(\{j\})$ as the singleton coefficients and $\tau$ as the threshold. Denote the resulting large and small configurations by $\mathcal L_i$ and $\mathcal S_i$, respectively, and continue to write $y_{i,S}$ for their masses. Every $S\in\mathcal S_i$ satisfies $u^{i,S}(\{j\})<\tau$ for all $j\in S$.

Define $b_j$ and $f_j=1-b_j$ by \eqref{eq:mm-residual} for the resulting solution. For each small configuration $(i,S)$ and each $j\in S$, \eqref{eq:mm-residual} gives $f_j\ge y_{i,S}>0$. We can therefore apply \cref{lem:mm-weighted-support} to $u^{i,S}$ with weights $(f_j)_{j\in S}$. Let $a^{i,S}$ and $\mu_{i,S}$ be the resulting additive lower bound and baseline. Then $a^{i,S}(Q)\le u^{i,S}(Q)\le v_i(Q)$ for every $Q\subseteq S$, and $a^{i,S}_j\le u^{i,S}(\{j\})<\tau$. Thus the resulting solution satisfies the setting of \cref{lem:mm-rounding}.

\paragraph{Bounding the contribution of a small configuration.}
Set $H=4+\lceil\log_2(m+1)\rceil$. Fix a small configuration $(i,S)$ and write
\[
  q=\frac{\mu_{i,S}}{T},
  \qquad
  P(S)=\sum_{j\in S}\frac{b_j}{f_j}
      =\sum_{j\in S}\frac{1-f_j}{f_j}.
\]
\cref{lem:mm-weighted-support} gives $\mu_{i,S}>0$, while $\mu_{i,S}\le a^{i,S}(S)\le u^{i,S}(S)=T$, so $0<q\le1$. Applying \eqref{eq:mm-envelope} with $W=T$, $s=|S|\le m$, and $\nu=\max_{j\in S}u^{i,S}(\{j\})<T/\kappa$, and dividing by $T$, we obtain
\[
  \begin{aligned}
    \frac34
    &\le q\left[2+\log_2\frac{s\nu}{Tq}\right]
       +4\sqrt{\frac{\nu P(S)q}{T}} \le q\left[2+\log_2\frac{m}{q}\right]
       +4\sqrt{\frac{P(S)q}{\kappa}}.
  \end{aligned}
\]
The second inequality uses $\nu/T<1/\kappa\le1$. To verify \eqref{eq:mm-budget}, we consider two cases according to the value of $q$.

\emph{Case 1: $q\ge1/(16H)$.} Since $\tau=T/\kappa$ and $\mu_{i,S}=Tq$, we have
\[
  \min\left\{1,\left(\frac{32\tau}{\mu_{i,S}}\right)^3\right\}
  =\min\left\{1,\left(\frac{32}{\kappa q}\right)^3\right\}
  \le\left(\frac{512H}{\kappa}\right)^3.
\]

\emph{Case 2: $q<1/(16H)$.} The expression $q[2+\log_2(m/q)]$ is increasing in $q$ on $(0,1]$. Using $\log_2m\le H-4$ and $\log_2H\le H$, we obtain
\[
  q\left[2+\log_2\frac{m}{q}\right]
  <\frac{6+\log_2m+\log_2H}{16H}
  \le\frac{2H+2}{16H}
  <\frac38.
\]
Consequently, $4\sqrt{P(S)q/\kappa}\ge3/8$. Thus $P(S)>0$ and
\[
  q\ge\frac{9\kappa}{1024P(S)}.
\]
It follows that
\[
  \min\left\{1,\left(\frac{32\tau}{\mu_{i,S}}\right)^3\right\}
  =\min\left\{1,\left(\frac{32}{\kappa q}\right)^3\right\}
  \le\min\left\{1,\left(\frac{32768P(S)}{9\kappa^2}\right)^3\right\}
  \le\frac{32768P(S)}{9\kappa^2},
\]
where the last inequality uses $\min\{1,x^3\}\le x$ for $x\ge0$.

\paragraph{Verifying the budget condition.}
By \eqref{eq:mm-row}, the total mass of small configurations is at most $n$. The argument for \eqref{eq:mm-penalty-budget} uses only \eqref{eq:mm-row} and \eqref{eq:mm-capacity}, so it also applies to the present solution. Thus,
\[
  \sum_i\sum_{S\in\mathcal S_i}y_{i,S}\le n,
  \qquad
  \sum_i\sum_{S\in\mathcal S_i}y_{i,S}P(S)\le n.
\]
We apply the appropriate bound from the two cases to each small configuration and sum with weights $y_{i,S}$. Since all summands are nonnegative, we may extend each of the two resulting sums to all small configurations. This gives
  \begin{align}
    \sum_i\sum_{S\in\mathcal S_i}y_{i,S}
       \min\left\{1,\left(\frac{32\tau}{\mu_{i,S}}\right)^3\right\}
    &\le
       \left(\frac{512H}{\kappa}\right)^3
       \sum_i\sum_{S\in\mathcal S_i}y_{i,S}
       +\frac{32768}{9\kappa^2}
       \sum_i\sum_{S\in\mathcal S_i}y_{i,S}P(S)\nonumber\\
    &\le n(512H/\kappa)^3+\frac{32768n}{9\kappa^2}. \label{eq:mm-subadd-budget}
  \end{align}

Since $m\le128n^2$, we have $H\le13+2\log_2n\le32n^{1/6}$. Together with $\kappa\ge2^{20}\sqrt n$, this gives
\[
  n(512H/\kappa)^3+\frac{32768n}{9\kappa^2}
  \le2^{-18}+\frac{1}{9\cdot2^{25}}
  <\frac1{128}.
\]
Thus \eqref{eq:mm-budget} holds with $1/128$ in place of $1/4$. We retain this stronger bound to obtain the failure probability needed in \cref{subsec:mm-general}, as stated in \cref{cor:mm-local} below.

\paragraph{Empty configurations.}
\cref{subsec:mm-general} applies the following corollary to configuration solutions in which agents may have the empty configuration, so we first extend \cref{lem:mm-rounding} to such solutions. We treat $\emptyset$ as a small configuration without coefficients, skip the constructions of $u^{i,\emptyset}$ and $a^{i,\emptyset}$ above, omit it from the left side of \eqref{eq:mm-budget}, and require $a^{i,S_i}(A_i)\ge\tau$ only when $S_i\ne\emptyset$.

The matching construction represents $\emptyset$ by a private vertex $u_{i,\emptyset}$, just as for any other small configuration, so the marginal and product bounds in \cref{lem:mm-matching} still hold. If $S_i=\emptyset$, we set $A_i=\emptyset$ and skip the subsequent assignment steps for $i$, and this agent contributes nothing to any $C_j$. Moreover, in the proof of \cref{lem:mm-bound-Ai} for a nonempty small configuration $S$, we have $a(S\cap\emptyset)=0$ and hence $c_{i',\emptyset}=0$. Empty configurations therefore contribute nothing to $Y$, and that proof remains valid for every nonempty small configuration.

Although $\mu_{i,\emptyset}=0$, selecting $\emptyset$ is not counted as a failure. The final union bound therefore only includes nonempty small configurations:
\[
  \Pr\bigl[\exists i\in N:\ S_i\ne\emptyset
    \text{ and }a^{i,S_i}(A_i)<\tau\bigr]\le\sum_{i\in N}\sum_{\substack{S\in\mathcal S_i\\S\ne\emptyset}}
    y_{i,S}\min\left\{1,\left(\frac{32\tau}{\mu_{i,S}}\right)^3\right\},
\]
which is the left side of \eqref{eq:mm-budget} with the empty configurations omitted.

\begin{corollary}\label{cor:mm-local}
  Suppose $m\le128n^2$, and a feasible configuration solution with objective value $T$ for subadditive valuations is given. There is a rounding procedure that selects original configurations with their prescribed marginals and, with probability at least $127/128$, gives every agent value at least $T/\kappa$. If empty configurations are allowed, the value guarantee applies only to agents selecting a nonempty configuration.
\end{corollary}
\begin{proof}
  Apply the rounding procedure of \cref{lem:mm-rounding} to the solution constructed above, and recover the original configurations using the proportional rule of \cref{subsec:mm-algorithm}. This preserves their prescribed marginals without changing the allocation.

  Since the constructed coefficients are dominated by the valuations, \eqref{eq:mm-local-tail}, \eqref{eq:mm-subadd-budget}, and the union bound imply that the probability of any agent receiving value less than $T/\kappa$ is at most $1/128$.

  When empty configurations are present, the extension above leaves the estimates for nonempty configurations unchanged, and the union bound includes only nonempty configurations.
\end{proof}

\subsection{Sampling and rounding tools}
\label{subsec:mm-tools}
In this subsection and the next one, we design the $O(\sqrt{n})$-approximation for max-min allocation for subadditive valuations, with arbitrary number of items. This subsection presents three tools used in \cref{subsec:mm-general}. \cref{lem:mm-sampling} gives a value guarantee for independent item sampling when singleton values are sufficiently small. \cref{lem:mm-service} provides a conditional version of Feige's guiding-tree rounding~\cite{Feige09}. \cref{lem:mm-colored-service} combines independent runs of the rounding procedure in \cref{lem:mm-service} with a random coloring of the items to obtain an exponential bound on each agent's failure probability.

Feige and Huang~\cite{FH25}, building on Seddighin and Seddighin~\cite{SS24MMS}, combined independent trials of Feige's welfare rounding~\cite{Feige09} with random per-item contention resolution for maximin shares. We use their independent-sampling bound and adapt the repeated rounding to arbitrary feasible configuration solutions. We condition on the configuration chosen by each agent and assign each item to a random trial. This gives a failure bound that depends on the agent's total nonempty configuration mass.

\paragraph{Independent sampling.}
Let $v$ be a normalized monotone subadditive valuation with singleton values at most $\nu$. Let $Q$ be a random set that includes each item independently, and write $\xi=v(Q)$. By monotonicity and subadditivity, adding or removing an item changes the value by at most its singleton value. Thus, the Lipschitz condition in Feige and Huang~\cite[Lemma~3.1]{FH25} holds with constant $\nu$. Applying their lemma with both thresholds set to $a$ gives
\begin{equation}\label{eq:mm-product}
  \Pr[\xi>2a+k\nu]\Pr[\xi\le a]^2\le2^{-k}
  \qquad(a>0,\ k\in\mathbb Z_{\ge0}).
\end{equation}

\begin{lemma}\label{lem:mm-sampling}
  Let $v$ be a normalized monotone subadditive valuation, and let $S$ satisfy $v(S)=W>0$. Fix an integer $h\ge1$ and suppose $v(\{j\})\le W/(8h)$ for every $j\in S$. If $Q\subseteq S$ is obtained by retaining each item independently with probability at least $1/h$, then
  \[
    \Pr\left[v(Q)\ge\frac{W}{4h}\right]\ge\frac12.
  \]
\end{lemma}
\begin{proof}
  Apply Feige and Huang~\cite[Lemma~2.2]{FH25} with $t=h$. Their singleton bound is $8W/(23h)$, which is at least $W/(8h)$, and the value they guarantee with probability at least $1/2$ is $8W/(23h)\ge W/(4h)$.
\end{proof}

\paragraph{Conditional guiding-tree rounding.}
Feige~\cite{Feige09} orients the components of a randomly labeled guiding graph and compares the bundles at the two endpoints of an edge with a fixed label. We apply this endpoint argument on a tree. The item-mass bound of $1/2$ lets us truncate exploration in polynomial time while preserving a positive success probability conditional on each configuration.

\begin{lemma}\label{lem:mm-service}
  Let $(v_i)_{i\in N}$ be normalized monotone subadditive valuations. For each agent $i$, we are given an explicit distribution over configurations, where configuration $S$ has probability $y_{i,S}$, and the empty configuration is allowed. Suppose that
  \[
    \sum_{i\in N}\sum_{S\ni j}y_{i,S}\le\frac12
    \qquad(j\in M).
  \]
  Then there is a randomized guiding-tree rounding procedure that selects configurations $S_i$ with $\Pr[S_i=S]=y_{i,S}$ and returns a feasible allocation $G$ with $G_i\subseteq S_i$. For every $i\in N$ and every $S$ with $y_{i,S}>0$, it satisfies
  \[
    \Pr[v_i(G_i)\ge v_i(S)/2\mid S_i=S]\ge1/4.
  \]
  The procedure uses no valuation queries and requires only polynomially many combinatorial operations and sampling calls.
\end{lemma}
\begin{proof}
  If there is only one agent, sample its configuration and assign it all items in that configuration. The claim then holds with probability $1$, henceforth assume $n\ge2$. We first describe the tree construction and prove a conditional success probability of at least $1/2$, then limit the number of operations to obtain the stated guarantee.

  \emph{The labeled tree.}
  Consider an infinite tree in which every vertex has exactly one incident edge of color $i$ for each agent $i\in N$, and fix a root $r$. Each edge of color $i$ is independently labeled with a configuration $S$ chosen with probability $y_{i,S}$. Let $S_i$ be the label of the root's color-$i$ edge, so $\Pr[S_i=S]=y_{i,S}$.

  For each item $j$, retain precisely the edges whose labels contain $j$, then a color-$i$ edge is retained with probability $\sum_{S\ni j}y_{i,S}$. The total mass on $j$ is $\sum_{i\in N}\sum_{S\ni j}y_{i,S}$, which is at most $1/2$ by assumption. Hence, the expected number of retained edges from a vertex to new vertices is at most $1/2$. Since different edges are labeled independently, then for any fixed vertex, its retained component has expected size at most $\sum_{d\ge0}2^{-d}=2$. Thus, these components are finite almost surely. To construct the allocation, we only need to explore the retained component containing $r$ for each item $j$.

  \emph{The allocation rule.}
  A finite tree has either a central vertex or a central edge, given by the middle of a longest path. Orient the edges of each retained component away from its center. If the center is an edge, orient that edge in both directions. Every edge then points toward at least one endpoint, and every vertex has at most one incoming edge.

  For each vertex $w$ and agent $i$, consider the color-$i$ edge incident to $w$. We include item $j$ in $\widehat G_i(w)$ precisely when the edge's label contains $j$ and the edge points toward $w$ in the orientation for $j$. For a fixed $w$, these bundles are pairwise disjoint, since the orientation for each item gives $w$ at most one incoming edge. At the root, the color-$i$ edge has label $S_i$, so $\widehat G_i(r)\subseteq S_i$. Thus, the bundles $\widehat G_i(r)$ form a feasible allocation.


  The allocation rule and the endpoint comparison below are illustrated in \cref{fig:guiding-tree}.

  \begin{figure}[htbp]
    \centering
    \input{figures/guiding-tree}
    \caption{Guiding-tree rounding and the endpoint argument in \cref{lem:mm-service}, following Feige~\cite{Feige09}. Vertices are auxiliary tree positions; edge colors identify agents. (a) For a fixed item $j$, orient its retained component away from its center, with a central edge oriented both ways. At the fixed root $r$, assign $j$ to the agent whose edge points inward, if any. Different items may have different retained components and orientations. (b) Conditional on $S_i=S$, the two endpoint bundles for agent $i$ cover $S$ and have the same conditional distribution. These bundles are compared only in the analysis; the allocation is read at $r$.}
    \label{fig:guiding-tree}
  \end{figure}
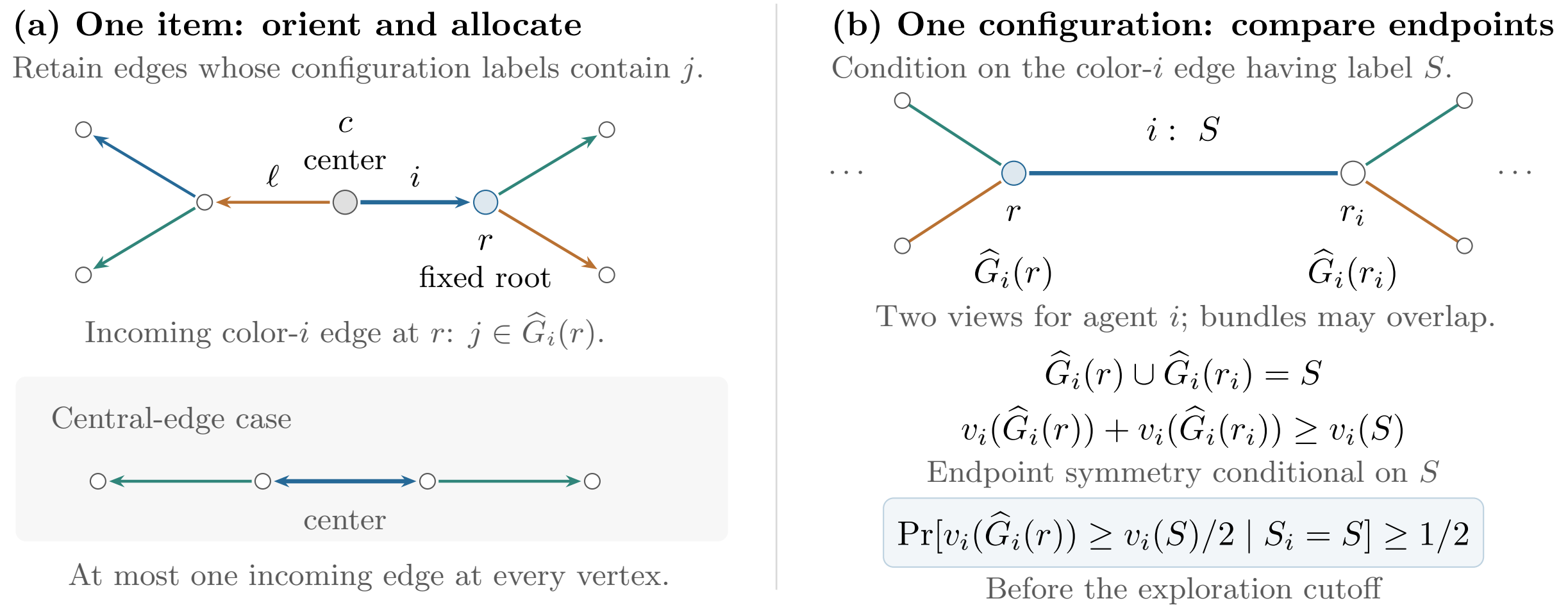

  \emph{The conditional value guarantee.}
  Fix an agent $i$ and a configuration $S$ with $y_{i,S}>0$, and condition on $S_i=S$. Let $r_i$ be the other endpoint of the root's color-$i$ edge. For every $j\in S$, this edge is retained and points toward at least one of $r$ and $r_i$. Consequently, $\widehat G_i(r)\cup\widehat G_i(r_i)=S$, and subadditivity gives
  \[
    v_i(\widehat G_i(r))+v_i(\widehat G_i(r_i))\ge v_i(S).
  \]
  Hence at least one of these two bundles has value at least $v_i(S)/2$.

  The two sides of the edge $\{r,r_i\}$ have the same tree structure, with corresponding edges having the same color. Exchanging the two sides swaps $r$ and $r_i$ while leaving their common edge and its label $S$ unchanged. Since all other edge labels are independent and have the same distribution on edges of the same color, this exchange preserves the conditional distribution of the labeled tree. The orientation rule depends only on each component's shape, so it is also preserved. Thus $\widehat G_i(r)$ and $\widehat G_i(r_i)$ have the same conditional distribution. The two events of receiving value at least $v_i(S)/2$ therefore have equal conditional probabilities. Since their union has probability $1$, we obtain
  \[
    \Pr[v_i(\widehat G_i(r))\ge v_i(S)/2\mid S_i=S]\ge1/2.
  \]

  \emph{Bounding the number of operations.}
  We first sample the labels of all edges incident to the root $r$. For each item $j$, we then explore the retained component containing $r$. An edge label is sampled only when first needed and is reused in all later explorations. Once these components have been explored, their centers and the bundles at the root can be computed.

  We count both combinatorial operations and sampling calls. Conditional on $S_i=S$, only the label of $\{r,r_i\}$ is fixed; all other labels retain their original independent distributions. For $j\in S$, removing this edge splits the retained component containing $r$ into two parts, one at each endpoint. Each part has expected size at most $2$ by the preceding bound; for $j\notin S$, the edge is absent, so only the part containing $r$ is explored. Each item's exploration therefore visits at most $4$ vertices in conditional expectation.

  Over all $m$ items, the conditional expected total number of vertex visits is therefore at most $4m$. Each visit requires at most $n$ edge examinations, and each new label is sampled from an explicit input distribution. Finding the centers and orienting the retained edges takes linear time in the explored component sizes. Together with the initial sampling and construction of the root bundles, these bounds give a polynomial $\omega$ in the input size that bounds the conditional expected number of operations, for all $i$ and $S$ with $y_{i,S}>0$.

  We stop the exploration after $4\omega$ operations. If this limit is reached before the exploration is complete, we return $G_i=\emptyset$ for every agent while keeping the already sampled configurations $S_i$. Otherwise, we return $G_i=\widehat G_i(r)$. By Markov's inequality, the conditional probability of reaching the limit is at most $1/4$. Subtracting this possible failure probability from the preceding $1/2$ bound gives
\begin{flalign*}
    &&\Pr[v_i(G_i)\ge v_i(S)/2\mid S_i=S]
    \ge\frac12-\frac14=\frac14. && \qedhere
\end{flalign*}
\end{proof}

\begin{lemma}\label{lem:mm-colored-service}
  Let $V>0$ and let $h\ge1$ be an integer. Consider a feasible configuration solution for subadditive valuations. For each agent $i$, let $r_i$ be the total mass of its nonempty configurations, with the remaining mass $1-r_i$ on the empty configuration. Suppose that, for every agent $i$, each nonempty configuration $S$ of agent $i$ satisfies $v_i(S)\ge V$ and $v_i(\{j\})\le V/(16h)$ for every $j\in S$. Then there is a randomized procedure that returns a feasible allocation $\bar G$ such that, for every $i\in N$,
  \[
    \Pr[v_i(\bar G_i)<V/(8h)]\le e^{-hr_i/16}.
  \]
\end{lemma}
\begin{proof}
  \emph{Constructing the allocation.}
  We halve the mass of every configuration and add mass $1/2$ to the empty configuration of every agent. The total mass of every agent remains $1$, and the total mass on every item is now at most $1/2$. We apply the guiding-tree rounding procedure of \cref{lem:mm-service} independently $h$ times, obtaining allocations $G^1,\ldots,G^h$.

  Independently of these runs, assign each item an independent uniformly random color in $\{1,\ldots,h\}$, and let $M_\ell$ be the set of items of color $\ell$. For each agent $i$, return $\bar G_i=\bigcup_{\ell=1}^h(G_i^\ell\cap M_\ell)$. Each $G^\ell$ is feasible, and the color classes are disjoint, so $\bar G$ is feasible.

  \emph{Success in one run.}
  Fix an agent $i$ and a run $\ell$. After halving the masses, agent $i$ selects a nonempty configuration with probability $r_i/2$, every such configuration has value at least $V$. The conditional guarantee of guiding-tree rounding therefore gives $v_i(G_i^\ell)\ge V/2$ with probability at least $(r_i/2)(1/4)=r_i/8$.

  Now fix an output with $W=v_i(G_i^\ell)\ge V/2$. Since $G_i^\ell$ is a subset of the selected configuration, every item in it has value at most $V/(16h)\le W/(8h)$. Meanwhile, each item belongs to $M_\ell$ independently with probability $1/h$. \cref{lem:mm-sampling} therefore gives a probability of at least $1/2$ that $v_i(G_i^\ell\cap M_\ell)\ge W/(4h)\ge V/(8h)$. Consequently,
  \[
    \Pr[v_i(G_i^\ell\cap M_\ell)\ge V/(8h)]
    \ge \frac{r_i}{8}\cdot\frac12
    =\frac{r_i}{16}.
  \]

  \emph{Combining the runs.}
  Let $g_\ell$ be the indicator of the event $v_i(G_i^\ell\cap M_\ell)<V/(8h)$. The preceding bound gives $\E[g_\ell]\le1-r_i/16$. To bound the probability that all runs fail, we prove that $\E[\prod_{\ell=1}^h g_\ell]\le\prod_{\ell=1}^h\E[g_\ell]$.

  Fix $G^1,\ldots,G^h$. For each item $j$, its color indicators $(\mathbf{1}_{\{j\in M_\ell\}})_{\ell=1}^h$ have a multinomial distribution with one trial and hence are negatively associated. The vectors for distinct items are independent, so their union is negatively associated~\cite{JDP83}. Each $g_\ell$ is a nonnegative decreasing function of the indicators for color $\ell$, and different colors use disjoint coordinates. The defining covariance inequality also holds for two decreasing functions, since negating both preserves their covariance. Applying it successively to products of the nonnegative $g_\ell$ gives
  \[
    \E\left[\prod_{\ell=1}^h g_\ell\,\middle|\,G^1,\ldots,G^h\right]
    \le\prod_{\ell=1}^h\E[g_\ell\mid G^\ell].
  \]
  Each factor on the right depends only on $G^\ell$. Since the guiding-tree rounding runs are independent, averaging over their outputs gives $\E[\prod_{\ell=1}^h g_\ell]\le\prod_{\ell=1}^h\E[g_\ell]$.

  Finally, $\bar G_i$ contains $G_i^\ell\cap M_\ell$ for every $\ell$. By monotonicity, if $v_i(\bar G_i)<V/(8h)$, then every run fails after coloring. Hence,
  \begin{flalign*}
    && \Pr[v_i(\bar G_i)<V/(8h)]
    \le \E\left[\prod_{\ell=1}^h g_\ell\right]
    \le \prod_{\ell=1}^h\E[g_\ell]
    \le (1-r_i/16)^h
    \le e^{-hr_i/16}. &&\qedhere
  \end{flalign*}
\end{proof}

\subsection{Subadditive rounding with arbitrarily many items}
\label{subsec:mm-general}

We now extend \cref{cor:mm-local} to an arbitrary number of items and prove \cref{thm:mm-subadditive}. For larger instances, we construct a core $K$ of $128n^2$ items through repeated rounds of greedy singleton selection and use one uniformly chosen round to provide backup items; some agents already meet the value requirement at this stage. For the remaining agents, we apply the rounding procedure of \cref{cor:mm-local} to the rest of $K$. We randomly divide the items outside $K$ into two parts. On one part, we apply repeated guiding-tree rounding with random colors, using the tools of \cref{subsec:mm-tools}. The other part is reserved for the few agents whose allocation outside $K$ falls short of the required value.

\paragraph{Core construction and backup selection.}
By \cref{lem:mm-lp-solution}, we obtain a configuration solution with objective value $T>T_{\mathrm{LP}}/6$. As before, set $\kappa=2^{20}\lceil\sqrt n\rceil$. If $m<128n^2$, we apply \cref{cor:mm-local}. Therefore we assume $m\ge128n^2$ in the following construction.

We construct the core in $128n$ rounds, at each round, we process the agents in a fixed order. Each agent chooses a remaining item of maximum singleton value, and the chosen item is removed from further selection. Let $K_t$ be the set of $n$ items chosen in round $t$, and define
\[
  K=\bigcup_{t=1}^{128n} K_t,\qquad \overline K=M\setminus K.
\]
Since no item is chosen twice, we have $|K|=128n^2$.

Choose $J$ uniformly from $\{1,\ldots,128n\}$ and let $B=K_J$. For each agent $i$, let $j_i$ be the item it chose in round $J$, which we use as its backup. Every item in $\overline K$ was available when $i$ chose $j_i$, so the greedy selection rule gives
\begin{equation}\label{eq:mm-backbone-domination}
  v_i(\{j_i\})\ge\max_{j\in \overline K}v_i(\{j\}).
\end{equation}
For any fixed agent $i$ and set $S\subseteq M$, we also have
\begin{equation}\label{eq:mm-backbone-loss}
  \Pr[v_i(S\setminus B)<v_i(S)/2]\le1/(128n).
\end{equation}
To see this, consider two distinct rounds $s$ and $t$. Since $K_s$ and $K_t$ are disjoint, $(S\setminus K_s)\cup(S\setminus K_t)=S$. Subadditivity therefore gives
\[
  v_i(S)\le v_i(S\setminus K_s)+v_i(S\setminus K_t).
\]
Thus, at most one round $t$ can satisfy $v_i(S\setminus K_t)<v_i(S)/2$. Since $J$ is uniform among the $128n$ rounds, \eqref{eq:mm-backbone-loss} follows.

\paragraph{The allocation algorithm.}
We set
\[
  \tau=\frac{T}{4096\kappa},\qquad
  V=\frac{T}{64}=64\kappa\tau,\qquad
  N_0=\{i:v_i(\{j_i\})<\tau\}.
\]
Here, $N_0$ is the set of agents whose backup items have value below $\tau$. By \eqref{eq:mm-backbone-domination}, every item in $\overline K$ is worth less than $\tau$ to each agent in $N_0$. The algorithm below allocates items from the four disjoint sets $B$, $K\setminus B$, $R_1$, and $R_2$, where $R_1$ and $R_2$ form a partition of $\overline K$.

\begin{enumerate}
  \item\label[step]{step:mm-backup} Every agent $i\notin N_0$ receives its backup item $j_i$.

  \item\label[step]{step:mm-classify} For each original configuration $(i,S)$ with $i\in N_0$, let $S_K=S\cap K$ and $S_{\overline K}=S\cap\overline K$. We call $(i,S)$ core-type if $v_i(S_K)\ge T/2$, and residual-type otherwise. By subadditivity, every residual-type configuration satisfies $v_i(S_{\overline K})>T/2$. Let $r_i$ be the total original mass of residual-type configurations of agent $i$.

  \item\label[step]{step:mm-core-configurations} We construct a configuration solution on $K\setminus B$. For each agent $i\in N_0$, we replace a core-type configuration $S$ by $S_K\setminus B$ if $v_i(S_K\setminus B)\ge v_i(S_K)/2$. We replace all other configurations, including those of agents outside $N_0$, by the empty set. We then merge identical resulting sets by adding their masses and record their original source configurations, as described in \cref{subsec:mm-algorithm}.

  \item\label[step]{step:mm-core-rounding} We run the rounding procedure of \cref{cor:mm-local} once on this solution with objective value $T/4$, allowing empty configurations. This selects a configuration for each agent and produces disjoint bundles contained in $K\setminus B$. For each agent $i$, we use the proportional rule to recover an original source configuration corresponding to its selected set and denote it by $S_i$. Each agent $i\in N_0$ with a core-type $S_i$ receives its bundle from this procedure. If $S_i$ is residual-type, its bundle is empty because its source configuration was replaced by the empty set in \cref{step:mm-core-configurations}; we defer its allocation to \cref{step:mm-residual-rounding}.

  \item\label[step]{step:mm-residual-configurations} We independently assign each item of $\overline K$ to $R_1$ or $R_2$, each with probability $1/2$. We construct a configuration solution on $R_1$ by replacing each original residual-type configuration $(i,S)$ with $i\in N_0$ by $S\cap R_1$ if $v_i(S\cap R_1)\ge V$, and by the empty set otherwise. All other configurations are also replaced by the empty set. We merge identical resulting sets by adding their masses.

  \item\label[step]{step:mm-residual-rounding} Using randomness independent of that used in \cref{step:mm-core-rounding}, we apply \cref{lem:mm-colored-service} to the solution from \cref{step:mm-residual-configurations} with $h=\kappa$ to obtain a candidate allocation $\bar G$ on $R_1$. We now consider only agents $i\in N_0$ whose source configuration $S_i$ is residual-type. If $v_i(\bar G_i)\ge\tau$, agent $i$ receives $\bar G_i$; otherwise, we put $i$ in a queue $\mathcal Q$. If $|\mathcal Q|\le\kappa$, we attempt the reserve repair described below using $R_2$.

  \item\label[step]{step:mm-combine} We combine the bundles assigned in \cref{step:mm-backup,step:mm-core-rounding,step:mm-residual-rounding}, including those supplied by the reserve repair. If every agent receives value at least $\tau$, we return the allocation; otherwise, we declare failure.
\end{enumerate}

\paragraph{Bounding the queue and allocating the reserve.}
For every fixed $B$, the source configurations recovered in \cref{step:mm-core-rounding} have their original marginals. In particular, each agent $i\in N_0$ selects a residual-type configuration with probability $r_i$. We first bound the probability that an agent in $N_0$ selects a core-type source that was replaced by the empty set in \cref{step:mm-core-configurations}.

Fix an original configuration $(i,S)$ with $v_i(S\cap K)\ge T/2$. Whenever $i\in N_0$, this source is discarded only if $v_i((S\cap K)\setminus B)<v_i(S\cap K)/2$. By \eqref{eq:mm-backbone-loss}, the probability of this event over the choice of $B$ is at most $1/(128n)$. Since $\Pr[S_i=S\mid B]=y_{i,S}$ for every $B$, the probability that $i\in N_0$ selects this source and the source is discarded is at most $y_{i,S}/(128n)$. A union bound over the agents and their core-type source configurations therefore bounds the probability of any such failure by
\[
  \frac{1}{128n}\sum_{i\in N}\sum_S y_{i,S}
  =\frac{1}{128}.
\]
Here, we extended the sum to all original configurations and used \eqref{eq:mm-row}.

We next analyze the random partition. Fix $B$, so that $N_0$ is also fixed. Recall that every item in $\overline K$ has value less than $\tau$ to every agent in $N_0$. For each $i\in N_0$, let $r'_i\le r_i$ be the total mass of the nonempty configurations retained in \cref{step:mm-residual-configurations}, and define
\[
  \Delta=\sum_{i\in N_0}(r_i-r'_i),\qquad
  \epsilon_s=2^{1-32\kappa}.
\]
Thus, $\Delta$ measures the total mass of residual-type configurations discarded because their restrictions to $R_1$ have value below $V$. We prove that this loss is small in expectation and that, with high probability, $R_2$ has value at least $V$ to every agent that may enter the queue:
\begin{equation}\label{eq:mm-half-reserve}
  \E[\Delta\mid B]\le n\epsilon_s,\qquad
  \Pr[\exists i\in N_0:\ r_i>0,\ v_i(R_2)<V\mid B]
       \le n\epsilon_s.
\end{equation}

To establish both bounds, fix $i\in N_0$ and $Q\subseteq\overline K$ with $v_i(Q)=W>T/2$. By subadditivity,
\[
  v_i(Q\cap R_1)+v_i(Q\cap R_2)\ge W.
\]
At least one of the two values is therefore at least $W/2$. Since the two values have the same distribution, each is at least $W/2$ with probability at least $1/2$. For either $\ell\in\{1,2\}$, write $\xi=v_i(Q\cap R_\ell)$. Applying \eqref{eq:mm-product} with $a=V=T/64$, $k=64\kappa$, and singleton bound $\tau$ gives
\[
  \Pr[\xi>2V+64\kappa\tau\mid B]
  \Pr[\xi\le V\mid B]^2\le2^{-64\kappa}.
\]
Since $2V+64\kappa\tau=3T/64<W/2$, the first probability is at least $1/2$. Consequently,
\[
  \Pr[v_i(Q\cap R_\ell)<V\mid B]
  \le\sqrt{2}\,2^{-32\kappa}
  \le\epsilon_s
  \qquad(\ell=1,2).
\]

For the first bound in \eqref{eq:mm-half-reserve}, apply this estimate to $Q=S_{\overline K}$ for every original residual-type configuration $(i,S)$. Since $S\cap R_1=S_{\overline K}\cap R_1$, we obtain
\begin{align*}
  \E[r_i-r'_i\mid B]
  &=\sum_{S:\,(i,S)\text{ is residual-type}}y_{i,S}
    \Pr[v_i(S\cap R_1)<V\mid B]
  \le r_i\epsilon_s\le\epsilon_s.
\end{align*}
Summing over $i\in N_0$ proves $\E[\Delta\mid B]\le n\epsilon_s$. For the second bound, $r_i>0$ implies that $i$ has a residual-type configuration and hence, by monotonicity, $v_i(\overline K)>T/2$. Apply the same estimate to $Q=\overline K$ and $\ell=2$, and take a union bound over these agents.

We now bound the number of agents entering the queue. To do so, we first check that \cref{lem:mm-colored-service} applies in \cref{step:mm-residual-rounding}. The solution constructed in \cref{step:mm-residual-configurations} is feasible: replacing configurations by subsets preserves \eqref{eq:mm-row} and cannot increase the total mass on any item, and merging identical sets preserves both. Every nonempty configuration has value at least $V$ by construction. Moreover, every item in such a configuration has value less than $\tau=V/(64\kappa)<V/(16\kappa)$ to its agent. Thus, all hypotheses of the lemma hold with $h=\kappa$ and nonempty mass $r'_i$ for agent $i$. Its value guarantee is $V/(8\kappa)=8\tau$, so in particular
\[
  \Pr[v_i(\bar G_i)<\tau\mid B,R_1,R_2]
  \le e^{-\kappa r'_i/16}.
\]

An agent $i\in N_0$ enters $\mathcal Q$ exactly when $S_i$ is residual-type and $v_i(\bar G_i)<\tau$. Conditioned on $(B,R_1,R_2)$, these two events are independent: \cref{step:mm-residual-rounding} uses randomness independent of \cref{step:mm-core-rounding} and is run on the solution from \cref{step:mm-residual-configurations} regardless of the recovered source configurations. The probability of the first event is $r_i$, because the partition is independent of \cref{step:mm-core-rounding} and source recovery preserves the marginals. Therefore,
\begin{align*}
  \E[|\mathcal Q|\mid B,R_1,R_2]
  &=\sum_{i\in N_0}r_i
    \Pr[v_i(\bar G_i)<\tau\mid B,R_1,R_2]
   \le\sum_{i\in N_0}r_i e^{-\kappa r'_i/16}\\
  &\le\sum_{i\in N_0}r'_i e^{-\kappa r'_i/16}+\Delta
   \le\frac{16n}{e\kappa}+\Delta.
\end{align*}
The second inequality uses $r_i=r'_i+(r_i-r'_i)$ and $e^{-\kappa r'_i/16}\le1$. The last inequality follows from $t e^{-\kappa t/16}\le16/(e\kappa)$ for $t\ge0$, with equality at $t=16/\kappa$.

Averaging over the partition and then over $B$, and using \eqref{eq:mm-half-reserve}, gives $\E[|\mathcal Q|]\le16n/(e\kappa)+n\epsilon_s$. Markov's inequality now yields
\[
  \Pr[|\mathcal Q|>\kappa]
  \le\frac{\E[|\mathcal Q|]}{\kappa}
  \le\frac{16n}{e\kappa^2}+\frac{n\epsilon_s}{\kappa}
  <\frac1{128},
\]
where the final inequality uses $\kappa=2^{20}\lceil\sqrt n\rceil$ and $\epsilon_s=2^{1-32\kappa}$.

Finally, we describe the deterministic reserve repair and prove that it succeeds whenever $|\mathcal Q|\le\kappa$ and $v_i(R_2)\ge V$ for every $i\in\mathcal Q$. Every queued agent has $r_i>0$, so the latter condition holds outside the reserve failure event in \eqref{eq:mm-half-reserve}.

Scan the items of $R_2$ in a fixed order, adding them one at a time to a bundle until some waiting agent values it at least $\tau$. Assign the bundle to one such agent, remove the agent from the queue, and repeat with the remaining items. Before the last item was added to an allocated bundle, every waiting agent valued that bundle below $\tau$. The last item also has value below $\tau$ to each of these agents. By subadditivity, the allocated bundle therefore has value below $2\tau$ to every agent still waiting.

To see that the procedure can continue until the queue is empty, suppose that $t$ bundles $P_1,\ldots,P_t$ have been allocated and let $U=R_2\setminus\bigcup_{\ell=1}^t P_\ell$ be the remaining items. For any agent $i$ still waiting, we have $t\le\kappa-1$ and
\[
  v_i(U)
  \ge v_i(R_2)-\sum_{\ell=1}^t v_i(P_\ell)
  \ge V-2t\tau
  \ge64\kappa\tau-2(\kappa-1)\tau
   >\tau.
\]
Thus, scanning $U$ must eventually produce a bundle of value at least $\tau$ to a waiting agent. Repeating this argument shows that every agent in the queue receives such a bundle. The bundles are disjoint because each assigned bundle is removed before the next scan. Each item is added to a bundle at most once, and each addition requires at most $n$ value queries, so the procedure uses $O(n|R_2|)$ value queries.

\begin{theorem}\label{thm:mm-subadditive}
  For normalized monotone subadditive valuations and any number of items, there is a randomized polynomial-time $O(\sqrt n)$-approximation algorithm for max-min allocation using ordinary demand queries. A single run succeeds with probability at least $3/4$, and independent repetitions can make the failure probability arbitrarily small.
\end{theorem}
\begin{proof}
  We first consider $m\ge128n^2$ and analyze one run assuming exact sampling from explicitly given rational distributions. Fix the backup set $B$, then every core-type source configuration $(i,S)$ retained in \cref{step:mm-core-configurations} satisfies
  \[
    v_i((S\cap K)\setminus B)
    \ge\frac12v_i(S\cap K)
    \ge\frac{T}{4}.
  \]
  Thus, the solution constructed in \cref{step:mm-core-configurations} is a feasible configuration solution with objective value $T/4$ on at most $128n^2$ items, with empty configurations allowed. By \cref{cor:mm-local}, the rounding in \cref{step:mm-core-rounding} gives every agent selecting a nonempty configuration value at least $T/(4\kappa)$, except with probability at most $1/128$. This bound holds for every fixed $B$, and hence also after averaging over $B$.

  If an agent $i\in N_0$ recovers a core-type source $S_i$ that was not discarded, its selected core configuration is nonempty. On the success event of \cref{step:mm-core-rounding}, every such agent therefore receives a bundle $A_i$ satisfying
  \[
    v_i(A_i)\ge\frac{T}{4\kappa}
    =1024\tau>\tau.
  \]
  In particular, the guarantee applies simultaneously to all agents with retained core-type sources.

  The preceding analysis bounds three other failure events. The probability that some agent in $N_0$ selects a discarded core-type source is at most $1/128$. By \eqref{eq:mm-half-reserve}, after averaging over $B$, the probability that $v_i(R_2)<V$ for some $i\in N_0$ with $r_i>0$ is at most $n\epsilon_s<1/128$. Finally, $\Pr[|\mathcal Q|>\kappa]<1/128$.

  If none of these four events occurs, every agent receives value at least $\tau$. Agents outside $N_0$ receive backup items of value at least $\tau$. Agents in $N_0$ with core-type sources satisfy the bound above. Each remaining agent either receives a bundle $\bar G_i$ of value at least $\tau$ in \cref{step:mm-residual-rounding} or enters $\mathcal Q$. Every queued agent has $r_i>0$, so the reserve has value at least $V$ to it; since $|\mathcal Q|\le\kappa$, the reserve repair succeeds for all queued agents. The resulting bundles are disjoint because each allocation is feasible within its item set, and $B$, $K\setminus B$, $R_1$, and $R_2$ are disjoint.

  A union bound therefore gives success probability at least $1-4/128=31/32$. Since $T>T_{\mathrm{LP}}/6\ge\mathrm{OPT}_{\mathrm{MM}}/6$, the resulting allocation satisfies
  \[
    \min_i v_i(A_i)\ge\tau
      >\frac{\mathrm{OPT}_{\mathrm{MM}}}{24576\kappa}
      =\frac{\mathrm{OPT}_{\mathrm{MM}}}{O(\sqrt n)}.
  \]
  For $m<128n^2$, \cref{cor:mm-local} directly gives every agent value at least $T/\kappa$ with probability at least $127/128$, so both the value and probability guarantees are stronger.

  It remains to obtain a worst-case polynomial-time implementation. The limits on tree exploration are already included in \cref{lem:mm-service}. All other combinatorial operations and the number of sampling calls are polynomially bounded. As explained in \cref{subsec:rational-sampling}, we can cap the rational samplers so that the total abort probability is at most $1/16$. This adds at most $1/16$ to the failure probability, giving success probability at least $31/32-1/16>3/4$. All value evaluations used by the algorithm can be implemented using ordinary demand queries by \cref{app:demand-values}.

  Finally, we repeat the entire algorithm independently and keep the returned allocation with the largest minimum agent value. After $q$ runs, the probability that all runs fail is at most $4^{-q}$.
\end{proof}

\paragraph{Proof of \cref{thm:intro-maxmin}.}
If \cref{lem:mm-lp-solution} detects $T_{\mathrm{LP}}=0$, every allocation is optimal. Otherwise, \cref{part:mm-additive,part:mm-submodular} follow from \cref{thm:mm-exact}, applied to the configuration solution of \cref{lem:mm-lp-solution} with the supporting clauses constructed in \cref{subsec:mm-exact}. With probability at least $3/4$, every agent receives value at least $T/(128\lceil\sqrt n\rceil)>\mathrm{\mathrm{OPT}_{\mathrm{MM}}}/(768\lceil\sqrt n\rceil)$. \cref{part:mm-subadditive} is \cref{thm:mm-subadditive}.

%% file: figures/guiding-tree.tex
\begingroup
\definecolor{GTblue}{RGB}{38,105,151}
\definecolor{GTteal}{RGB}{48,133,122}
\definecolor{GTorange}{RGB}{185,113,50}
\begin{tikzpicture}[
  x=1cm,y=1cm,font=\small,
  vertex/.style={circle,draw=black!65,fill=white,inner sep=0pt,minimum size=4.5pt},
  edge/.style={line width=.85pt},
  flow/.style={edge,-{Stealth[length=4.5pt,width=3.5pt]}},
  agenti/.style={draw=GTblue},
  agentk/.style={draw=GTteal},
  agentl/.style={draw=GTorange},
  note/.style={font=\footnotesize,text=black!65,align=center},
  every node/.style={outer sep=1pt}
]
\path[use as bounding box] (0,.15) rectangle (16.3,-6.3);
\draw[black!15,line width=.5pt] (8.05,-.1)--(8.05,-6.15);

\node[anchor=north west,font=\small\bfseries] at (0,0)
  {(a) One item: orient and allocate};
\node[note,anchor=north west] at (0,-.48)
  {Retain edges whose configuration labels contain $j$.};
\node[vertex,minimum size=7pt,fill=black!12] (c) at (3.6,-2.15) {};
\node[vertex] (u) at (2.15,-2.15) {};
\node[vertex,minimum size=7pt,fill=GTblue!15,draw=GTblue] (r) at (5.05,-2.15) {};
\node[vertex] (ul) at (.9,-1.4) {};
\node[vertex] (dl) at (.9,-2.9) {};
\node[vertex] (ur) at (6.3,-1.4) {};
\node[vertex] (dr) at (6.3,-2.9) {};
\draw[flow,agentl] (c)--node[above,font=\footnotesize] {$\ell$}(u);
\draw[flow,agenti,line width=1.25pt] (c)--node[above,font=\footnotesize] {$i$}(r);
\draw[flow,agenti] (u)--(ul);
\draw[flow,agentk] (u)--(dl);
\draw[flow,agentk] (r)--(ur);
\draw[flow,agentl] (r)--(dr);
\node[above=5pt,align=center] at (c) {$c$\\[-1pt]\footnotesize center};
\node[below=5pt,align=center] at (r) {$r$\\[-1pt]\footnotesize fixed root};
\node[note] at (3.6,-3.48)
  {Incoming color-$i$ edge at $r$: $j\in\widehat G_i(r)$.};

\path[fill=black!3,rounded corners=3pt] (.2,-3.95) rectangle (7.55,-5.65);
\node[note,anchor=north west] at (.4,-4.1) {Central-edge case};
\node[vertex] (a0) at (1.05,-5.03) {};
\node[vertex] (a1) at (2.75,-5.03) {};
\node[vertex] (a2) at (4.45,-5.03) {};
\node[vertex] (a3) at (6.15,-5.03) {};
\draw[flow,agentk] (a1)--(a0);
\draw[edge,agenti,line width=1.25pt,
  {Stealth[length=4.5pt,width=3.5pt]}-{Stealth[length=4.5pt,width=3.5pt]}]
  (a1)--node[below=4pt,note] {center}(a2);
\draw[flow,agentk] (a2)--(a3);
\node[note] at (3.85,-6.02) {At most one incoming edge at every vertex.};

\begin{scope}[xshift=8.45cm]
\node[anchor=north west,font=\small\bfseries] at (0,0)
  {(b) One configuration: compare endpoints};
\node[note,anchor=north west] at (0,-.48)
  {Condition on the color-$i$ edge having label $S$.};
\node[vertex,minimum size=7pt,fill=GTblue!15,draw=GTblue] (b0) at (2.05,-1.85) {};
\node[vertex,minimum size=7pt] (b1) at (5.55,-1.85) {};
\draw[edge,agenti,line width=1.25pt] (b0)--node[above=5pt] {$i:\ S$}(b1);
\foreach \yy/\sty in {-1.1/agentk,-2.6/agentl}{
  \draw[edge,\sty] (b0)--(.9,\yy);
  \draw[edge,\sty] (b1)--(6.7,\yy);
  \node[vertex] at (.9,\yy) {};
  \node[vertex] at (6.7,\yy) {};
}
\node[note] at (.35,-1.85) {$\cdots$};
\node[note] at (7.25,-1.85) {$\cdots$};
\node[below=5pt] at (b0) {$r$};
\node[below=5pt] at (b1) {$r_i$};
\node at (2.05,-2.85) {$\widehat G_i(r)$};
\node at (5.55,-2.85) {$\widehat G_i(r_i)$};
\node[note,text width=7.3cm] at (3.8,-3.35) {Two views for agent $i$; bundles may overlap.};
\node at (3.8,-3.9) {$\widehat G_i(r)\cup\widehat G_i(r_i)=S$};
\node at (3.8,-4.47)
  {$v_i(\widehat G_i(r))+v_i(\widehat G_i(r_i))\ge v_i(S)$};
\node[note] at (3.8,-4.96) {Endpoint symmetry conditional on $S$};
\node[fill=GTblue!6,draw=GTblue!30,rounded corners=3pt,inner sep=4pt]
  at (3.8,-5.56)
  {$\Pr[v_i(\widehat G_i(r))\ge v_i(S)/2\mid S_i=S]\ge1/2$};
\node[note] at (3.8,-6.16) {Before the exploration cutoff};
\end{scope}
\end{tikzpicture}
\endgroup

%% file: nsw.tex
\section{Weighted Nash social welfare}\label{sec:nsw}
In this section, we present an approximation algorithm for weighted Nash social welfare with subadditive valuations using ordinary demand queries. The approximation factor depends on the entropy of the agents' weights but not on the number of items. We first prove a rounding theorem for an explicit configuration solution and then show how to construct such a solution using demand queries.

\subsection{The configuration setting}\label{subsec:nsw-setting}
We use the weighted Nash social welfare objective and notation of \cref{sec:prelim} and assume throughout this section that the valuations are subadditive. We define the entropy of the weight vector and the associated approximation factor by
\begin{equation}\label{eq:nsw-entropy}
  H(w)=-\sum_{i\in N}w_i\ln w_i,
  \qquad
  \Gamma(w)=\prod_{i\in N}(nw_i)^{w_i/2}
    =\exp\!\left(\frac{\ln n-H(w)}2\right).
\end{equation}
Since $0\le H(w)\le\ln n$, we have $1\le\Gamma(w)\le\sqrt n$. For equal weights, $\Gamma(w)=1$.

Consider an explicit configuration solution $(\mathcal C_i,y_{i,S})_{i\in N}$ satisfying \eqref{eq:mm-row} and \eqref{eq:mm-capacity}. We assume that $v_i(S)>0$ for every $S\in\mathcal C_i$ and call such a solution positive. Configurations of the same agent may have different values. Unless stated otherwise, sums indexed by $(i,S)$ range over $i\in N$ and $S\in\mathcal C_i$. Constraint \eqref{eq:mm-row} and the normalization of the weights imply
\[
  \sum_{i,S}y_{i,S}=n,
  \qquad
  \sum_{i,S}w_i y_{i,S}=1.
\]
We use $L(y)$ from \eqref{eq:nsw-benchmark} as the logarithmic benchmark of this solution.

\begin{theorem}\label{thm:nsw-rounding}
  Given an explicit positive configuration solution as above, there is a randomized polynomial-time algorithm that returns a feasible allocation $A$ satisfying $v_i(A_i)>0$ for every agent and
  \begin{equation}\label{eq:nsw-rounding}
    \E\ln\mathrm{NSW}_w(A)
      \ge L(y)-\ln\Gamma(w)-C,
  \end{equation}
  where $C$ is an absolute constant.
\end{theorem}

For instances with positive optimum, \cref{subsec:nsw-application} constructs a positive feasible configuration solution whose benchmark is within an additive constant of $\ln\mathrm{OPT}_{\mathrm{NSW}}$. Together with the theorem above, this gives an $O(\Gamma(w))$ approximation.

Throughout this section, $C$ denotes an absolute constant whose value may change between statements.

\subsection{The rounding algorithm}\label{subsec:nsw-algorithm}
If $n=1$, we assign all items to the only agent. Henceforth, we assume $n\ge2$.

\paragraph{Reserving one singleton from each configuration.}
Following the large-item rule of Feng, Hu, Li, and Zhang~\cite{FHLZ25} for submodular valuations, we reserve one item per configuration. For each $(i,S)$, choose an item of maximum singleton value, breaking ties according to a fixed rule, and define
\begin{equation}\label{eq:nsw-heads}
  g_{i,S}\in\argmax_{j\in S}v_i(\{j\}),
  \qquad
  \nu_{i,S}=v_i(\{g_{i,S}\}),
  \qquad
  T_{i,S}=S\setminus\{g_{i,S}\}.
\end{equation}
We call $g_{i,S}$ the \emph{head} of the configuration and $T_{i,S}$ its \emph{tail}. Since $v_i(S)>0$, subadditivity implies $\nu_{i,S}>0$. By construction, every item in $T_{i,S}$ has singleton value at most $\nu_{i,S}$.

For each agent $i$ and item $j$, let $x_{ij}$ be the total mass of configurations of agent $i$ whose head is $j$. Define
\begin{equation}\label{eq:nsw-head-marginals}
  x_{ij}=\sum_{S:\,g_{i,S}=j}y_{i,S},
  \qquad
  b_j=\sum_{i\in N}x_{ij},
  \qquad f_j=1-b_j.
\end{equation}
The quantities $b_j$ and $f_j$ play the same roles as in \cref{subsec:mm-setting}. The configuration constraints imply that every row of $x$ sums to one and every column sums to at most one. In particular, $\sum_{j\in M}b_j=n$.

For each item $j$, the configurations containing $j$ have total mass at most one. Of this mass, $b_j$ comes from configurations whose head is $j$, the rest comes from configurations whose tail contains $j$. Thus,
\begin{equation}\label{eq:nsw-tail-capacity}
  \sum_{i,S:\,j\in T_{i,S}}y_{i,S}\le f_j=1-b_j
  \qquad(j\in M).
\end{equation}
In the analysis, we use this inequality and $\sum_{j\in M}b_j=n$ to bound the total loss from deleting reserved items.

\paragraph{Matching the heads and allocating the tails.}
We first sample a covering matching $\pi:N\to M$ that assigns each agent $i$ a distinct item $\pi(i)$ with $x_{i,\pi(i)}>0$. We use the distribution in \cref{lem:nsw-matching}, which has edge marginals $\Pr[\pi(i)=j]=x_{ij}$ and provides the negative-correlation and coupling properties needed in the analysis. Let
\[
  B=\{\pi(i):i\in N\}
\]
be the set of matched items.

Dobzinski, Li, Rubinstein, and Vondr{\'a}k~\cite{DLRV24} used repeated welfare rounding and random filtering for unweighted subadditive NSW. Feige and Huang~\cite{FH25} used independent runs of Feige's rounding~\cite{Feige09} with uniform contention resolution for maximin shares. Here we allocate the tails independently of the head matching, using the conditional version of guiding-tree rounding from \cref{lem:mm-service} and geometrically decreasing retention probabilities. As in the proof of \cref{lem:mm-colored-service}, we first halve each tail mass and add mass $1/2$ to an empty index for every agent, so that the total mass on every item is at most $1/2$. We then run the procedure independently for
\[
  k=0,\ldots,K,\qquad K=\lceil\log_2(m+1)\rceil,
\]
and denote the allocation from run $k$ by $G^k$. We then color the items independently, assigning each item color $k$ with probability $2^{-k-1}$ and leaving it uncolored with the remaining probability. The coloring is independent of the matching and the rounding runs. Let $M_k$ be the set of items of color $k$, the combined tail allocation is
\begin{equation}\label{eq:nsw-tail-output}
  Z_i=\bigcup_{k=0}^K(G_i^k\cap M_k).
\end{equation}
Finally, we remove the matched items from the tail bundles and give each agent its matched head together with the remaining tail items:
\begin{equation}\label{eq:nsw-output}
  A_i=\{\pi(i)\}\cup(Z_i\setminus B).
\end{equation}
Each $G^k$ is feasible, and the color classes are disjoint, so $Z$ is feasible. Removing $B$ from the tail bundles prevents any overlap with the matched heads, thus $A$ is feasible.

Following the maximum-entropy approach to prescribed marginals of Singh and Vishnoi~\cite{SV14}, we use product-weight (Gibbs) distributions on matchings. \cref{subsec:nsw-matching} obtains exact edge marginals $x_{ij}$ by taking a limit. In the finite implementation, we calibrate rational edge weights to approximate $x$ and use the permanent approximation of Jerrum, Sinclair, and Vigoda~\cite{JSV04} to sample a covering matching. \cref{app:nsw-finite} establishes the required guarantees for this implementation and shows that one calibration can be reused across repeated trials.

\paragraph{Overview of the analysis.}
We now outline how to compare the configuration benchmark $L(y)$ with the expected logarithmic welfare of the final allocation. For each fixed configuration $(i,S)$, define its \emph{buffered residual value} by
\begin{equation}\label{eq:nsw-buffer}
  D_{i,S}(B)=\nu_{i,S}+v_i(T_{i,S}\setminus B).
\end{equation}
This quantity adds the value of the configuration's head to the value remaining in its tail after removing $B$. It is used only for comparison: the head actually assigned to agent $i$ may come from another configuration, and $D_{i,S}(B)$ may exceed $v_i(S)$. In all expectations $\E_B$ below, we use the matching distribution without conditioning on any selected configuration.

First, we bound how much the configuration values can exceed these auxiliary values, in aggregate:
\begin{equation}\label{eq:nsw-loss-roadmap}
  \sum_{i,S}y_{i,S}\E_B
    \left(\frac{v_i(S)}{D_{i,S}(B)}\right)^2
    \le Cn.
\end{equation}
Second, we show that the expected logarithmic welfare of $A$ is at least the weighted average of $\ln D_{i,S}(B)$, up to an additive constant:
\begin{equation}\label{eq:nsw-utility-roadmap}
  \E\ln\mathrm{NSW}_w(A)
    \ge\sum_{i,S}w_i y_{i,S}\E_B\ln D_{i,S}(B)-C.
\end{equation}
The first estimate and Jensen's inequality show that this weighted average is at least $L(y)-\ln\Gamma(w)-O(1)$, the second estimate then gives \cref{thm:nsw-rounding}.

\subsection{Singleton matching}\label{subsec:nsw-matching}
\begin{lemma}\label{lem:nsw-matching}
  Let $x$ be the matrix in \eqref{eq:nsw-head-marginals}. There exists a random covering matching $\pi:N\to M$ using only edges $(i,j)$ with $x_{ij}>0$. This matching and its set of matched items $B=\{\pi(i):i\in N\}$ satisfy the following properties.
  \begin{enumerate}[label=(\roman*),ref=\roman*]
    \item\label[matchingproperty]{prop:nsw-marginals} For every $i\in N$ and $j\in M$, $\Pr[\pi(i)=j]=x_{ij}$, hence $\Pr[j\in B]=b_j$ for every $j\in M$.
    \item\label[matchingproperty]{prop:nsw-negative-correlation} For every set of items $J\subseteq M$,
      \begin{equation}\label{eq:nsw-cylinder}
        \Pr[J\subseteq B]\le\prod_{j\in J}b_j.
      \end{equation}
    \item\label[matchingproperty]{prop:nsw-reference-coupling} For each fixed agent $i$, $\pi$ can be coupled with a random reference set $\widehat B_i\subseteq M$. This set has the same distribution as $B$ and is independent of $\pi(i)$. The coupling also guarantees that
      \begin{equation}\label{eq:nsw-one-item}
        |B\setminus\widehat B_i|\le1,
        \qquad |\widehat B_i\setminus B|\le1.
      \end{equation}
  \end{enumerate}
  These couplings are used only in the analysis and may be chosen separately for different agents.
\end{lemma}
\begin{proof}
  We first prove the negative-correlation and coupling properties for any positive choice of edge weights. We then choose weights whose edge marginals approach $x$ and take a limit to obtain the required distribution.

  \emph{Gibbs distributions.}
  Give each edge $(i,j)$ with $x_{ij}>0$ a positive weight $\lambda_{ij}$, and let $\mathcal M$ be the set of covering matchings using these edges. Feasibility of $x$ ensures that $\mathcal M$ is nonempty. Define
  \begin{equation}\label{eq:nsw-gibbs}
    \mu_\lambda(P)=\frac{\prod_{(i,j)\in P}\lambda_{ij}}{Z_\lambda},
    \qquad
    Z_\lambda=\sum_{P\in\mathcal M}\prod_{(i,j)\in P}\lambda_{ij}.
  \end{equation}

  \emph{The reference coupling.}
  Fix an agent $i$, let $p_{ij}$ be the probability that a matching drawn from $\mu_\lambda$ assigns $j$ to $i$. For each $j$ with $p_{ij}>0$, we will couple a matching conditioned on assigning $j$ to $i$ with a matching whose distribution is $\mu_\lambda$ and does not depend on $j$.

  Draw $P$ according to $\mu_\lambda$, and independently of $P$, draw $Q$ according to $\mu_\lambda$ conditioned on containing $(i,j)$. As in M\'esz\'aros~\cite{Meszaros23}, we use the alternating components of $P\mathbin\triangle Q$. These components are paths and cycles, and every path has items as its endpoints because both matchings cover every agent. If $P$ contains $(i,j)$, set $C=\emptyset$; otherwise, let $C$ be the component containing $i$, viewed as a set of edges. Define
  \begin{equation}\label{eq:nsw-switch}
    (P,Q)\longmapsto(P^*,Q^*)
      =(Q\mathbin\triangle C,\ P\mathbin\triangle C).
  \end{equation}
  Thus $P^*$ consists of the edges of $P$ in $C$ and the edges of $Q$ outside $C$. Similarly, $Q^*$ consists of the edges of $Q$ in $C$ and the edges of $P$ outside $C$.

  We claim that this switch preserves the joint distribution of $(P,Q)$. First, we check that the second matching still contains $(i,j)$ after the switch and that applying the same rule again returns $(P,Q)$. If $C=\emptyset$, both $P$ and $Q$ contain $(i,j)$, and the switch simply exchanges the two matchings. Both new matchings therefore still contain $(i,j)$, so the rule again selects $C=\emptyset$, then a second swap returns $(P,Q)$.

  Suppose now that $C\ne\emptyset$. The edge $(i,j)$ lies in $C$ and belongs to $Q$ but not $P$. Since each new matching keeps its original edges on $C$, this edge belongs to $Q^*$ but not $P^*$. Moreover,
  \[
    P^*\mathbin\triangle Q^*
      =(Q\mathbin\triangle C)\mathbin\triangle(P\mathbin\triangle C)
      =P\mathbin\triangle Q.
  \]
  When the rule is applied to $(P^*,Q^*)$, it again selects the component containing $i$. The equality above shows that this component is still $C$. Then a second switch therefore gives
  \[
    (Q^*\mathbin\triangle C,\ P^*\mathbin\triangle C)=(P,Q).
  \]
  Thus, in both cases, $Q^*$ contains $(i,j)$, and applying the switch twice returns $(P,Q)$. Hence the switch is a bijection on the set of pairs $(P,Q)$ with $(i,j)\in Q$.

  Second, we compare the probability of a fixed outcome before and after the switch. Fix $R,T\in\mathcal M$ with $(i,j)\in T$, since the switch is a bijection, there is a unique pair $(R',T')$ with $(i,j)\in T'$ that it transforms into $(R,T)$. Thus the output equals $(R,T)$ exactly when the input equals $(R',T')$.

  The switch does not change the total number of occurrences of any edge in the two matchings. By the definition of $\mu_\lambda$, this implies
  \[
    \mu_\lambda(R')\mu_\lambda(T')
      =\mu_\lambda(R)\mu_\lambda(T).
  \]
  Using the sampling rule, we obtain
  \begin{align*}
    \Pr[P^*=R,\ Q^*=T]
      &=\Pr[P=R',\ Q=T']\\
      &=\frac{\mu_\lambda(R')\mu_\lambda(T')}{p_{ij}}\\
      &=\frac{\mu_\lambda(R)\mu_\lambda(T)}{p_{ij}}\\
      &=\Pr[P=R,\ Q=T].
  \end{align*}
  This equality holds for every such pair $(R,T)$, so $(P^*,Q^*)$ has the same joint distribution as $(P,Q)$. In particular, $P^*$ has distribution $\mu_\lambda$.

  For a matching $R$, let $B(R)$ be its set of matched items. Since the matchings $P^*$ and $Q$ differ only on $C$, if $C$ is empty or a cycle, then $B(P^*)=B(Q)$; if $C$ is a path, these sets differ only by exchanging its two endpoint items. Thus
  \[
    |B(P^*)\setminus B(Q)|\le1,
    \qquad |B(Q)\setminus B(P^*)|\le1.
  \]

  So far, the construction has been carried out for a fixed item $j$. To complete the coupling, first choose $j$ with probability $p_{ij}$ and then perform the construction for this choice. Since $p_{ij}$ is the marginal probability of assigning $j$ to $i$ under $\mu_\lambda$, the resulting matching $Q$ has distribution $\mu_\lambda$. For every possible choice of $j$, the conditional distribution of $P^*$ is the same distribution $\mu_\lambda$, thus $P^*$ and $B(P^*)$ are independent of $j=Q(i)$.

  Set $\pi=Q$ and $\widehat B_i=B(P^*)$. Since both $P^*$ and $Q$ have distribution $\mu_\lambda$, their matched-item sets have the same distribution. Moreover, $\widehat B_i$ is independent of $\pi(i)$, and the set-difference bounds above hold for every outcome. This proves \cref{prop:nsw-reference-coupling} for every fixed Gibbs distribution.

  \emph{Negative correlation.}
  We apply M\'esz\'aros's conditional BK inequality for product-weight bipartite matchings~\cite[Theorem~1.4]{Meszaros23}. The random vertex set in that theorem contains the unmatched agents and the matched items. Requiring this set to contain no agent is therefore equivalent to requiring the matching to cover every agent. Under this conditioning, the matching has distribution $\mu_\lambda$, and the random vertex set is exactly $B$.

  Let $U,V\subseteq M$ be disjoint. The events $U\subseteq B$ and $V\subseteq B$ are increasing, and checking them only requires the items in $U$ and $V$, respectively. Since these sets are disjoint, the BK inequality gives
  \[
    \Pr[U\cup V\subseteq B]
      \le \Pr[U\subseteq B]\Pr[V\subseteq B].
  \]
  For any nonempty $J\subseteq M$, choose $j\in J$ and apply this bound with $U=J\setminus\{j\}$ and $V=\{j\}$. We obtain
  \[
    \Pr[J\subseteq B]
      \le \Pr[J\setminus\{j\}\subseteq B]\Pr[j\in B].
  \]
  Induction on $|J|$ now gives
  \[
    \Pr[J\subseteq B]\le\prod_{j\in J}\Pr[j\in B].
  \]

  \emph{Obtaining the prescribed marginals.}
  By \cref{lem:nsw-finite-matching} in \cref{app:nsw-finite}, for every $\eta>0$ we can choose positive rational edge weights whose Gibbs marginals $x'$ satisfy $\|x'-x\|_1\le\eta$. Choose a sequence of tolerances tending to zero and a corresponding Gibbs distribution for each tolerance. Since $\mathcal M$ is finite, we can pass to a subsequence along which the probability of every matching converges. The limiting probabilities define a distribution on $\mathcal M$ with edge marginals exactly $x$, proving \cref{prop:nsw-marginals}. Every event probability in the negative-correlation inequalities is a finite sum of matching probabilities, so these inequalities remain valid in the limit. Together with $\Pr[j\in B]=b_j$, this proves \cref{prop:nsw-negative-correlation}.

  To obtain \cref{prop:nsw-reference-coupling}, consider the couplings of the matching and its reference set constructed above for each Gibbs distribution in this subsequence. There are finitely many matchings, reference sets, and agents. We can therefore choose a further subsequence along which all joint probabilities converge for every agent. Each limiting coupling has the matching distribution obtained above.

  For every $U\subseteq M$ and $j\in M$, each coupling in the sequence satisfies $\Pr[\widehat B_i=U]=\Pr[B=U],\Pr[\widehat B_i=U,\pi(i)=j]=\Pr[\widehat B_i=U]\Pr[\pi(i)=j]$.
  These equalities remain valid when we take limits. Thus the limiting reference set has the same distribution as $B$ and is independent of $\pi(i)$.

  Finally, every coupling in the sequence assigns probability zero to outcomes violating either set-difference bound. Those outcomes still have probability zero in the limit, so both bounds continue to hold. This proves \cref{prop:nsw-reference-coupling} and completes the proof.
\end{proof}

\subsection{Rounding the tails}\label{subsec:nsw-service}
In this subsection, we analyze the tail allocation described in \cref{subsec:nsw-algorithm}. The construction combines independent runs of the guiding-tree rounding procedure of \cref{lem:mm-service} with geometrically decreasing color probabilities. The main estimate, proved in \cref{lem:nsw-geometric}, is used together with singleton matching to bound each agent's expected logarithmic value.

\paragraph{Tail configurations and rounding runs.}
We use the tail sets $T_{i,S}$ and singleton values $\nu_{i,S}$ defined in \eqref{eq:nsw-heads}. For each agent, we halve the mass of every tail configuration and add an empty configuration of mass $1/2$. We index the added configuration by $\bot$ and define
\[
  \bar y_{i,S}=\frac{y_{i,S}}2\quad(S\in\mathcal C_i),
  \qquad \bar y_{i,\bot}=\frac12,
  \qquad T_{i,\bot}=\emptyset,
  \qquad \nu_{i,\bot}=0.
\]
The modified masses sum to one for each agent, and the total mass on each item is at most $1/2$ by \eqref{eq:nsw-tail-capacity}.
Thus these indexed sets satisfy the load condition of \cref{lem:mm-service}.

As in \cref{subsec:nsw-algorithm}, run the procedure independently for $k=0,\ldots,K$, where $K=\lceil\log_2(m+1)\rceil$. Write $S_i^k$ for the index selected for agent $i$ in run $k$, and $G^k$ for the resulting allocation. Then
\[
  \Pr[S_i^k=S]=\bar y_{i,S},
  \qquad G_i^k\subseteq T_{i,S_i^k}.
\]
Independently of these runs, each item receives color $k$ with probability $p_k=2^{-k-1}$ and is left uncolored with the remaining probability, different items are colored independently. Let $M_k$ be the color-$k$ class, so the tail allocation is $Z_i=\bigcup_{k=0}^K(G_i^k\cap M_k)$, as in \eqref{eq:nsw-tail-output}.

\paragraph{Guiding-tree rounding after a fixed deletion.}
Fix one run of the guiding-tree rounding procedure, let $G$ be the resulting allocation and $S_i$ the index selected for agent $i$. Fix a set $D\subseteq M$, for every agent $i$ and every index $S$ with $\bar y_{i,S}>0$, we have
\begin{equation}\label{eq:nsw-conditional-service}
  \Pr\!\left[
    v_i(G_i\setminus D)\ge\tfrac12v_i(T_{i,S}\setminus D)
    \,\middle|\,S_i=S\right]\ge\frac14.
\end{equation}
Indeed, replace each valuation $v_i$ by $Q\mapsto v_i(Q\setminus D)$. These modified valuations remain normalized, monotone, and subadditive. Guiding-tree rounding makes no valuation queries, so the same random choices produce the same allocation and selected indices under either set of valuations. Applying \cref{lem:mm-service} to the modified valuations gives \eqref{eq:nsw-conditional-service}. The claim also includes the empty index, for which both values in the inequality are zero.

\paragraph{Combining the rounding runs.}
The rounding runs are independent, but the color classes are not. The following lemma extends the joint-failure bound used in \cref{lem:mm-colored-service} to the unequal probabilities $p_k$.

\begin{lemma}\label{lem:nsw-failure-product}
  Fix an agent, for each $k=0,\ldots,K$, let $F_k$ be an event determined by the history of rounding run $k$ and its color class $M_k$. Assume that, for each fixed history, the indicator of $F_k$ is nonincreasing as items are added to $M_k$. If the rounding histories are independent across runs and independent of the coloring, then
  \begin{equation}\label{eq:nsw-failure-product}
    \Pr\!\left[\bigcap_{k=0}^K F_k\right]
      \le\prod_{k=0}^K\Pr[F_k].
  \end{equation}
  If the deleted set $D$ is independent of the rounding runs and the coloring, the bound also holds conditional on $D$.
\end{lemma}
\begin{proof}
  We use the argument in the \emph{Combining the runs} part of the proof of \cref{lem:mm-colored-service}. That argument also applies to unequal color probabilities and to items that remain uncolored. To see this, regard being uncolored as an additional color. Each item's color indicators still have a multinomial distribution with one trial and hence are negatively associated~\cite{JDP83}. Since different items are colored independently, the same negative-association argument applies to all the color indicators.

  Fix the random choices made in all rounding runs, and let $g_k$ be the indicator of $F_k$. By assumption, $g_k$ is nonnegative and nonincreasing as items are added to $M_k$. Different factors depend on disjoint sets of color indicators, so the product bound from that proof gives
  \[
    \E_{\mathrm{col}}\!\left[\prod_{k=0}^K g_k\right]
      \le\prod_{k=0}^K\E_{\mathrm{col}}[g_k],
  \]
  where $\E_{\mathrm{col}}$ denotes expectation over the coloring. Each factor on the right depends only on the random choices in run $k$. Averaging over the independent runs therefore gives
  \[
    \Pr\!\left[\bigcap_{k=0}^K F_k\right]
      =\E\!\left[\prod_{k=0}^K g_k\right]
      \le\prod_{k=0}^K\E[g_k]
      =\prod_{k=0}^K\Pr[F_k],
  \]
  proving \eqref{eq:nsw-failure-product}.

  If $D$ is independent of the rounding runs and the coloring, conditioning on $D$ leaves their joint distribution unchanged. The same proof therefore applies conditional on $D$.
\end{proof}

\paragraph{Bounding the expected logarithmic value.}
For agent $i$, run $k$ contributes the items of $G_i^k$ that receive color $k$, namely $G_i^k\cap M_k$. We evaluate this bundle after removing a fixed set $D$, and then subtract $\nu_{i,S_i^k}$ from its value. By subadditivity, deleting one additional item reduces the value by at most its singleton value, which is bounded by $\nu_{i,S_i^k}$. The next lemma takes the maximum of these adjusted values and an independent singleton value, and gives a lower bound on the expected logarithm of this maximum. For a real number $z$, write $[z]_+=\max\{0,z\}$.

\begin{lemma}\label{lem:nsw-geometric}
  Fix an agent $i$ and a deleted set $D\subseteq M$. For each $S\in\mathcal C_i$, let $W_S=v_i(T_{i,S}\setminus D)$. Independently of all rounding runs and the coloring, choose a configuration $S\in\mathcal C_i$ with probability $y_{i,S}$. Set $H=\nu_{i,S}>0$ and define
  \begin{equation}\label{eq:nsw-readout}
    Z^*=\ln\max\!\left\{
      H,\ \max_{0\le k\le K}
      \bigl[v_i((G_i^k\cap M_k)\setminus D)
            -\nu_{i,S_i^k}\bigr]_+
    \right\}.
  \end{equation}
  Then
  \begin{equation}\label{eq:nsw-geometric}
    \E Z^*\ge\sum_{S\in\mathcal C_i}y_{i,S}
      \ln(\nu_{i,S}+W_S)-C_0,
    \qquad C_0=\ln33+16\ln2.
  \end{equation}
\end{lemma}
\begin{proof}
  For each run, abbreviate its contribution to the maximum in \eqref{eq:nsw-readout} by
  \[
    R_k=\bigl[v_i((G_i^k\cap M_k)\setminus D)
                    -\nu_{i,S_i^k}\bigr]_+.
  \]
  Thus $Z^*=\ln\max\{H,R_0,\ldots,R_K\}$.

  \emph{A bound for one run.}
  Fix $k$ and put $h=2^{k+1}$, so each item belongs to $M_k$ independently with probability $1/h$. Condition on selecting an original index $S$ with $y_{i,S}>0$, and suppose $W_S>16h\nu_{i,S}$. By \eqref{eq:nsw-conditional-service}, with conditional probability at least $1/4$ the set $Q=G_i^k\setminus D$ satisfies $v_i(Q)\ge W_S/2$. For every such output and every $j\in Q$,
  \[
    v_i(\{j\})\le\nu_{i,S}
      <\frac{W_S}{16h}
      \le\frac{v_i(Q)}{8h}.
  \]
  The coloring is independent of the rounding history. Applying \cref{lem:mm-sampling} to $Q$ therefore shows that, with probability at least $1/2$ over the coloring,
  \[
    v_i(Q\cap M_k)\ge\frac{v_i(Q)}{4h}
      \ge\frac{W_S}{8h}.
  \]
  On this event,
  \[
    R_k\ge\frac{W_S}{8h}-\nu_{i,S}
      >\frac{W_S}{16h}
      =\frac{W_S}{32\cdot2^k}.
  \]
  Since the probability of selecting $S$ is $\bar y_{i,S}=y_{i,S}/2$, the joint probability of selecting this index and obtaining the displayed bound is at least
  \[
    \frac{y_{i,S}}2\cdot\frac14\cdot\frac12
      =\frac{y_{i,S}}{16}.
  \]

  To express this estimate in terms of the original configuration distribution, draw an auxiliary index $S\in\mathcal C_i$ with probability $y_{i,S}$ and define
  \[
    U=\ln\max\{\nu_{i,S},W_S/32\},
    \qquad V=\ln\nu_{i,S}.
  \]
  These variables are used only for the comparison. For any real $t$, the event $U>\max\{V,t\}+k\ln2$ implies
  \[
    W_S>32\cdot2^k\max\{\nu_{i,S},e^t\}.
  \]
  It therefore ensures both $W_S>16h\nu_{i,S}$ and $W_S/(32\cdot2^k)>e^t$. Summing the preceding success probabilities over these original indices gives
  \begin{equation}\label{eq:nsw-scale-opportunity}
    \Pr[R_k>e^t]
      \ge\frac1{16}\Pr[U>\max\{V,t\}+k\ln2].
  \end{equation}

  \emph{Combining the bounds across runs.}
  Define
  \[
    \phi(t)=\E[U-\max\{V,t\}]_+,
    \qquad \lambda=\frac1{16\ln2}.
  \]
  For a fixed auxiliary index, put $d=[U-\max\{V,t\}]_+$. There are exactly $\lceil d/\ln2\rceil$ nonnegative integers $k$ satisfying $k\ln2<d$, so their number is at least $d/\ln2$. We next show that these integers are all at most $K$. The set $T_{i,S}\setminus D$ contains at most $m$ items, each with singleton value at most $\nu_{i,S}$. Subadditivity gives $W_S\le m\nu_{i,S}$, so
  \[
    d\le U-V
      =\ln\max\!\left\{1,\frac{W_S}{32\nu_{i,S}}\right\}
      \le\ln m<K\ln2.
  \]
  Taking the expectation of this count and summing \eqref{eq:nsw-scale-opportunity} over $k$ yields
  \begin{align*}
    \sum_{k=0}^K\Pr[R_k>e^t]
      &\ge\frac1{16}\E\!\left[
        \#\{0\le k\le K:k\ln2<d\}\right]\\
      &\ge\frac{\E d}{16\ln2}
       =\lambda\phi(t).
  \end{align*}

  For any random variable $Y$, write $F_Y(t)=\Pr[Y\le t]$. The event $R_k\le e^t$ depends only on run $k$ and its color class. For a fixed rounding history, the subtracted value $\nu_{i,S_i^k}$ is fixed, and adding items to $M_k$ can only increase $R_k$. Thus the failure-product bound of \cref{lem:nsw-failure-product} applies to these events. Also, $H$ is independent of all the runs and colors, and $\ln H$ has the same distribution as $V$. Consequently,
  \begin{align}
    F_{Z^*}(t)
      &=F_V(t)\Pr\!\left[\bigcap_{k=0}^K\{R_k\le e^t\}\right]\notag\\
      &\le F_V(t)\prod_{k=0}^K\bigl(1-\Pr[R_k>e^t]\bigr)\notag\\
      &\le F_V(t)\exp\!\left(-\sum_{k=0}^K\Pr[R_k>e^t]\right)
       \le F_V(t)e^{-\lambda\phi(t)}.
       \label{eq:nsw-cdf}
  \end{align}

  \emph{From distribution functions to expectations.}
  Fix values $U=u$ and $V=v$, where $u\ge v$. Then
  \[
    [u-\max\{v,t\}]_+
    =\begin{cases}
      u-v, & t\le v,\\
      u-t, & v<t<u,\\
      0, & t\ge u.
    \end{cases}
  \]
  Its derivative with respect to $t$ is therefore $-1$ when $v<t<u$ and $0$ outside this interval, except possibly at $t=u$ and $t=v$.

  The expectation defining $\phi(t)$ is a finite weighted sum over the original configurations. We may therefore differentiate it term by term away from the finitely many possible values of $U$ and $V$, obtaining
  \[
    -\phi'(t)=\Pr[V\le t<U]=F_V(t)-F_U(t)
      \quad\text{for almost every }t.
  \]
  The last equality follows because $U\ge V$: subtracting $\Pr[U\le t]$ from $\Pr[V\le t]$ leaves exactly $\Pr[V\le t<U]$. We also have
  \[
    \lim_{t\to-\infty}\phi(t)=\E(U-V),
    \qquad \lim_{t\to\infty}\phi(t)=0.
  \]
  The variables $U,V,Z^*$ are bounded: the ground set and configuration supports are finite, and $H$ has finite positive support. We may therefore use the distribution-function identity for the difference of two expectations. By \eqref{eq:nsw-cdf},
  \begin{align*}
    \E U-\E Z^*
      &=\int_{\mathbb R}\bigl(F_{Z^*}(t)-F_U(t)\bigr)\,dt\\
      &\le\int_{\mathbb R}
        \bigl(F_V(t)e^{-\lambda\phi(t)}-F_U(t)\bigr)\,dt\\
      &\le\int_{\mathbb R}
        (F_V(t)-F_U(t))e^{-\lambda\phi(t)}\,dt.
  \end{align*}
  The last inequality uses $e^{-\lambda\phi(t)}\le1$. To evaluate the remaining integral, observe that
  \[
    \frac{d}{dt}e^{-\lambda\phi(t)}
      =\lambda(F_V(t)-F_U(t))e^{-\lambda\phi(t)}
      \quad\text{for almost every }t.
  \]
  Since $\phi$ is continuous and piecewise linear, integration and the two limits above give
  \[
    \int_{\mathbb R}(F_V(t)-F_U(t))e^{-\lambda\phi(t)}\,dt
      =\frac{1-e^{-\lambda\E(U-V)}}{\lambda}
      \le\frac1\lambda=16\ln2.
  \]

  \emph{Comparing with the configuration values.}
  For every original index $S$,
  \[
    \nu_{i,S}+W_S
      \le33\max\{\nu_{i,S},W_S/32\}.
  \]
  Taking logarithms and averaging with probabilities $y_{i,S}$ shows that
  \[
    \E U\ge\sum_{S\in\mathcal C_i}y_{i,S}
       \ln(\nu_{i,S}+W_S)-\ln33.
  \]
  Combining this with $\E U-\E Z^*\le16\ln2$ proves \eqref{eq:nsw-geometric}.
\end{proof}

\subsection{Analysis of the rounding algorithm}\label{subsec:nsw-analysis}
\paragraph{Bounding the loss from deleting items.}
We first bound how much the configuration value $v_i(S)$ can exceed $D_{i,S}(B)$. The following lemma gives a general estimate for the loss caused by random deletions. We then apply it to each tail $T_{i,S}$ and sum the bounds over all configurations.

\begin{lemma}\label{lem:nsw-squared-loss}
  Let $v$ be a normalized monotone subadditive valuation, and let $Q$ be a set with $W=v(Q)>0$. Suppose that $v(\{j\})\le\nu$ for every $j\in Q$, where $\nu>0$. Let $D\subseteq Q$ be a random set of deleted items. Suppose there are numbers $b_j\in[0,1)$ for $j\in Q$ such that
  \[
    \Pr[J\subseteq D]\le\prod_{j\in J}b_j
      \qquad\text{for every }J\subseteq Q.
  \]
  Let $V=v(Q\setminus D)$ be the value remaining after deletion, and set $P=\sum_{j\in Q}b_j/(1-b_j)$. Then
  \begin{equation}\label{eq:nsw-squared-loss}
    \E\left(\frac{W}{V+\nu}\right)^2\le48(1+P).
  \end{equation}
\end{lemma}
\begin{proof}
  Dividing all values by $\nu$ leaves the ratio in \eqref{eq:nsw-squared-loss} unchanged, so we may assume $\nu=1$. If $W<4$, then $W^2/(V+1)^2<16$. If $P=0$, then $b_j=0$ for every $j\in Q$, so $D=\emptyset$ almost surely and $V=W$. The desired bound holds in both cases. Henceforth, assume $W\ge4$ and $P>0$.

  \emph{Bounding the probability of a small remaining value.}
  Write $q(t)=\Pr[V\le t]$. Fix $1\le t\le W/4$ and set $k=\lfloor W/(2t)\rfloor$. Since $W/(2t)\ge2$, we have
  \[
    \frac{W}{4t}\le k\le\frac{W}{2t}.
  \]
  Let $D^{(1)},\ldots,D^{(k)}$ be independent random sets, each with the same distribution as $D$, and define
  \[
    R_k=\left|\bigcap_{r=1}^kD^{(r)}\right|.
  \]
  Thus, $R_k$ counts the items deleted in every copy. Suppose that $v(Q\setminus D^{(r)})\le t$ for every $r$. Every item of $Q$ either remains in at least one copy or is deleted in all copies. Subadditivity and the bound of $1$ on singleton values therefore give
  \[
    W\le\sum_{r=1}^k v(Q\setminus D^{(r)})
        +v\!\left(\bigcap_{r=1}^kD^{(r)}\right)
      \le kt+R_k.
  \]
  Since $kt\le W/2$, this event implies $R_k\ge W/2$. Its probability is $q(t)^k$ by independence, so
  \[
    q(t)^k\le\Pr[R_k\ge W/2].
  \]

  We now bound the probability on the right. Let $I_j$ indicate that item $j$ is deleted in every copy, so that $R_k=\sum_{j\in Q}I_j$. For every $J\subseteq Q$, independence of the copies and the assumed deletion bound imply
  \[
    \E\left[\prod_{j\in J}I_j\right]
      =\prod_{r=1}^k\Pr[J\subseteq D^{(r)}]
      \le\prod_{j\in J}b_j^k.
  \]
  For $s\ge0$, all coefficients in the following expansion are nonnegative. Applying this product bound gives
  \begin{align*}
    \E e^{sR_k}
      &=\E\prod_{j\in Q}\bigl(1+(e^s-1)I_j\bigr)\\
      &=\sum_{J\subseteq Q}(e^s-1)^{|J|}
          \E\left[\prod_{j\in J}I_j\right]\\
      &\le\prod_{j\in Q}\bigl(1+b_j^k(e^s-1)\bigr)\\
      &\le\exp\!\left(\mu_k(e^s-1)\right),
      \qquad \mu_k=\sum_{j\in Q}b_j^k.
  \end{align*}
  The last inequality uses $1+x\le e^x$. Moreover, $kb^k\le b+b^2+\cdots+b^k\le b/(1-b)$ for $0\le b<1$, so
  \[
    \mu_k\le\frac1k\sum_{j\in Q}\frac{b_j}{1-b_j}
      =\frac Pk.
  \]

  If $2eP/(Wk)<1$, take $s=\ln(Wk/(2P))>0$. Markov's inequality and the exponential-moment bound give
  \begin{align*}
    q(t)^k
      &\le \Pr[R_k\ge W/2]\\
      &\le \exp\!\left(-\frac{sW}{2}+\frac Pk(e^s-1)\right)\\
      &=e^{-P/k}\left(\frac{2eP}{Wk}\right)^{W/2}\\
      &\le\left(\frac{2eP}{Wk}\right)^{W/2}.
  \end{align*}
  Taking the $k$th root and using $W/(2k)\ge t\ge1$, we obtain
  \[
    q(t)\le\left(\frac{2eP}{Wk}\right)^{W/(2k)}
      \le\frac{2eP}{Wk}.
  \]
  If $2eP/(Wk)\ge1$, the same bound follows from $q(t)\le1$. Thus, using $k\ge W/(4t)$ and $8e<24$, we conclude that
  \begin{equation}\label{eq:nsw-deletion-tail}
    q(t)\le\min\{1,24Pt/W^2\}
      \qquad(1\le t\le W/4).
  \end{equation}
  For $0\le t<1$, we also have $q(t)\le q(1)\le24P/W^2$. Thus, throughout $0\le t\le W/4$,
  \[
    q(t)\le\frac{24P(t+1)}{W^2}.
  \]

  \emph{Bounding the expected squared ratio.}
  Since $0\le V\le W$, we can write
  \[
    \frac{W^2}{(V+1)^2}
      =\frac{W^2}{(W+1)^2}
       +\int_V^W\frac{2W^2}{(t+1)^3}\,dt.
  \]
  Taking expectations and using $\Pr[V\le t]=q(t)$ gives
  \[
    \E\frac{W^2}{(V+1)^2}
      =\frac{W^2}{(W+1)^2}
       +\int_0^W\frac{2W^2q(t)}{(t+1)^3}\,dt.
  \]
  Split the integral at $W/4$. For the part above $W/4$, the bound $q(t)\le1$ gives
  \[
    \frac{W^2}{(W+1)^2}
      +\int_{W/4}^W\frac{2W^2q(t)}{(t+1)^3}\,dt
      \le\frac{W^2}{(W/4+1)^2}
      \le16.
  \]
  For the remaining part, apply the bound $q(t)\le24P(t+1)/W^2$ to obtain
  \[
    \int_0^{W/4}\frac{2W^2q(t)}{(t+1)^3}\,dt
      \le48P\int_0^{W/4}\frac{1}{(t+1)^2}\,dt
      \le48P.
  \]
  Adding these bounds proves
  \[
    \E\frac{W^2}{(V+1)^2}
      \le16+48P\le48(1+P),
  \]
  as required.
\end{proof}

\paragraph{Bounding the total loss.}
For the following estimate, let $B$ be the set of items matched by the random matching in \cref{lem:nsw-matching}. We use \cref{lem:nsw-squared-loss} to bound the loss for each configuration. We then multiply each bound by the corresponding mass $y_{i,S}$ and sum over all configurations, using \eqref{eq:nsw-tail-capacity} to obtain a bound proportional to $n$.

\begin{lemma}\label{lem:nsw-aggregate}
  For $D_{i,S}(B)$ defined in \eqref{eq:nsw-buffer}, we have
  \begin{equation}\label{eq:nsw-aggregate}
    \sum_{i,S}y_{i,S}\E_B
      \left(\frac{v_i(S)}{D_{i,S}(B)}\right)^2\le196n.
  \end{equation}
\end{lemma}
\begin{proof}
  Configurations with zero mass do not contribute to the sum, so we consider only those with $y_{i,S}>0$. For every such configuration and every $j\in T_{i,S}$, \eqref{eq:nsw-tail-capacity} gives $f_j\ge y_{i,S}>0$. Thus, the quantity
  \[
    P_{i,S}=\sum_{j\in T_{i,S}}\frac{b_j}{f_j}
  \]
  is well defined.

  Fix a configuration $(i,S)$ with $v_i(T_{i,S})>0$. Apply \cref{lem:nsw-squared-loss} to the set $Q=T_{i,S}$, with deleted set $B\cap T_{i,S}$ and singleton bound $\nu_{i,S}$. The required deletion condition follows from \eqref{eq:nsw-cylinder}, and $b_j<1$ for every item in the tail because $f_j>0$. Since
  \[
    D_{i,S}(B)=\nu_{i,S}+v_i(T_{i,S}\setminus B),
  \]
  the lemma gives
  \[
    \E_B\left(\frac{v_i(T_{i,S})}{D_{i,S}(B)}\right)^2
      \le48(1+P_{i,S}).
  \]

  To obtain a bound for the entire configuration, use subadditivity to write $v_i(S)\le\nu_{i,S}+v_i(T_{i,S})$. Together with $D_{i,S}(B)\ge\nu_{i,S}$, this gives
  \[
    \frac{v_i(S)}{D_{i,S}(B)}
      \le1+\frac{v_i(T_{i,S})}{D_{i,S}(B)}.
  \]
  Applying $(1+z)^2\le2+2z^2$ and taking expectations, we obtain
  \begin{align*}
    \E_B\left(\frac{v_i(S)}{D_{i,S}(B)}\right)^2
      &\le2+2\E_B
        \left(\frac{v_i(T_{i,S})}{D_{i,S}(B)}\right)^2\\
      &\le2+96(1+P_{i,S})\\
      &\le98(1+P_{i,S}).
  \end{align*}
  If $v_i(T_{i,S})=0$, monotonicity and subadditivity imply $v_i(S)=\nu_{i,S}$ and $D_{i,S}(B)=\nu_{i,S}$. The ratio is then one, so the same bound holds.

  It remains to sum these bounds over all configurations. Changing the order of summation and applying \eqref{eq:nsw-tail-capacity}, we have
  \begin{equation}\label{eq:nsw-odds-budget}
    \begin{aligned}
      \sum_{i,S}y_{i,S}P_{i,S}
        &=\sum_{j:\,f_j>0}\frac{b_j}{f_j}
          \sum_{i,S:\,j\in T_{i,S}}y_{i,S}\\
        &\le\sum_{j:\,f_j>0}\frac{b_j}{f_j}f_j\\
        &\le\sum_{j\in M}b_j=n.
    \end{aligned}
  \end{equation}
  Therefore, using $\sum_{i,S}y_{i,S}=n$, we conclude that
  \[
    \sum_{i,S}y_{i,S}\E_B
      \left(\frac{v_i(S)}{D_{i,S}(B)}\right)^2
      \le98\left(
        \sum_{i,S}y_{i,S}
        +\sum_{i,S}y_{i,S}P_{i,S}
      \right)
      \le196n.
  \]
\end{proof}

\paragraph{Bounding the expected logarithmic welfare.}
The preceding lemma compares the configuration values $v_i(S)$ with the buffered residual values $D_{i,S}(B)$. We now use these buffered values to obtain a lower bound on the expected logarithmic welfare of the final allocation.

\begin{lemma}\label{lem:nsw-realize}
  Let $A$ be the allocation constructed in \cref{subsec:nsw-algorithm} using the matching distribution of \cref{lem:nsw-matching}. Then
  \begin{equation}\label{eq:nsw-realize}
    \E\ln\mathrm{NSW}_w(A)
      \ge\sum_{i,S}w_i y_{i,S}\E_B\ln D_{i,S}(B)-C_0.
  \end{equation}
  Here $C_0$ is the constant in Lemma~\ref{lem:nsw-geometric}.
\end{lemma}
\begin{proof}
  Fix an agent $i$. Let $\widehat B_i$ be the reference set from \cref{prop:nsw-reference-coupling} of \cref{lem:nsw-matching}, it has the same distribution as $B$ and is independent of $\pi(i)$. Moreover, $|B\setminus\widehat B_i|\le1$. Take the random choices for the guiding-tree rounding runs and the coloring independently of $(\pi,\widehat B_i)$.

  For each run $k$, compare the items left in $G_i^k\cap M_k$ after deleting $B$ with those left after deleting $\widehat B_i$. Deleting $B$ can remove at most one additional item, because $|B\setminus\widehat B_i|\le1$. Since $G_i^k\subseteq T_{i,S_i^k}$, this item's singleton value is at most $\nu_{i,S_i^k}$. Monotonicity and subadditivity therefore give
  \begin{equation}\label{eq:nsw-one-item-value}
    v_i((G_i^k\cap M_k)\setminus B)
      \ge v_i((G_i^k\cap M_k)\setminus\widehat B_i)
         -\nu_{i,S_i^k}.
  \end{equation}
  If $S_i^k$ is the added empty index, then $G_i^k=\emptyset$ and $\nu_{i,S_i^k}=0$, so the inequality also holds.

  Let $H=v_i(\{\pi(i)\})>0$. By \eqref{eq:nsw-output}, $A_i$ contains both its matched item $\pi(i)$ and every set $(G_i^k\cap M_k)\setminus B$. Its value is therefore at least $H$. Moreover, \eqref{eq:nsw-one-item-value} and nonnegativity give, for every $k$,
  \[
    v_i(A_i)\ge
      \bigl[v_i((G_i^k\cap M_k)\setminus\widehat B_i)
            -\nu_{i,S_i^k}\bigr]_+.
  \]
  Taking the maximum of these lower bounds and then its logarithm gives $\ln v_i(A_i)\ge Z^*$, where $Z^*$ is the quantity in \eqref{eq:nsw-readout} with $D=\widehat B_i$ and this choice of $H$.

  We now check that \cref{lem:nsw-geometric} applies after conditioning on $\widehat B_i$. Since $\widehat B_i$ is independent of $\pi(i)$, \cref{prop:nsw-marginals} of \cref{lem:nsw-matching} and \eqref{eq:nsw-head-marginals} give
  \[
    \Pr[\pi(i)=j\mid\widehat B_i]
      =x_{ij}=\sum_{S:\,g_{i,S}=j}y_{i,S}.
  \]
  Thus $H$ has the same distribution as $\nu_{i,S}$ when $S$ is chosen with probabilities $y_{i,S}$. Conditional on $\widehat B_i$, $H$ is also independent of the rounding runs and the coloring, because their random choices are independent of $(\pi,\widehat B_i)$. \cref{lem:nsw-geometric} therefore yields
  \[
    \E[\ln v_i(A_i)\mid\widehat B_i]
      \ge\sum_{S\in\mathcal C_i}y_{i,S}
        \ln\!\left(\nu_{i,S}+v_i(T_{i,S}\setminus\widehat B_i)\right)-C_0.
  \]
  Averaging over $\widehat B_i$ and using the fact that it has the same distribution as $B$, we obtain
  \[
    \E\ln v_i(A_i)
      \ge\sum_{S\in\mathcal C_i}y_{i,S}
        \E_B\ln D_{i,S}(B)-C_0.
  \]
  Finally, multiply by $w_i$ and sum over all agents. Since $\ln\mathrm{NSW}_w(A)=\sum_i w_i\ln v_i(A_i)$ and $\sum_iw_i=1$, this gives \eqref{eq:nsw-realize}.
\end{proof}

\paragraph{Combining the estimates.}
We now use Lemmas~\ref{lem:nsw-aggregate} and~\ref{lem:nsw-realize} to complete the proof of Theorem~\ref{thm:nsw-rounding}. Applying Jensen's inequality to the squared-ratio bound gives a logarithmic bound that depends on the entropy of the agents' weights.

\begin{proof}[Proof of \cref{thm:nsw-rounding}]
  If $n=1$, giving all items to the agent yields at least the value of every configuration, so the claim follows from monotonicity. Suppose $n\ge2$, and let $R$ denote the left side of \eqref{eq:nsw-aggregate}. \cref{lem:nsw-aggregate} gives $R\le196n$.

  Sample $(i,S)$ with probability $w_i y_{i,S}$, independently of the matching in Lemma~\ref{lem:nsw-matching}, and let $B$ be the set of matched items. Then
  \[
    \E_{i,S,B}\frac{v_i(S)^2}{w_iD_{i,S}(B)^2}
      =\sum_{i,S}y_{i,S}\E_B\frac{v_i(S)^2}{D_{i,S}(B)^2}
      =R.
  \]
  Using the definitions of $L(y)$ and $H(w)$ and applying Jensen's inequality to the logarithm, we obtain
  \begin{align}
    L(y)-\sum_{i,S}w_i y_{i,S}\E_B\ln D_{i,S}(B)
      &=\frac12\left(
        \E_{i,S,B}\ln\frac{v_i(S)^2}{w_iD_{i,S}(B)^2}
        -H(w)\right)\notag\\
      &\le\frac12\bigl(\ln R-H(w)\bigr)
       \le\ln\Gamma(w)+\frac12\ln196.
        \label{eq:nsw-entropy-average}
  \end{align}
  The last inequality uses $R\le196n$ and the expression for $\Gamma(w)$ in \eqref{eq:nsw-entropy}.

  Combining \eqref{eq:nsw-entropy-average} with \cref{lem:nsw-realize} proves \eqref{eq:nsw-rounding} with $C=C_0+\tfrac12\ln196$ for the matching distribution of \cref{lem:nsw-matching}.

  \cref{app:nsw-finite} implements the matching step using rational edge weights and approximate sampling, and shows that the two estimates used above remain valid up to absolute constants. This yields the stated polynomial-time algorithm. Choosing the heads is the only step that requires valuation information, the remaining operations use the explicit sets and masses and random sampling.
\end{proof}

\subsection{Computing configurations and applying rounding}
\label{subsec:nsw-application}

We now use ordinary demand queries to construct a positive configuration solution with a near-optimal logarithmic benchmark. We first determine when $\mathrm{OPT}_{\mathrm{NSW}}$ is positive.

\begin{lemma}\label{lem:nsw-positive}
  Consider the bipartite graph with vertex sets $N$ and $M$ and an edge $(i,j)$ whenever $v_i(\{j\})>0$. Then $\mathrm{OPT}_{\mathrm{NSW}}>0$ if and only if this graph has a matching covering every agent.
\end{lemma}
\begin{proof}
  By subadditivity, $v_i(S)\le\sum_{j\in S}v_i(\{j\})$. Thus, any bundle of positive value contains an item of positive singleton value. If an allocation gives every agent positive value, choosing one such item from each bundle gives a matching covering every agent, since the bundles are disjoint. Conversely, assigning each agent its matched item in such a matching gives every agent positive value, so $\mathrm{OPT}_{\mathrm{NSW}}>0$.
\end{proof}

We use the reduction in \cref{app:demand-values} to evaluate the singleton values and then test whether such a matching exists. If no such matching exists, every allocation has zero Nash social welfare. Otherwise, we keep the corresponding singleton allocation as a fallback.

\paragraph{The logarithmic configuration program.}
We use the configuration LP for weighted Nash social welfare from \cref{sec:prelim}, with objective $L(y)$ and optimal value $L_{\mathrm{LP}}$. Its dual has a free variable $z_i$ for each agent and a nonnegative price $p_j$ for each item:
\begin{equation}\label{eq:nsw-dual}
  \begin{aligned}
    \text{minimize}\quad
      &U(z,p)=\sum_{i\in N}z_i+\sum_{j\in M}p_j,\\
    \text{subject to}\quad
      &z_i+p(S)\ge w_i\ln v_i(S)
        &&(i\in N,\ v_i(S)>0),\\
      &p\ge0,
  \end{aligned}
  \qquad p(S)=\sum_{j\in S}p_j.
\end{equation}

\begin{lemma}\label{lem:nsw-producer}
  Suppose $\mathrm{OPT}_{\mathrm{NSW}}>0$. Using ordinary demand queries, we can compute an explicit positive configuration solution $y$ with polynomial-size support and a feasible solution $(z,p)$ to the dual program \eqref{eq:nsw-dual}. The solution $y$ satisfies \eqref{eq:mm-row} and \eqref{eq:mm-capacity} exactly, and the two solutions satisfy
  \begin{equation}\label{eq:nsw-producer-gap}
    0\le U(z,p)-L(y)\le3.
  \end{equation}
  In particular,
  $L(y)\ge L_{\mathrm{LP}}-3
    \ge\ln\mathrm{OPT}_{\mathrm{NSW}}-3$.
\end{lemma}

We prove the lemma in \cref{app:nsw-producer}, adapting the approximate dual-separation approach used by Feng, Hu, Li, and Zhang~\cite{FHLZ25} for submodular valuations. For subadditive valuations, we combine the exact value-evaluation procedure of \cref{app:demand-values} with the budgeted demand-query algorithm of Badanidiyuru, Dobzinski, and Oren~\cite{BDO19} (\cref{lem:demand-budget}). We then use rational dual separation to recover a primal solution supported on finitely many configurations.

As in \cref{app:configuration-producer}, the resulting solution satisfies \eqref{eq:mm-row} and \eqref{eq:mm-capacity} exactly, only the objective is approximated. All oracle calls are ordinary demand queries for the original valuations $v_i$.

\begin{theorem}\label{thm:nsw-approximation}
  For normalized monotone subadditive valuations and positive rational weights summing to $1$, there is a randomized polynomial-time algorithm using ordinary demand queries that achieves an $O(\Gamma(w))$ approximation to weighted Nash social welfare. If $\mathrm{OPT}_{\mathrm{NSW}}>0$, its output $A$ satisfies
  \begin{align}
    \E\ln\mathrm{NSW}_w(A)
      &\ge\ln\mathrm{OPT}_{\mathrm{NSW}}-\ln\Gamma(w)-C,
        \label{eq:nsw-final-log}\\
    \E\mathrm{NSW}_w(A)
      &\ge\frac{\mathrm{OPT}_{\mathrm{NSW}}}{C\Gamma(w)}.
        \label{eq:nsw-final-welfare}
  \end{align}
  Here $C>0$ is an absolute constant. Moreover, for any $\delta\in(0,1)$, an $O(\Gamma(w))$-approximate allocation can be obtained with probability at least $1-\delta$. This requires computing the configuration solution once, calibrating the matching distribution once, and performing
  \[
    O\!\left((1+\log\Gamma(w))\log(2/\delta)\right)
  \]
  rounding trials. Each trial uses fresh randomness to sample the matching, run guiding-tree rounding, and color the items.
\end{theorem}
\begin{proof}
  \cref{lem:nsw-positive} allows us to identify instances with zero optimum, for which any allocation is optimal. Assume now that $\mathrm{OPT}_{\mathrm{NSW}}>0$. Compute the primal and dual solutions from \cref{lem:nsw-producer} and round the resulting configuration solution using \cref{thm:nsw-rounding}. Combining the rounding guarantee with \eqref{eq:nsw-producer-gap} gives \eqref{eq:nsw-final-log}. Jensen's inequality for the exponential function then gives \eqref{eq:nsw-final-welfare}, after adjusting the absolute constant.

  To obtain success probability at least $1-\delta$, use the implementation in \cref{app:nsw-finite} and set the calibration failure probability to at most $\min\{\eta,\delta/2\}$, where $\eta$ is the accuracy parameter specified there. Fix any successful calibration outcome, and consider one rounding trial conditional on this outcome. The implementation satisfies the rounding bound of \cref{thm:nsw-rounding} with an absolute constant independent of the calibration outcome. Let
  \[
    D=U(z,p)-\ln\mathrm{NSW}_w(A)\ge0.
  \]
  Dual feasibility gives $U(z,p)\ge\ln\mathrm{OPT}_{\mathrm{NSW}}\ge\ln\mathrm{NSW}_w(A)$, so $D$ is nonnegative. The rounding bound and \eqref{eq:nsw-producer-gap} imply $\E D\le a:=\ln\Gamma(w)+C$, after increasing the absolute constant $C>0$ if necessary. Markov's inequality therefore gives
  \[
    \Pr[D\le a+1]
      \ge1-\frac{\E D}{a+1}
      \ge\frac1{a+1}.
  \]
  Whenever $D\le a+1$, we have
  \[
    \mathrm{NSW}_w(A)
      \ge\exp\!\left(U(z,p)-a-1\right)
      \ge\frac{e^{-C-1}\mathrm{OPT}_{\mathrm{NSW}}}{\Gamma(w)}.
  \]

  Keeping the configuration solution and calibrated matching distribution fixed, perform $t$ rounding trials with independent random choices. Conditional on the fixed calibration outcome, the probability that none achieves this guarantee is at most
  \[
    \left(1-\frac1{a+1}\right)^t
      \le\exp\!\left(-\frac{t}{a+1}\right).
  \]
  Taking $t=O((a+1)\log(2/\delta))$ with a sufficiently large absolute constant makes this probability at most $\delta/2$. This bound holds for every successful calibration outcome. Since calibration itself fails with probability at most $\delta/2$, the total failure probability is at most $\delta$.

  Finally, evaluate the utilities of each candidate allocation exactly and compute its logarithmic welfare to a fixed additive accuracy. Return the candidate with the largest computed value. Its logarithmic welfare is within an absolute additive constant of the best trial, so the selection incurs only an absolute multiplicative loss in welfare and preserves the $O(\Gamma(w))$ guarantee. These comparisons take polynomial time.
\end{proof}

\paragraph{Proof of \cref{thm:intro-nsw}.}
If $\mathrm{OPT}_{\mathrm{NSW}}=0$, the claimed bound holds trivially. Otherwise, it follows from \eqref{eq:nsw-final-welfare} in \cref{thm:nsw-approximation}.

For equal weights, $\Gamma(w)=1$, so the approximation factor is constant. For general weights, $\Gamma(w)\le\sqrt n$, giving an $O(\sqrt n)$ approximation. The running time includes the $O(\log m)$ guiding-tree rounding calls per trial and the repetitions needed to achieve success probability $1-\delta$.

\subsection{The integrality gap for prescribed weights}
\label{subsec:nsw-gap}

We show that the dependence on $\Gamma(w)$ in the configuration LP bound is tight, even for XOS valuations.

\begin{theorem}\label{thm:nsw-gap}
  Fix a positive weight vector $w$ on $n$ agents with $\sum_{i\in N}w_i=1$. The worst-case ratio $e^{L_{\mathrm{LP}}}/\mathrm{OPT}_{\mathrm{NSW}}$ over XOS instances with positive optimum is $\Theta(\Gamma(w))$. More precisely, the worst-case gap is at least $\Gamma(w)/\sqrt8$ and at most $C\Gamma(w)$, where $C$ is an absolute constant. The same worst-case order holds for subadditive valuations.
\end{theorem}
\begin{proof}
  For the upper bound, apply \cref{thm:nsw-rounding} to an optimal basic solution of the configuration LP. The bound on expected logarithmic welfare implies that some allocation has welfare at least $e^{L_{\mathrm{LP}}}/(C\Gamma(w))$. This proves the upper bound for all subadditive valuations.

  For the lower bound, we adapt the groups of unit-demand agents in the submodular gap construction of Bei, Feng, Hu, Li, and Zhang~\cite{BFHLZ26}, described in Section~6 of the full version~\cite{BFHLZfull}. We give the special agent an XOS valuation and choose the group sizes according to the prescribed weights.

  We first describe a basic instance. Fix an integer $\ell\ge2$ and form $\ell$ disjoint rows $R_1,\ldots,R_\ell$, each containing $\ell$ items. For each row $R_r$, introduce $\ell-1$ unit-demand agents, called \emph{blockers}, each with valuation $\mathbf1_{\{S\cap R_r\ne\emptyset\}}$. One special agent has valuation
  \[
    v(S)=\frac1\ell\max_{r\in[\ell]}|S\cap R_r|.
  \]
  These valuations are normalized, monotone, and XOS.

  For the special agent, assign configuration mass $1/\ell$ to each full row. For each blocker, assign mass $1/\ell$ to every singleton in its row. The masses sum to one for each agent, and every item has total mass $1/\ell+(\ell-1)/\ell=1$, so this is a feasible configuration solution. Every configuration in its support has value one, giving $L(y)=0$. Since all valuations are at most one, this solution is optimal and $e^{L_{\mathrm{LP}}}=1$.

  In any allocation of positive welfare, every blocker must receive an item from its row. The blockers therefore use at least $\ell-1$ distinct items in each row, leaving at most one for the special agent. Its value is consequently at most $1/\ell$. This bound is attained by giving the special agent one item from each row and assigning the remaining items to the blockers, one per agent. If the special agent has weight $\omega$, the optimal weighted product of utilities is therefore $\ell^{-\omega}$, since every blocker has value one.

  Now consider the prescribed weight vector $w$ and set
  \[
    I=\{i\in N:nw_i\ge8\},
    \qquad
    \ell_i=\left\lfloor\sqrt{nw_i/2}\right\rfloor
      \quad(i\in I).
  \]
  For each $i\in I$, take a disjoint copy of the basic instance with parameter $\ell_i$ and use $i$ as its special agent. Each copy requires $\ell_i(\ell_i-1)$ blockers, and the total number needed is at most $n/2$, since
  \[
    \sum_{i\in I}\ell_i(\ell_i-1)
      \le\frac n2\sum_{i\in I}w_i\le\frac n2.
  \]
  Also, $|I|\le n/8$, so $|N\setminus I|\ge7n/8\ge n/2$. Thus there are enough agents outside $I$ to assign a distinct agent to every blocker role. Each unused agent values only a private item of value one, and agents in each copy ignore items outside that copy. Every agent retains its prescribed weight.

  Combine the configuration solutions of these copies and assign mass one to each unused agent's private singleton. The resulting instance has $e^{L_{\mathrm{LP}}}=1$, while its integral optimum is $\prod_{i\in I}\ell_i^{-w_i}$. For $i\in I$, we have $\sqrt{nw_i/2}\ge2$, so taking the floor reduces this quantity by at most a factor of two. Hence $\ell_i\ge\frac12\sqrt{nw_i/2}=\sqrt{nw_i/8}$. For $i\notin I$, we have $1\ge\sqrt{nw_i/8}$. Consequently, the gap is at least
  \[
    \prod_{i\in I}\ell_i^{w_i}
      \ge\prod_{i\in N}(nw_i/8)^{w_i/2}
      =\frac{\Gamma(w)}{\sqrt8}.
  \]
  This also covers $I=\emptyset$, when every agent has a private item and the gap is one. Since XOS valuations are subadditive, the same lower bound applies to subadditive valuations.
\end{proof}

The basic instance also illustrates how an individual agent's loss affects the logarithmic objective. The special agent has value one in each of its fractional configurations and value $1/\ell$ in an optimal integral allocation, while every blocker has value one in both. The loss in logarithmic welfare is therefore $\omega\ln\ell$. With equal weights, this loss is $(\ln\ell)/n$, consistent with the constant gap when $\Gamma(w)=1$.

%% file: oracle-lower-bounds.tex
\section{Oracle Query Lower Bounds}
\label{sec:oracle-lower-bounds}

The integrality gap in \cref{subsec:nsw-gap} limits comparisons with the configuration LP. We now prove lower bounds on the number of oracle queries, without restricting the algorithm's computation or requiring it to use that LP. The hard instances have XOS valuations and remain hard when value, demand, and XOS supporting-clause queries are all available. For each prescribed weight vector, the construction keeps every agent's weight unchanged.

We use the oracle definitions in \cref{sec:prelim}, with one extension local to this section: demand queries may use arbitrary real prices, including negative prices. A lower bound in this stronger model also applies to ordinary nonnegative-price demand queries. The algorithm is given the weights and may know all parameters and the baseline construction below, but it accesses the planted valuations only through the stated oracles. In particular, it is not given their full XOS representations or the hidden sets. The oracles use a fixed, legal rule that prefers a baseline answer whenever that answer remains valid. Our lower bounds concern these oracle implementations; they do not assert hardness for every legal tie-breaking rule.

\subsection{Row gadgets and a hidden additive clause}
\label{subsec:oracle-gadgets}

We use a rectangular version of the row construction in \cref{subsec:nsw-gap}, retaining its group of unit-demand blockers and the attribution discussed there. Fix integers $R\ge L\ge2$. There are $R$ disjoint rows $G_1,\ldots,G_R$, each containing $L$ items. Each row has $L-1$ distinct blocker agents, each with valuation $u_r(S)=\mathbf1_{\{S\cap G_r\ne\emptyset\}}$. One special agent has baseline valuation
\begin{equation}\label{eq:oracle-baseline}
  v_B(S)=\frac1L\max_{r\in[R]}|S\cap G_r|.
\end{equation}
All agents ignore items outside their own gadget. Thus, the gadget has $1+R(L-1)$ agents and $RL$ items. Each row defines a nonnegative additive clause with coefficient $1/L$ on its items and zero elsewhere; their maximum is $v_B$. Blockers are unit-demand and hence also XOS.

Independently in each row, choose an item $t_r$ uniformly from $G_r$, and let $T=\{t_1,\ldots,t_R\}$. Fix a positive integer $b$ such that
\begin{equation}\label{eq:oracle-parameters}
  \lambda=R/L\ge1,\qquad y=b/\lambda>1,\qquad 2b<R.
\end{equation}
Replace the special agent's valuation by
\begin{equation}\label{eq:oracle-planted}
  v_T(S)=\max\left\{v_B(S),\frac{|S\cap T|}{bL}\right\}.
\end{equation}
Here $b$ is a gadget parameter, unrelated to the reserved masses $b_j$ used in the rounding algorithms. The set $T$ is local to this construction. The second term in \eqref{eq:oracle-planted} is a hidden additive clause. The planted valuation is normalized, monotone, and XOS, and its value on all items is one because $R/(bL)=1/y<1$.

\begin{lemma}\label{lem:oracle-gadget-optimum}
  In every allocation that gives all blockers positive value, the special agent's value is at most $1/y$. This value is attained while giving every blocker value one. Thus, the gadget's max-min optimum is $1/y$, and its optimal contribution to weighted Nash social welfare is $y^{-\omega}$ when the special agent has weight $\omega$.
\end{lemma}
\begin{proof}
  Every blocker must receive an item from its own row. Since the bundles are disjoint, the $L-1$ blockers in a row leave at most one item of that row for the special agent. Its baseline value is therefore at most $1/L$, and its hidden-clause value is at most $R/(bL)=1/y$. The condition $2b<R$ gives $1/y>2/L$, so the latter is the larger bound. Assigning $T$ to the special agent attains $1/y$; in each row, the remaining $L-1$ items can be assigned one per blocker. All blockers then have value one. An allocation that fails to give some blocker positive value has both max-min welfare and weighted Nash social welfare zero, proving the claims.
\end{proof}

We combine gadgets on disjoint item sets, using distinct agents for every special and blocker role. Any unused agent receives one private item and values a set at one if it contains that item and at zero otherwise. These agents do not affect the positive optimum. All agents retain their prescribed weights, including those used as blockers.

\paragraph{Fixed oracle implementations.}
Fix deterministic baseline answers for every query, including choices among maximizing rows and demand sets. For each fixed hidden set $T$, define the planted oracles as follows. A value query returns $v_T(S)$. A supporting-clause query returns the canonical maximizing baseline row clause whenever it attains $v_T(S)$, and otherwise returns the hidden clause. Every returned clause is nonnegative, is dominated by $v_T$ on all sets, and equals $v_T$ on the query set. For a demand query, return the canonical baseline demand set if it still maximizes $v_T(S)-p(S)$; otherwise, return a fixed maximizing set for the hidden clause. The latter is valid because an XOS valuation's maximum profit is the maximum of its clauses' optimized profits. These rules define consistent, legal oracles for each fixed valuation.

For value and supporting-clause queries, an answer can differ from the baseline answer only if the hidden clause strictly exceeds every baseline row clause on the query set. For demand queries, a different answer requires the optimized hidden profit to strictly exceed the largest optimized baseline profit. If these profits tie, the baseline demand set remains optimal. Even when the hidden profit is strictly larger, the baseline set might still be optimal; we only need the stated necessary condition for a different answer. We call a change from the baseline answer a \emph{revelation}.

\subsection{A one-query revelation bound}
\label{subsec:oracle-one-query}

For $z>0$, define $h(z)=\ln z-1+1/z$. In particular, $h(z)>0$ for $z>1$.

\begin{lemma}\label{lem:oracle-revelation}
  For any fixed value, demand, or XOS supporting-clause query to the special agent, the probability of a revelation, over the random hidden set $T$, is at most $\exp(-bh(y))$. The query may involve items outside the gadget, and demand prices may be negative.
\end{lemma}
\begin{proof}
  We first establish a tail bound shared by all three query types. Fix numbers $d_j\in[0,1]$ for items in the gadget, and write
  \[
    a=\max_{r\in[R]}\sum_{j\in G_r}d_j,\qquad
    u=\max_j d_j,\qquad X=\sum_{r=1}^R d_{t_r}.
  \]
  If $u=0$, then $X=0$ and $X>ba$ is impossible. Otherwise, $a\ge u>0$, and the variables $d_{t_r}/u$ are independent and lie in $[0,1]$. Their sum has expectation
  \[
    \E[X/u]=\frac1{Lu}\sum_jd_j\le\frac{\lambda a}{u}.
  \]
  For $s\ge0$, convexity gives $e^{sx}\le1+x(e^s-1)$ on $[0,1]$. Applying this to each summand, using independence and $1+x\le e^x$, gives
  \[
    \E e^{sX/u}\le\exp\left(\frac{\lambda a}{u}(e^s-1)\right).
  \]
  Exponential Markov with $s=\ln y>0$ and $\lambda y=b$ now yields
  \begin{equation}\label{eq:oracle-common-tail}
    \Pr[X>ba]\le
    \exp\left(-\frac{ba}{u}\ln y+\frac{\lambda a}{u}(y-1)\right)
    =\exp\left(-\frac{ba}{u}h(y)\right)
    \le e^{-bh(y)}.
  \end{equation}

  For a value or supporting-clause query on $S$, take $d_j=\mathbf1_{\{j\in S\}}$ on the gadget. A revelation requires $|S\cap T|>b\max_r|S\cap G_r|$, which is exactly $X>ba$. If $S$ contains no gadget item, both valuations are zero on $S$ and the baseline answer remains valid.

  For a demand query with prices $p_j$, scale profits by $bL$ and put $q_j=bLp_j$. For any additive clause with scaled coefficients $c_j$, the optimized scaled profit is $\sum_j\max\{c_j-q_j,0\}$. Subtract the common contribution of negative prices,
  \[
    C=\sum_{j\in M}\max\{-q_j,0\}.
  \]
  For each gadget item define
  \begin{align*}
    d_j&=\max\{1-q_j,0\}-\max\{-q_j,0\},\\
    e_j&=\max\{b-q_j,0\}-\max\{-q_j,0\}.
  \end{align*}
  These quantities satisfy $0\le d_j\le1$ and $e_j\ge bd_j$. To check the latter, if $q_j\le0$ then $(d_j,e_j)=(1,b)$; if $0<q_j<1$ then $e_j=b-q_j\ge b(1-q_j)=bd_j$; and if $q_j\ge1$ then $d_j=0$ and $e_j\ge0$. Here $b>1$ follows from \eqref{eq:oracle-parameters}.

  Above the common contribution $C$, the optimized hidden profit is $X=\sum_r d_{t_r}$, while the largest optimized baseline profit is
  \[
    \max_r\sum_{j\in G_r}e_j\ge b\max_r\sum_{j\in G_r}d_j=ba.
  \]
  A revelation therefore requires $X>ba$, so \eqref{eq:oracle-common-tail} applies. Items outside the gadget have zero coefficient in every clause and contribute only to $C$. Zero prices, empty maximizing sets, and ties are all covered by these formulas and the baseline-preferring rule.
\end{proof}

\subsection{Adaptive queries, outputs, and composition}
\label{subsec:oracle-composition}

An algorithm may produce a valuable bundle without revealing the hidden clause in a query. We first bound this possibility for a fixed output.

\begin{lemma}\label{lem:oracle-fixed-output}
  For a fixed set $S$ that contains at most one item from each row and is independent of $T$,
  \begin{equation}\label{eq:oracle-output-tail}
    \Pr[v_T(S)\ge2/L]\le e^{-2bh(2y)}.
  \end{equation}
\end{lemma}
\begin{proof}
  The baseline value is at most $1/L$, so the event requires $Z=|S\cap T|\ge2b$. The contributions to $Z$ from different rows are independent indicators, each of mean at most $1/L$, and hence $\E Z\le R/L=\lambda$. As in the preceding proof, $\E e^{sZ}\le\exp(\lambda(e^s-1))$. Taking $s=\ln(2y)$ gives
  \[
    \Pr[Z\ge2b]\le\exp\bigl(-2b\ln(2y)+\lambda(2y-1)\bigr)
      =e^{-2bh(2y)}.
  \]
\end{proof}

Choose a nonempty set $I$ of special agents, and give each $i\in I$ a gadget with parameters $R_i,L_i,b_i$ satisfying \eqref{eq:oracle-parameters}. Choose all hidden transversals independently, use distinct blockers, and pad with private-item agents. Define
\begin{equation}\label{eq:oracle-composition-parameters}
  y_i=\frac{b_iL_i}{R_i},\qquad
  \rho=\prod_{i\in I}\left(\frac{R_i}{2b_i}\right)^{w_i}>1.
\end{equation}
By \cref{lem:oracle-gadget-optimum}, every hidden instance has the same positive optimum
\[
  \mathrm{OPT}_{\mathrm{NSW}}=\prod_{i\in I}y_i^{-w_i}.
\]

\begin{lemma}\label{lem:oracle-adaptive}
  For any randomized algorithm making at most $Q$ queries in total across all agents and all three query types, the composed distribution satisfies
  \begin{equation}\label{eq:oracle-adaptive-bound}
    \Pr\!\left[\mathrm{NSW}_w(A)\ge\frac{\mathrm{OPT}_{\mathrm{NSW}}}{\rho}\right]
      \le Q\max_{i\in I}e^{-b_ih(y_i)}
         +\sum_{i\in I}e^{-2b_ih(2y_i)}.
  \end{equation}
  The probability includes the hidden transversals and the algorithm's random choices. For one gadget, the same bound holds for max-min welfare with $\mathrm{OPT}_{\mathrm{MM}}=1/y$ and $\rho=R/(2b)$.
\end{lemma}
\begin{proof}
  Fix the algorithm's random choices and run it against the baseline oracles. This determines a sequence of at most $Q$ queries and a fixed output allocation $A^0$. The planted execution follows the same sequence until its first revelation: by induction, identical earlier answers give identical internal states and next queries. If a revelation occurs, it therefore occurs at a query on the fixed baseline sequence. Queries to blockers and private-item agents never reveal anything. Applying \cref{lem:oracle-revelation} to each query to a special agent and taking a union bound gives the first term in \eqref{eq:oracle-adaptive-bound}.

  If the fixed allocation $A^0$ gives zero value to a blocker or a private-item agent, it has zero welfare for every hidden instance. Otherwise, every special bundle $A_i^0$ contains at most one item from each of its rows. If all special agents have value strictly below $2/L_i$, then
  \[
    \frac{\mathrm{NSW}_w(A^0)}{\mathrm{OPT}_{\mathrm{NSW}}}
      <\prod_{i\in I}\left(\frac{2y_i}{L_i}\right)^{w_i}
      =\rho^{-1},
  \]
  since every blocker and private-item agent has value one. Thus, success without a revelation implies that $v_{T_i}(A_i^0)\ge2/L_i$ for at least one $i\in I$. Apply \cref{lem:oracle-fixed-output} to the fixed bundles and take a union bound to obtain the second term. This is an unconditional bound on the fixed output; we do not condition the hidden sets on the absence of a revelation.

  Averaging over the algorithm's random choices proves \eqref{eq:oracle-adaptive-bound}. For one gadget, every nonspecial agent has value one whenever the welfare is positive, so success at factor $R/(2b)$ requires special value at least $2/L$. The same argument proves the max-min assertion.
\end{proof}

\paragraph{Fixed instances and expected welfare.}
Let $\varepsilon$ denote the right-hand side of \eqref{eq:oracle-adaptive-bound}, and let $W$ be the returned objective, either weighted Nash social welfare or the single-gadget max-min welfare. Every output allocation satisfies $0\le W\le\mathrm{OPT}$. Splitting according to whether $W\ge\mathrm{OPT}/\rho$ gives
\begin{equation}\label{eq:oracle-expectation}
  \E_{T,\mathrm{coins}}[W/\mathrm{OPT}]\le\rho^{-1}+\varepsilon.
\end{equation}
The optimum is the same for every hidden instance. Moreover, some fixed choice of the hidden sets has conditional success probability at most $\varepsilon$, since the average of these probabilities is at most $\varepsilon$. On this same fixed instance, splitting over the algorithm's random choices gives expected normalized welfare at most $\rho^{-1}+\varepsilon$. Thus the bound produces a fixed hard instance and legal fixed oracles for each randomized algorithm. It concerns the expectation of welfare, not the expectation of its logarithm.

\subsection{Polynomial-query lower bounds and consequences}
\label{subsec:oracle-polynomial}

Fix $D>0$. For $n\ge3$, take
\begin{equation}\label{eq:oracle-polynomial-parameters}
  \lambda=\lceil(D+4)\ln n\rceil,\qquad R_i=\lambda L_i,\qquad b_i=5\lambda.
\end{equation}
Whenever $L_i\ge20$, these integers satisfy \eqref{eq:oracle-parameters}, with $y_i=5$ and $R_i/(2b_i)=L_i/10$. Since $5h(5)>4$ and $10h(10)>14$, for $Q\le n^D$ and $|I|\le n$ the right-hand side of \eqref{eq:oracle-adaptive-bound} is at most
\begin{equation}\label{eq:oracle-error}
  \varepsilon_n=n^{-3D-16}+n^{-14D-55}.
\end{equation}
For example, the query term is at most $n^De^{-4\lambda}\le n^{-3D-16}$, and the output term is at most $ne^{-14\lambda}\le n^{-14D-55}$. Also, $\lambda\le2(D+4)\ln n$ for $n\ge3$. We may use the common constant
\[
  c_D=\frac1{80\sqrt{D+4}}
\]
for the two theorems below, taking the threshold on $n$ large enough that the stated construction exists, $\varepsilon_n<2/3$, and $\varepsilon_n\sqrt n<1$.

\begin{theorem}\label{thm:oracle-maxmin}
  For every fixed $D>0$, there are constants $c_D>0$ and $n_D$, depending only on $D$, with the following property. For every $n\ge n_D$ and every randomized algorithm making at most $n^D$ adaptive queries in total, there is a normalized monotone XOS instance with exactly $n$ agents, $n\le m\le20n/19$ items, and positive optimum, together with legal baseline-preferring oracles, on which its output $A$ satisfies
  \begin{align*}
    \Pr\!\left[\min_i v_i(A_i)\ge
      \frac{\mathrm{OPT}_{\mathrm{MM}}}{c_D\sqrt{n/\ln n}}\right]&<\frac23,\\
    \E\bigl[\min_i v_i(A_i)\bigr]&<
      \frac{\mathrm{OPT}_{\mathrm{MM}}}{c_D\sqrt{n/\ln n}}.
  \end{align*}
  Queries may be value, demand at arbitrary real prices, or XOS supporting-clause queries, and computation between queries is unrestricted. The probability and expectation above are over the algorithm's random choices on the fixed instance.
\end{theorem}
\begin{proof}
  Use one special agent with the parameters in \eqref{eq:oracle-polynomial-parameters} and
  \[
    L=\left\lfloor\sqrt{\frac{n-1}{\lambda}}\right\rfloor.
  \]
  For all sufficiently large $n$, depending only on $D$, we have $L\ge20$. There are $\lambda L(L-1)\le n-1$ blockers, so private-item agents complete the instance to exactly $n$ agents. Its number of items is $m=n-1+R$. Since $R\ge1$ and
  \[
    R=\lambda L\le\frac{\lambda L(L-1)}{19}\le\frac{n-1}{19},
  \]
  we have $n\le m\le20n/19$.

  Every planted instance has $\mathrm{OPT}_{\mathrm{MM}}=1/5$. Let $\rho=L/10$, so that $\mathrm{OPT}_{\mathrm{MM}}/\rho=2/L$. Taking the floor loses at most a factor of two. Using $n-1\ge n/2$ and $\lambda\le2(D+4)\ln n$, we obtain
  \[
    \rho\ge\frac1{20}\sqrt{\frac{n-1}{\lambda}}
      \ge\frac1{40\sqrt{D+4}}\sqrt{\frac n{\ln n}}
      =2c_D\sqrt{\frac n{\ln n}}.
  \]
  Also, $1<\rho\le\sqrt n$. By \cref{lem:oracle-adaptive} and \eqref{eq:oracle-error}, some fixed hidden instance has success probability at most $\varepsilon_n$ at factor $\rho$. The displayed target factor in the theorem is at most $\rho/2$, so attaining it is no more likely. This proves the probability claim. On the same instance, \eqref{eq:oracle-expectation} and $\varepsilon_n\rho<1$ give
  \[
    \E\bigl[\min_i v_i(A_i)/\mathrm{OPT}_{\mathrm{MM}}\bigr]
       \le\rho^{-1}+\varepsilon_n<2/\rho
       \le\frac1{c_D\sqrt{n/\ln n}},
  \]
  proving the expectation claim. All coefficients are among $0,1,1/L,1/(bL)$, with polynomial encoding length.
\end{proof}

\begin{theorem}\label{thm:oracle-weighted-nsw}
  For every fixed $D>0$, there are constants $c_D>0$ and $n_D$, depending only on $D$, such that the following holds for every $n\ge n_D$ and every prescribed positive rational weight vector $w$ with $\sum_iw_i=1$. For every randomized algorithm making at most $n^D$ adaptive value, demand, and XOS supporting-clause queries in total, there is a normalized monotone XOS instance with these exact weights, exactly $n$ agents, $n\le m\le20n/19$ items, and positive optimum, together with legal baseline-preferring oracles, for which
  \begin{align*}
    \Pr\!\left[\mathrm{NSW}_w(A)\ge
      \frac{\mathrm{OPT}_{\mathrm{NSW}}}{c_D\Gamma(w)/\sqrt{\ln n}}\right]&<\frac23,\\
    \E\bigl[\mathrm{NSW}_w(A)\bigr]&<
      \frac{\mathrm{OPT}_{\mathrm{NSW}}}{c_D\Gamma(w)/\sqrt{\ln n}}.
  \end{align*}
  The algorithm may depend on $w$ and the public parameters and may use arbitrary real demand prices and unrestricted computation. The displayed lower bound on the approximation factor is vacuous when it is below one.
\end{theorem}
\begin{proof}
  Use \eqref{eq:oracle-polynomial-parameters} and select
  \begin{equation}\label{eq:oracle-weighted-selection}
    I=\{i:nw_i\ge800\lambda\},\qquad
    L_i=\left\lfloor\sqrt{\frac{nw_i}{2\lambda}}\right\rfloor
      \quad(i\in I).
  \end{equation}
  Every selected agent has $L_i\ge20$ and becomes the special agent of a separate gadget. The number of blocker roles is
  \[
    \sum_{i\in I}\lambda L_i(L_i-1)
       \le\frac n2\sum_{i\in I}w_i\le\frac n2.
  \]
  Since $|I|\le n/(800\lambda)$, at least $n/2$ agents lie outside $I$. We can therefore fill all blocker roles with distinct agents from outside $I$. Give each remaining agent a private item. No weight changes. For nonempty $I$, the item count is
  \[
    m=n-|I|+\sum_{i\in I}R_i.
  \]
  Each $R_i\ge1$, and $R_i\le\lambda L_i(L_i-1)/19$. Consequently $m\ge n$ and $m\le n+n/38\le20n/19$.

  Suppose first that $I\ne\emptyset$, and put $\rho=\prod_{i\in I}(L_i/10)^{w_i}$. Every hidden instance has optimum $\mathrm{OPT}_{\mathrm{NSW}}=5^{-\sum_{i\in I}w_i}$. For $i\in I$, the floor loses at most a factor of two, giving $L_i/10\ge\sqrt{nw_i/(800\lambda)}$. For $i\notin I$, this square root is below one. Thus
  \begin{equation}\label{eq:oracle-entropy-comparison}
    \rho\ge\prod_{i\in N}\left(\frac{nw_i}{800\lambda}\right)^{w_i/2}
      =\frac{\Gamma(w)}{\sqrt{800\lambda}}
      \ge\frac{\Gamma(w)}{40\sqrt{(D+4)\ln n}}
      =\frac{2c_D\Gamma(w)}{\sqrt{\ln n}}.
  \end{equation}
  Moreover, $1<\rho\le\sqrt n$, since every $L_i/10\le\sqrt n$ and $\sum_{i\in I}w_i\le1$. By \cref{lem:oracle-adaptive}, some fixed hidden instance has success probability at most $\varepsilon_n$ at factor $\rho$. Its expected normalized welfare is at most $(1+\varepsilon_n\rho)/\rho<2/\rho$. The two claims follow from \eqref{eq:oracle-entropy-comparison}, exactly as in the max-min proof.

  If $I=\emptyset$, then $nw_i<800\lambda$ for every agent, so $\Gamma(w)\le\sqrt{800\lambda}$. Our choice of $c_D$ gives $c_D\Gamma(w)/\sqrt{\ln n}\le1/2$. An instance of $n$ private-item agents has $m=n$ and positive optimum, and no allocation can exceed that optimum. Both inequalities in the theorem are therefore automatic; in this case the displayed approximation lower bound is only a vacuous bound below one.

  The valuations and their returned clauses have polynomial encoding length, independently of the numerical sizes of the given weights: their nonzero coefficients are $1$, $1/L_i$, and $1/(5\lambda L_i)$. The prescribed weights themselves are not altered or rounded.
\end{proof}

\paragraph{A more detailed dependence on the weights.}
The construction retains more information than the uniform comparison in \eqref{eq:oracle-entropy-comparison}. With $\ln^+x=\max\{0,\ln x\}$, it gives
\begin{equation}\label{eq:oracle-weight-profile}
  \rho\ge\exp\left(\frac12\sum_{i\in N}w_i
    \ln^+\!\left(\frac{nw_i}{800\lceil(D+4)\ln n\rceil}\right)\right)
  \qquad(I\ne\emptyset).
\end{equation}
For an index outside $I$, retain the factor one rather than replace it by $\sqrt{nw_i/(800\lambda)}<1$; the displayed inequality then follows from the same floor bound. The probability estimate holds at factor $\rho$, and the expectation estimate gives the same order within an absolute factor, because $\varepsilon_n\rho=o(1)$ uniformly over the weights. If $I=\emptyset$, the right-hand side of \eqref{eq:oracle-weight-profile} is one, but the composition lemma does not apply and supplies no nontrivial lower bound. In particular, equal weights have $\Gamma(w)=1$, so \cref{thm:oracle-weighted-nsw} does not exclude a constant-factor approximation in that case.

\begin{corollary}\label{cor:oracle-consequences}
  For any fixed query exponent $D$ and any sequence of prescribed weight vectors with $H(w)=O(1)$, the weighted NSW query lower bound is $\Omega(\sqrt{n/\ln n})$. Max-min allocation has the same lower bound. These bounds also apply to subadditive valuations with ordinary demand-query access.
\end{corollary}
\begin{proof}
  If $H(w)$ is bounded by an absolute constant, then $\Gamma(w)=\exp((\ln n-H(w))/2)=\Theta(\sqrt n)$. Apply \cref{thm:oracle-weighted-nsw}; the max-min statement is \cref{thm:oracle-maxmin}. XOS valuations form a subclass of subadditive valuations. Restricting an algorithm to ordinary demand queries cannot remove a lower bound already proved when all three types of queries are allowed. This implication does not assume that arbitrary subadditive valuations admit an XOS oracle.
\end{proof}

For example, $w_1=1-1/n$ and $w_i=1/[n(n-1)]$ for $i>1$ have bounded entropy. The obstruction is not determined by the largest-weight exponent $n^{w_{\max}/2}$ alone. For $n\ge3$, give two agents weight $(1-n^{-2})/2$ each and divide the remaining weight $n^{-2}$ equally among the other agents. Writing $\delta=n^{-2}$, their entropy is
\[
  H(w)=(1-\delta)\ln\frac2{1-\delta}
       +\delta\ln\frac{n-2}{\delta}=\ln2+o(1).
\]
Hence the query lower bound is $\Omega(\sqrt{n/\ln n})$, whereas $n^{w_{\max}/2}=n^{1/4+o(1)}$. This example only compares those two expressions; it does not rule out other analyses using the largest weight together with additional information.

\paragraph{Scope of the query bound.}
Every construction uses $m=\Theta(n)$ items. Thus any fixed polynomial query budget in $n$ and $m$ is covered by increasing $D$. A polynomial budget in the full public input length is covered as well when that length, including the encoding of $w$, is polynomial in $n$. For weights of unrestricted encoding length, the theorems assert the explicit $n^D$ query bound and do not identify it with every polynomial in that larger encoding length. These are information-theoretic oracle bounds, not NP-hardness, communication lower bounds, or lower bounds for explicitly listed XOS clauses. Together with the algorithms, they leave a factor $O(\sqrt{\ln n})$ between the upper and query-lower bounds, whereas the configuration-LP gaps match the upper bounds up to constants.

%% file: appendix.tex
\section{Computing configuration solution for max-min allocation}
\label[appendix]{app:configuration-producer}

Bansal and Sviridenko~\cite{BS06} used knapsack separation for the additive max-min configuration LP. Bamas, Garg, and Rohwedder~\cite[Theorem~2]{BGR21} used approximate knapsack separation for restricted submodular valuations, as shown in their full version~\cite[Appendix~B.1]{BGR20full}. We adapt this argument to agent-specific valuations and an arbitrary budgeted value-maximization routine. The rational item prices let us enforce a strict budget without losing a further constant factor. This gives \cref{lem:mm-lp-solution} by using the submodular knapsack algorithm of Sviridenko~\cite{Sviridenko04} or the subadditive demand-query algorithm of Badanidiyuru, Dobzinski, and Oren~\cite{BDO19}. In both cases the resulting solution satisfies \eqref{eq:mm-row} and \eqref{eq:mm-capacity} exactly.

\begin{lemma}\label{lem:configuration-producer}
  Fix a rational constant $\beta\in(0,1]$. Suppose there is a polynomial-time routine that takes nonnegative rational item costs, a budget, and a target $t>0$ as input and returns a set within the budget. The returned set has value at least $\beta t$ whenever some set within the budget has value at least $t$. Assume also that $v_i(S)$ can be evaluated exactly for every agent $i$ and every set $S\subseteq M$.

  Then, for every rational $\eta>0$, we can either detect that $T_{\mathrm{LP}}=0$ or compute an explicit configuration solution supported on polynomially many configurations at a target
  \[
    T>\frac{\beta}{1+\eta}T_{\mathrm{LP}}.
  \]
  The running time is polynomial in the input encoding length and $1/\eta$ for fixed $\beta$.
\end{lemma}

\begin{proof}[Proof of \cref{lem:configuration-producer}]
  We first construct a procedure that, for a given target $t>0$, either returns an exactly feasible configuration solution with objective value $t$ or certifies that the configuration LP is infeasible at target $t/\beta$. We then use this procedure to find a target satisfying the claimed guarantee.

  Fix $t>0$. Replace the constraint $\sum_{S}y_{i,S}=1$ of each agent $i$ by the lower bound $\sum_{S}y_{i,S}\ge\lambda$, and maximize $\lambda$. Recall that $\mathcal C_i(t)=\{S\subseteq M:v_i(S)\ge t\}$, this gives the packing LP
  \begin{align*}
    \text{maximize}\quad &\lambda\\
    \text{subject to}\quad
      &\sum_{S\in\mathcal C_i(t)}y_{i,S}\ge\lambda
        &&(i\in N),\\
      &\sum_{i\in N}\sum_{\substack{S\in\mathcal C_i(t)\\j\in S}}y_{i,S}\le1
        &&(j\in M),\\
      &y\ge0,\qquad \lambda\in\mathbb R.
  \end{align*}
  The configuration LP at target $t$ is feasible if and only if the packing optimum is at least $1$. To obtain a configuration solution from a packing solution with $\lambda\ge1$, divide the masses of each agent by their sum. This makes the total mass of every agent equal to $1$ without increasing the total mass on any item.

  Writing $p(S)=\sum_{j\in S}p_j$, the dual of the packing LP is
  \begin{align}
    \text{minimize}\quad &\sum_{j\in M}p_j
      \label{eq:producer-dual}\\
    \text{subject to}\quad
      &\sum_{i\in N}z_i=1,\qquad p\ge0,\ z\ge0,\notag\\
      &p(S)\ge z_i\qquad(i\in N,\ S\in\mathcal C_i(t)).\notag
  \end{align}

  We next describe how to test a rational candidate $(z,p)$ satisfying the explicit constraints $\sum_i z_i=1$ and $p,z\ge0$. Let $Q$ be a common denominator of its coordinates and set $\zeta=1/Q$. The number $Q$ has polynomial bit length in the encoding of $(z,p)$. Since all coordinates are multiples of $\zeta$, we have
  \[
    p(S)<z_i\quad\Longrightarrow\quad p(S)\le z_i-\zeta.
  \]
  For each agent $i$ with $z_i>0$, call the budgeted routine with item costs $p_j$, budget $z_i-\zeta$, and value target $t/\beta$, the budget is nonnegative because $z_i\ge\zeta$. Use exact value evaluation to check the returned set $D$. If $v_i(D)\ge t$, then $D\in\mathcal C_i(t)$ and
  \[
    p(D)\le z_i-\zeta<z_i,
  \]
  so $D$ identifies a violated configuration constraint in \eqref{eq:producer-dual}.

  Suppose that no call returns a set of value at least $t$. If a set $S\in\mathcal C_i(t/\beta)$ violated $p(S)\ge z_i$, it would lie within the budget $z_i-\zeta$. The guarantee of the budgeted routine would then force it to return a set of value at least $\beta(t/\beta)=t$, a contradiction. For agents with $z_i=0$, all configuration constraints hold automatically because $p\ge0$. Thus, these tests either identify a violated constraint at target $t$ or certify that $(z,p)$ is feasible for the dual at target $t/\beta$.

  To recover an exactly feasible configuration solution, we use a uniform bound on the denominators of optima of finite restricted packing LPs. Set $d=n+m+2$ and $H_0=2^{d^2}$. For any finite collection of configurations, the corresponding restricted packing LP has a standard-form constraint matrix with entries in $\{-1,0,1\}$, integral right-hand sides, and at most $n+m+1$ independent rows. Cramer's rule and Hadamard's bound therefore imply that its optimal value has denominator at most $H_0$. Consequently, if this optimum is below $1$, it is at most $1-1/H_0$.

  Run the rational feasibility algorithm of Gr{\"o}tschel, Lov{\'a}sz, and Schrijver~\cite[Theorem~6.4.1]{GLS88}, using the explicit dual constraints, the configuration tests above, and the additional bound
  \begin{equation}
    \sum_{j\in M}p_j\le1-\frac{1}{2H_0}.
    \label{eq:producer-cutoff}
  \end{equation}
  These explicit constraints place all coordinates in $[0,1]$. We stop on the first candidate accepted by the tests and record every configuration cut otherwise. We use a common encoding-length bound for inequalities in every subsystem. The rational feasibility procedure handles lower-dimensional feasible sets, including the equality $\sum_i z_i=1$.

  If the procedure finds a candidate accepted by all tests, this candidate is feasible for the dual at target $t/\beta$ and has objective value below $1$. By weak duality, the packing optimum at that target is below $1$, so the configuration LP at target $t/\beta$ is infeasible.

  Otherwise the feasibility algorithm reports an empty polyhedron. The recorded cuts and the explicit constraints, including \eqref{eq:producer-cutoff}, already form an infeasible system. Indeed, every oracle reply in this run strictly separated its query from this subsystem, so the same replies constitute a valid execution of the exact feasibility algorithm for the subsystem. Here and below we test the algorithm's final candidate as well if it returns one. Consider the restricted packing LP on exactly these configurations, and let $\lambda^*$ be its optimum. Its dual has no feasible solution with objective at most $1-1/(2H_0)$, so strong duality gives
  \[
    \lambda^*>1-\frac{1}{2H_0}.
  \]
  The denominator bound rules out $\lambda^*<1$, since that would imply $\lambda^*\le1-1/H_0$, hence $\lambda^*\ge1$. Solve this finite packing LP exactly and divide the masses of each agent by their sum. The resulting solution gives every agent total mass $1$ and every item total mass at most $1$, and has polynomial support. Every configuration in its support has value at least $t$ for its agent, so it is the required configuration solution with objective value $t$.

  It remains to choose the target. Let $\sigma$ be the encoding bound from \cref{subsec:encoding-convention}, and let $U=\min_i v_i(M)$. If $U=0$, then $T_{\mathrm{LP}}=0$. Otherwise, start with $t=U$ and apply the procedure above. If it returns a configuration solution, stop and set $T=t$. If it certifies infeasibility at target $t/\beta$, replace $t$ by $t/(1+\eta)$ and continue. Once $t\le\beta2^{-\sigma}$, such a certificate instead causes the algorithm to report $T_{\mathrm{LP}}=0$, we justify this stopping rule below.

  Suppose that the first configuration solution is returned with objective value $T<U$. The preceding call, at target $(1+\eta)T$, certified infeasibility at target $(1+\eta)T/\beta$. Therefore,
  \[
    T_{\mathrm{LP}}<\frac{(1+\eta)T}{\beta},
    \qquad\text{and hence}\qquad
    T>\frac{\beta}{1+\eta}T_{\mathrm{LP}}.
  \]
  If the first call returns a solution at $T=U$, the same guarantee follows from $T_{\mathrm{LP}}\le U$, $U>0$, and $\beta/(1+\eta)<1$.

  Finally, the encoding bound in \cref{subsec:encoding-convention} implies that every positive valuation is at least $2^{-\sigma}$ and that $U\le2^\sigma$. Thus, if $T_{\mathrm{LP}}>0$, then $T_{\mathrm{LP}}\ge2^{-\sigma}$. For $t\le\beta2^{-\sigma}$, the configuration LP at target $t/\beta$ is then feasible, so the procedure must return a configuration solution with objective value $t$. A certificate of infeasibility at such a target therefore proves that $T_{\mathrm{LP}}=0$. The number of calls is at most $1+\lceil\log_{1+\eta}(2^{2\sigma}/\beta)\rceil$, which is polynomial in $\sigma$, the encoding length of $\beta$, and $1/\eta$. All targets have polynomial encoding length, and each call runs in polynomial time. The total running time is therefore polynomial in the input encoding length and $1/\eta$ for fixed $\beta$.
\end{proof}

\paragraph{Submodular valuations.}
For monotone submodular valuations, Sviridenko~\cite[Theorem~1]{Sviridenko04} gives a $(1-1/e)$-approximation for maximization under a knapsack constraint using value queries. To apply it, we first include all zero-cost items and contract them, then multiply the remaining costs and the budget by a common denominator to obtain an equivalent instance with integer costs. Since $1-1/e>1/3$, this implements the routine of \cref{lem:configuration-producer} with $\beta=1/3$ using value queries.

\paragraph{Subadditive valuations.}
For subadditive valuations, we use ordinary demand queries. Under the bit bound, these can also implement restricted demand queries and exact value evaluation; see \cref{app:demand-values}.

\begin{lemma}\label{lem:demand-budget}
  For normalized monotone subadditive valuations, ordinary demand
  queries implement the budgeted routine of
  \cref{lem:configuration-producer} with $\beta=1/3$.
\end{lemma}
\begin{proof}
  Badanidiyuru, Dobzinski, and Oren~\cite{BDO19} give a $(2+\varepsilon)$-approximation for subadditive maximization under one knapsack constraint, using polynomially many demand queries for every fixed $\varepsilon>0$. Taking $\varepsilon=1/2$ gives value at least $2t/5$ whenever an affordable set has value at least $t$. For budget zero, monotonicity lets us return the set of all zero-cost items.
\end{proof}

\begin{proof}[Proof of \cref{lem:mm-lp-solution}]
  Apply \cref{lem:configuration-producer} with $\beta=1/3$ and $\eta=1$, using Sviridenko's algorithm for submodular valuations and \cref{lem:demand-budget} for subadditive valuations. This gives $T>T_{\mathrm{LP}}/6$. The exact value evaluations required by \cref{lem:configuration-producer} are value queries for submodular valuations. For subadditive valuations, they are implemented by ordinary demand queries under the bit bound; see \cref{app:demand-values}.
\end{proof}

\section{Restriction and exact value evaluation via ordinary demand queries}
\label[appendix]{app:demand-values}

Blumrosen and Nisan~\cite[Lemmas~1 and~2]{BN09} simulate a value query by recovering marginal values from demand queries and summing them. We give the finite prices and precision needed under the encoding convention of \cref{subsec:encoding-convention}, with $\sigma$ denoting the bound defined there.

\paragraph{Restricted demand queries.}
To restrict a query to $R\subseteq M$, retain the given nonnegative prices on $R$ and price every other item at $2^\sigma+1$. Every bundle containing such an item has negative utility because $v_i(M)\le 2^\sigma$. The empty set has utility zero, so every maximizing bundle lies in $R$.

\paragraph{Exact value evaluation.}
For $j\in S$, let $d=v_i(S)-v_i(S\setminus\{j\})$. Following the marginal-price search of Blumrosen and Nisan, restrict demand to $S$, price $j$ at $t$, and price all other items in $S$ at zero. By monotonicity we can adjoin all the zero-price items to the answer without changing its optimality. The resulting bundle contains $j$ only if $d\ge t$, and omits $j$ only if $d\le t$, so either answer is valid at a tie.

Under our bit bound, $0\le d\le 2^\sigma$ and $d$ has denominator at most $2^{2\sigma}$. Distinct rationals with denominators at most $2^{2\sigma}$ differ by at least $2^{-4\sigma}$. Thus $O(\sigma)$ binary-search steps give an interval containing $d$ of width less than $2^{-4\sigma-2}$. Its midpoint has $d$ as its unique closest rational of denominator at most $2^{2\sigma}$, which continued fractions find in polynomial time~\cite[Theorem~5.1.9]{GLS88}. Summing the successive marginal values along any ordering of $S$ recovers $v_i(S)$ using $O(|S|\sigma)$ ordinary demand queries, since $v_i(\emptyset)=0$.

\section{Finite implementation of NSW rounding}\label[appendix]{app:nsw-finite}

The proof in \cref{sec:nsw} uses exact matching marginals to simplify the analysis. Here we give a polynomial-time implementation using rational edge weights. We sample weighted matchings using a permanent approximation scheme~\cite{JSV04} and calibrate their edge weights using the maximum-entropy dual potential studied by Singh and Vishnoi~\cite{SV14}. Straszak and Vishnoi~\cite[Theorem~15]{SV19} also handle marginal vectors on the boundary of the feasible region, assuming an evaluation oracle for the generating polynomial. In our setting, however, the matching partition function is available only through randomized approximations. We therefore give a rational coordinate-update algorithm and account for the error in its sampling calls.

\begin{lemma}\label{lem:nsw-finite-matching}
  Let $x$ be a feasible rational covering-matching matrix, and let $\eta>0$ and $\rho>0$ be a tolerance and a failure bound. A randomized polynomial-time procedure returns positive rational weights on $E=\{(i,j):x_{ij}>0\}$ such that, with probability at least $1-\rho$, their exact Gibbs marginals $x'$ satisfy
  \begin{equation}\label{eq:nsw-calibration}
    \|x'-x\|_1\le\eta.
  \end{equation}
  For every weight vector returned by the procedure and every $\varepsilon>0$, a covering matching on $E$ can be sampled with total-variation error at most $\varepsilon$ in worst-case randomized polynomial time. The running times are polynomial in the input encoding length, $1/\eta$, $1/\varepsilon$, and $\log(1/\rho)$.
\end{lemma}

We first analyze the exact Gibbs distribution associated with a fixed weight vector returned by the procedure. We then account for the error introduced by approximate sampling.

\subsection{Approximate sampling and rational calibration}
\label[appendix]{subsec:nsw-calibration}

\paragraph{Sampling a Gibbs matching.}
We use the nonnegative-permanent approximation scheme of Jerrum, Sinclair, and Vigoda~\cite[Theorem~1.1]{JSV04}. Given a rational nonnegative matrix, it estimates the permanent to relative error $\alpha$ with failure probability at most $\zeta$, in time polynomial in the encoding length, $1/\alpha$, and $\log(1/\zeta)$. A matching computation detects exactly whether the permanent is zero. For an $r$-by-$s$ matrix with $r\le s$, we add $s-r$ rows whose entries are all one. The permanent of the resulting square matrix equals the covering-matching partition function multiplied by $(s-r)!$. Thus, the same approximation scheme applies to rectangular partition functions.

We adapt the counting-to-generation reduction of Jerrum, Valiant, and Vazirani~\cite[Theorem~6.3]{JVV86} using weighted residual partition functions. Fix the edge weights and process the agents one at a time. Conditional on the assignments already made, the probability of assigning an available item $j$ to the current agent $i$ is proportional to
\[
  \lambda_{ij}\,
  Z_{\text{remaining rows},\,\text{remaining columns}\setminus\{j\}}.
\]
A matching computation identifies the choices that can be extended to a covering matching. For each such choice, approximate its positive partition function to relative error $\varepsilon/(32n)$, and normalize the resulting weights to obtain the sampling probabilities. If all estimates are accurate, the total-variation error at each step is at most $\varepsilon/(8n)$. Coupling the choices successively therefore gives a total error of at most $\varepsilon/8$.

Choose the failure bounds so that the total failure probability of the polynomially many permanent estimates is at most $\varepsilon/4$. Also limit the rational sampling calls as in \cref{subsec:rational-sampling}, with total abort probability at most $\varepsilon/4$. Throughout the procedure, choose only edges that admit a covering completion. If a sampler aborts or an estimate is malformed, complete the current partial matching deterministically. Every output is therefore a covering matching on $E$, and the combined error bounds place its distribution within total-variation distance $\varepsilon$ of the exact Gibbs distribution. This procedure uses no valuation queries.

\paragraph{Calibrating the marginals.}
Write $d=|E|$ and assume $0<\eta,\rho<1$. Define
\begin{equation}\label{eq:nsw-calibration-parameters}
  \gamma=\frac{\eta}{4d},
  \qquad
  T_{\mathrm{cal}}=\left\lceil\frac{8nm}{\gamma^2}\right\rceil+1,
  \qquad
  s=\left\lceil\frac{32}{\gamma^2}\right\rceil
    \left\lceil\log_2\frac{2dT_{\mathrm{cal}}}{\rho}\right\rceil.
\end{equation}
Initialize $\lambda_e=1$ for every edge. In each of at most $T_{\mathrm{cal}}$ iterations, use the sampler above to draw $s$ independent matchings from a distribution within total-variation distance $\gamma/8$ of the current Gibbs distribution. Let $\widehat p_e$ be the empirical frequency of edge $e$. If $|\widehat p_e-x_e|\le\gamma$ for every edge, return the current weights. Otherwise, choose an edge that violates this condition. Multiply its weight by $1+\gamma/2$ if $\widehat p_e<x_e$, and by $1-\gamma/2$ if $\widehat p_e>x_e$. If the iteration limit is reached, return the current weights regardless of whether the stopping condition holds.

Let $p_e$ be the exact current marginal of edge $e$. Approximate sampling changes each edge probability by at most $\gamma/8$. Applying a Bernoulli tail bound to the empirical frequencies and taking a union bound over the edges gives
\[
  \Pr\!\left[\max_e|\widehat p_e-p_e|>\gamma/4\right]
    \le2d\exp(-s\gamma^2/32)\le\rho/T_{\mathrm{cal}}.
\]
This bound holds conditional on all preceding iterations. A union bound over the iterations therefore shows that all empirical estimates are accurate to within $\gamma/4$ with probability at least $1-\rho$.

To analyze progress, use the maximum-entropy dual objective~\cite{SV14}, evaluated at the logarithms of the edge weights, as the potential
\[
  \Phi(\lambda)=\ln Z_\lambda-\sum_e x_e\ln\lambda_e.
\]
By integrality of the bipartite matching polytope, $x$ is a convex combination of covering matchings on $E$. Thus, $\sum_e x_e\ln\lambda_e$ is an average of the logarithms of their weights and is at most $\ln Z_\lambda$. It follows that $\Phi\ge0$, while the initial potential satisfies $\Phi(1)=\ln|\mathcal M|\le n\ln m\le nm$. Multiplying one edge weight by $1+u$ changes the partition function to $Z_\lambda(1+up_e)$. For $|u|\le1/2$, the potential therefore changes by
\[
  \Delta\Phi
    =\ln(1+up_e)-x_e\ln(1+u)
    \le u(p_e-x_e)+u^2.
\]
Suppose the empirical estimates are accurate and the stopping condition is not satisfied. For the chosen edge, $p_e-x_e$ has the same sign as $\widehat p_e-x_e$ and magnitude greater than $3\gamma/4$. The update takes $u=\pm\gamma/2$ with the opposite sign, so it decreases $\Phi$ by at least $\gamma^2/8$. Since $\Phi$ starts at most $nm$ and remains nonnegative, there cannot be $T_{\mathrm{cal}}$ such updates. Hence, if all estimates are accurate, the procedure satisfies the stopping condition within the iteration limit. At that iteration,
\[
  \|p-x\|_1\le d(\gamma+\gamma/4)=5\eta/16<\eta.
\]
Together with the preceding accuracy bound, this proves that calibration succeeds with probability at least $1-\rho$.

Every weight returned by the procedure has the exact rational form
\[
  \lambda_e=(1+\gamma/2)^{u_e}(1-\gamma/2)^{v_e},
  \qquad u_e+v_e\le T_{\mathrm{cal}}.
\]
Its encoding length is polynomial even if some empirical estimates are inaccurate. This proves \cref{lem:nsw-finite-matching} and supplies the finite Gibbs distributions whose marginals converge to $x$ in the proof of \cref{lem:nsw-matching}.

\subsection{Stability of the rounding guarantee}
\label[appendix]{subsec:nsw-stability}

Let $\sigma$ be the encoding bound from \cref{subsec:encoding-convention}. For the rounding procedure in \cref{subsec:nsw-algorithm}, set
\begin{equation}\label{eq:nsw-eta}
  \eta=\frac1{64(\sigma+1)(m+1)^3}.
\end{equation}
Use calibration failure bound $\rho=\eta$ for the expectation guarantee, or $\rho=\min\{\eta,\delta/2\}$ for confidence $1-\delta$. In each trial, sample a matching from a distribution within total-variation distance $\eta$ of the calibrated Gibbs distribution.

\paragraph{The total squared loss.}
Fix a successfully calibrated weight vector, and let $x'$ be its exact edge marginals. Write $b'_j=\sum_i x'_{ij}$ and $f'_j=1-b'_j$. Then $|b'_j-b_j|\le\eta$ and $\sum_jb'_j=n$. Throughout this paragraph, $B$ is the set of matched items under the exact Gibbs distribution for these weights.

Call a configuration exceptional if its tail contains an item with $f_j<2\eta$. For each such item, \eqref{eq:nsw-tail-capacity} bounds the total mass of configurations containing it in their tails by $f_j<2\eta$. Summing over at most $m$ items shows that exceptional configurations have total mass at most $2m\eta$. Subadditivity gives $v_i(S)\le m\nu_{i,S}\le mD_{i,S}(B)$, so their total contribution to the squared-loss bound is at most
\[
  2m\eta\cdot m^2=2m^3\eta\le1.
\]
For every item in a nonexceptional tail, we have $f'_j\ge f_j-\eta\ge f_j/2>0$. Thus, with $P'_{i,S}=\sum_{j\in T_{i,S}}b'_j/f'_j$, the residual-capacity bound gives
\[
  \sum_{(i,S)\ \text{nonexceptional}}y_{i,S}P'_{i,S}
    \le\sum_{j:\,f_j\ge2\eta}\frac{b'_j}{f'_j}f_j
    \le2\sum_jb'_j=2n.
\]
The negative-correlation inequalities hold for this exact Gibbs distribution with item marginals $b'_j$. We can therefore apply \cref{lem:nsw-squared-loss} to the nonexceptional configurations and sum the resulting bounds as in the proof of \cref{lem:nsw-aggregate}. Adding the contribution of the exceptional configurations gives
\begin{equation}\label{eq:nsw-finite-aggregate}
  \sum_{i,S}y_{i,S}\E_B
    \left(\frac{v_i(S)}{D_{i,S}(B)}\right)^2\le Cn.
\end{equation}
This distinction is used only in the analysis; the algorithm rounds the full configuration solution.

\paragraph{The singleton buffer.}
The coupling with a reference set in \cref{lem:nsw-matching}, including its one-item difference bound, holds for every fixed Gibbs distribution. The proof of \cref{lem:nsw-realize} therefore still applies, except that agent $i$'s matched item now has distribution $x'_i$ instead of $x_i$. Condition on the reference set, the guiding-tree rounding runs, and the coloring. As the matched item varies, the logarithmic quantity $Z^*$ in \eqref{eq:nsw-readout} lies in $[-\sigma\ln2,\sigma\ln2]$. Replacing the distribution $x'_i$ of that item by $x_i$ changes the conditional expectation of $Z^*$ by at most $2\sigma\ln2\,\mathrm{TV}(x_i,x'_i)$, where $\mathrm{TV}$ denotes total-variation distance. With distribution $x_i$, the singleton value has the distribution required by \cref{lem:nsw-geometric}. Since
\[
  \sum_iw_i\mathrm{TV}(x_i,x'_i)
    \le\tfrac12\|x-x'\|_1\le\eta/2,
\]
applying that lemma and summing over the agents gives
\begin{equation}\label{eq:nsw-finite-realize}
  \E\ln\mathrm{NSW}_w(A)
    \ge\sum_{i,S}w_i y_{i,S}\E_B\ln D_{i,S}(B)
       -C_0-\sigma\ln2\,\eta.
\end{equation}
Combining \eqref{eq:nsw-finite-aggregate} with the Jensen argument in \eqref{eq:nsw-entropy-average} yields the bound on expected logarithmic welfare for the exact Gibbs distribution of every successfully calibrated weight vector.

\paragraph{Sampling errors and running time.}
Every output of the implementation gives each agent a head of positive value. Its logarithmic welfare therefore always lies in $[-\sigma\ln2,\sigma\ln2]$. Replacing exact Gibbs sampling by sampling with total-variation error at most $\eta$ changes the expected logarithmic welfare by at most $2\sigma\ln2\,\eta$. Calibration fails with probability $\rho\le\eta$, contributing at most another $2\sigma\ln2\,\eta$ to the loss. Together with the marginal error in \eqref{eq:nsw-finite-realize}, these losses total $O(\sigma\eta)=O(1)$.

The limit on tree exploration is already included in \cref{lem:mm-service}. We also limit its rational label-sampling calls as in \cref{subsec:rational-sampling}, so that the total abort probability across all guiding-tree rounding calls is at most $\eta$. Using the same random bits couples these calls with unlimited rejection sampling: the two procedures agree unless a call aborts. On an abort, return a fixed positive head matching. This adds at most $O(\sigma\eta)$ to the expected logarithmic loss. To generate an item's color, perform up to $K+1$ independent fair-bit tests. The first success determines its color; if all tests fail, leave the item uncolored.

All procedures run in worst-case polynomial time: $1/\eta$ is polynomial in $\sigma$ and $m$, there are $O(\log m)$ guiding-tree rounding calls, and calibration and matching sampling satisfy the bounds established above. Conditional on any successfully calibrated weight vector, the guarantees for each trial hold with the same absolute constants. This proves the finite implementation of \cref{thm:nsw-rounding} and justifies reusing one calibration across the trials in \cref{thm:nsw-approximation}.

\section{Computing a configuration solution for Nash social welfare}
\label[appendix]{app:nsw-producer}

Feng and Li~\cite{FL24} used approximate dual separation to solve the logarithmic configuration LP for additive valuations. Feng, Hu, Li, and Zhang~\cite{FHLZ25} implemented this approach for submodular valuations using a knapsack approximation. To prove \cref{lem:nsw-producer} for subadditive valuations, we use the demand-query knapsack routine of \cref{lem:demand-budget} and the exact value-evaluation procedure in \cref{app:demand-values}. We approximate the logarithmic coefficients by rationals and track the resulting error when recovering feasible primal and dual solutions.

\begin{proof}[Proof of \cref{lem:nsw-producer}]
  Retain the columns corresponding to the positive singleton matching found in \cref{lem:nsw-positive}. For each agent $i$ and each set $S$ with $v_i(S)>0$, fix a deterministic rational approximation satisfying
  \[
    \bigl|\widehat\ell_i(S)-w_i\ln v_i(S)\bigr|\le w_i/64.
  \]
  Reuse this approximation whenever the same pair $(i,S)$ appears. All generated configuration constraints then belong to a single rational LP, in which $w_i\ln v_i(S)$ is replaced by $\widehat\ell_i(S)$. Under the bit bound $\sigma$, these coefficients have polynomial encoding length.

  \emph{Approximate separation.}
  Consider a rational candidate $(z,p)$ with $p\ge0$. For each agent $i$ and each target $V=2^k$ with $-\sigma\le k\le\sigma$, compute a rational $q$ and define $\gamma$ so that
  \[
    |q-\ln(V/3)|\le1/32,
    \qquad \gamma=w_iq-z_i-w_i/16.
  \]
  If $\gamma\ge0$, apply the routine of \cref{lem:demand-budget} with item costs $p$, budget $\gamma$, and target $V$. If the returned set $D$ has value at least $V/3$, then $p(D)\le\gamma$ and
  \[
    \widehat\ell_i(D)-z_i-p(D)
      \ge w_i\left(\frac1{16}-\frac1{32}-\frac1{64}\right)>0.
  \]
  Thus, the constraint indexed by $(i,D)$ is violated in the fixed rational dual.

  Now suppose that the original dual constraint in \eqref{eq:nsw-dual} for some pair $(i,S)$ is violated by more than $2w_i$. Choose a target $V=2^k$ satisfying $v_i(S)/2<V\le v_i(S)$. The bit bound ensures that this target is among those considered. The corresponding budget satisfies
  \[
    \gamma-p(S)
      >w_i\left(2-\ln6-\frac3{32}\right)>0.
  \]
  Hence $S$ lies within the budget and has value at least $V$, so the routine returns a set of value at least $V/3$ and identifies a violated rational constraint. Consequently, if no violated constraint is found, replacing each $z_i$ by $z_i+2w_i$ makes every original dual constraint feasible. Since the weights sum to one, this shift increases the dual objective by exactly two.

  \emph{Recovering an exactly feasible primal solution.}
  Include the positive singleton-matching columns in every restricted primal, ensuring that it is feasible and bounded. Its constraint matrix has entries in $\{-1,0,1\}$, and its objective coefficients have uniformly polynomial encoding length. Standard bounds on basic solutions therefore give a common box with polynomially encoded bounds that contains an optimal dual solution for every restricted problem. The box can be chosen before any additional configurations are generated.

  We perform bisection on the unshifted dual objective level, using the rational feasibility algorithm~\cite[Theorem~6.4.1]{GLS88} and the recorded-cut argument of \cref{lem:configuration-producer}. At an accepted level, the separator returns a candidate that becomes feasible for the original dual after adding $2w_i$ to each $z_i$. At a rejected level, the infeasibility certificate consists of violated configuration constraints from the fixed rational LP and the explicit constraints. Since the box contains an optimal dual solution for every restricted problem, duality implies that the rational primal restricted to the recorded columns has optimal objective above the rejected level.

  We start with the accepted upper level $\sigma+1$, using $z_i=w_i(\sigma+1)$ and $p=0$; enlarge the box to include this candidate if necessary. The retained singleton columns certify rejection at the lower level $-\sigma-2$. Continue the bisection until the accepted and rejected levels differ by at most $1/64$. This requires only polynomially many separation calls.

  Collect all generated columns and solve the restricted rational primal exactly. Its objective is at least the final rejected level. By the row constraints \eqref{eq:mm-row} and $\sum_iw_i=1$, replacing the rational logarithmic coefficients by their true values changes the objective by at most $1/64$. The shifted dual is feasible for the original program and has objective at most the final accepted level plus two. Combining these bounds gives
  \[
    0\le U(z,p)-L(y)\le2+\frac2{64}<3,
  \]
  where $(z,p)$ now denotes the shifted dual solution. The lower bound follows from weak duality. Finally, discard all zero-mass columns. The remaining columns form a positive configuration solution with polynomial-size support satisfying \eqref{eq:mm-row} and \eqref{eq:mm-capacity} exactly, as required.
\end{proof}

\section{Computational conventions}\label[appendix]{app:implementation-conventions}

\subsection{Encoding}\label[appendix]{subsec:encoding-convention}
The finite-encoding convention in \cref{sec:prelim} gives a known bound $\sigma$ such that every value $v_i(S)$ is rational, with numerator and denominator at most $2^\sigma$. In particular, every positive value lies in $[2^{-\sigma},2^\sigma]$. Our running-time bounds are polynomial in $\sigma$ and the encoding lengths of the other numerical data, including the weights, configuration masses, and oracle-returned additive coefficients.

\subsection{Sampling from explicit rational distributions}\label[appendix]{subsec:rational-sampling}
In the rounding analyses, we sample exactly from explicitly given rational distributions. We describe how to implement these calls in worst-case polynomial time with an arbitrarily small additional failure probability.

For a distribution whose probabilities have common denominator $Q$, rejection sampling draws $\lceil\log_2Q\rceil$ independent uniform random bits and repeats if the resulting integer is at least $Q$. The accepted integer is uniform on $\{0,\ldots,Q-1\}$ and can be mapped to an outcome according to the prescribed probabilities. Each trial succeeds with probability at least $1/2$, and its running time is polynomial in the encoding length of the distribution.

For a procedure making at most $s\ge1$ sampling calls and any $0<\delta<1$, we cap each call at $\lceil\log_2(s/\delta)\rceil$ trials and abort if none succeeds. By the union bound, the abort probability is at most $\delta$. The truncated and untruncated procedures can be coupled to agree unless an abort occurs, so truncation adds at most $\delta$ to the failure probability. Each call uses $O(1+\log(s/\delta))$ trials.